\documentclass[review,10pt]{elsarticle}

\usepackage{lineno}
\modulolinenumbers[5]

\usepackage{hyperref} 
\hypersetup{ 
  colorlinks   = true, 
  urlcolor     = blue, 
  linkcolor    = blue, 
  citecolor    = blue, 
} 
\usepackage[fleqn]{amsmath}
\usepackage{amsfonts}
\usepackage{amssymb}
\usepackage{amsthm}

\usepackage{pifont}
\usepackage{natbib}
\usepackage{geometry}
\usepackage{graphicx}
\usepackage{txfonts}

\usepackage[utf8]{inputenc}
\usepackage{caption}
\usepackage{subcaption}
\usepackage[usenames, dvipsnames]{color}
\usepackage{color,soul}

\journal{International Journal of Multiphase Flow}

\biboptions{sort&compress}

\begin{document}

\begin{frontmatter}

\title{Self-similar solution of a supercritical two-phase laminar mixing layer}

\author{Jordi Poblador-Ibanez\fnref{myfootnote1}\corref{mycorrespondingauthor}}
\ead{poblador@uci.edu}
\author{Branson W. Davis\fnref{myfootnote2}}
\author{William A. Sirignano\fnref{myfootnote3}}
\address{University of California, Irvine, CA 92697-3975, United States}
\fntext[myfootnote1]{Graduate Student Researcher, Department of Mechanical and Aerospace Engineering.}
\fntext[myfootnote2]{Undergraduate Student Researcher, Department of Mechanical and Aerospace Engineering.}
\fntext[myfootnote3]{Professor, Department of Mechanical and Aerospace Engineering.}


\cortext[mycorrespondingauthor]{Corresponding author}


\begin{abstract}
Previous works for a liquid suddenly contacting a gas at a supercritical pressure show the coexistence of both phases and the generation of diffusion layers on both sides of the liquid-gas interface due to thermodynamic phase equilibrium. A related numerical study of a laminar mixing layer between the liquid and gas streams in the near field of the splitter plate suggests that mass, momentum and thermal diffusion layers evolve in a self-similar manner at very high pressures. In this paper, the high-pressure, two-phase, laminar mixing-layer equations are recast in terms of a similarity variable. A liquid hydrocarbon and gaseous oxygen are considered. Freestream conditions and proper matching conditions at the liquid-gas interface are applied. To solve the system of equations, a real-fluid thermodynamic model based on the Soave-Redlich-Kwong equation of state is selected. A comparison with results obtained by directly solving the laminar mixing-layer equations shows the validity of the similarity approach applied to non-ideal two-phase flows. Even when the gas is hotter than the liquid, condensation can occur at high pressures while heat conducts into the liquid. Finally, a generalized correlation is proposed to represent the evolution of the mixing layer thickness for different problem setups.
\end{abstract}

\begin{keyword}
similarity \sep laminar mixing layer \sep supercritical pressure \sep phase equilibrium \sep phase change \sep two-phase
\end{keyword}

\end{frontmatter}


\setlength\abovedisplayshortskip{0pt}
\setlength\belowdisplayshortskip{-5pt}
\setlength\abovedisplayskip{-5pt}

\section{Introduction}
\label{sec:intro}




Combustion chambers used in many engineering applications (e.g., power units, gas turbines or liquid-propellant rockets) are designed to operate at elevated pressures. That is, a thermodynamic regime is sought where less dissociation of the reaction products occurs and a better combustion efficiency and specific energy conversion are obtained. In many situations, the chamber pressure is well above the critical pressure of the liquid fuel that is being injected. Well-known fuels are based on hydrocarbon mixtures (e.g., diesel fuel, Jet A, RP-1) with critical pressures around 20 bar, while operating pressures may range from 25 to 40 bar in diesel engines or gas turbines and 70 to 200 bar in rocket engines. \par 

The performance of the combustion reaction also depends on how fast the liquid fuel vaporizes and mixes with the surrounding oxidizer. Therefore, understanding the physical phenomena involved in this process is necessary for a proper design of the injectors' shape and distribution, combustion chamber size, etc. At subcritical pressures, a clear distinction exists between liquid and gas phases, which allows for extensive experimental studies. However, for near-critical or supercritical pressures, experimental studies show that a thermodynamic transition occurs where the liquid and gas present similar fluid properties near the liquid-gas interface, which is suddenly immersed in a variable-density layer and is rapidly affected by turbulence~\cite{mayer1996propellant,h1998atomization,mayer2000injection,chehroudi2002visual,chehroudi2002cryogenic,oschwald2006injection,segal2008subcritical,chehroudi2012recent,falgout2016gas}. 

Past works have described this behavior as a very fast transition of the liquid phase to a supercritical state, assuming that a two-phase behavior cannot be sustained under these thermodynamic conditions as suggested by experimental results where a gas-like turbulent structure is observed~\cite{spalding1959theory,rosner1967liquid}. But evidence of a two-phase behavior at supercritical pressures in multicomponent fluids exists based on a requirement of thermodynamic phase equilibrium at the liquid-gas interface~\cite{hsieh1991droplet,delplanque1993numerical,yang1994vaporization,sirignano1997selected,juanos2015thermodynamic,poblador2018transient}.
At supercritical pressures, phase equilibrium enhances the dissolution of the gas into the liquid phase, thus generating diffusion layers on both sides of the liquid-gas interface where strong variations of fluid properties occur. Mixture critical properties near the interface differ from pure fluid critical properties and, in general, the new critical pressure is higher than the chamber pressure. However, this does not seem to be a requirement to sustain two phases under very high pressures as discussed in Poblador-Ibanez and Sirignano~\cite{poblador2018transient}. Some authors suggest caution at pressures near the critical pressure of the mixture, where the liquid-gas interface enters a continuum region of a few nanometers thickness~\cite{dahms2013transition,dahms2015liquid}. Accordingly, a two-phase behavior cannot exist. Nevertheless, diffusion around the interface occurs rapidly enough to reach diffusion layer thicknesses of the order of micrometers while phase equilibrium is well established~\cite{poblador2018transient}. Thus, the interface may still be treated as a discontinuity under phase equilibrium with a clear jump in fluid properties across it, similar to how a compressive shock wave is considered as a discontinuity in compressible fluids. In that case, the non-equilibrium layer thickness for the shock is an order of magnitude or more greater than the layer thickness for phase non-equilibrium. \par 

There are further reasons that support the fact that liquid injection at supercritical pressures is still a two-phase problem. Surface tension forces are reduced at very high pressures due to both phases presenting similar densities at the interface. Therefore, the aerodynamic effects on the liquid breakup process are enhanced~\cite{yang2000modeling}. Moreover, mixing causes the liquid viscosity to drop to gas-like values near the interface, reducing viscous damping of surface instabilities~\cite{poblador2019axisymmetric}. Altogether, fast growing instabilities associated with smaller wavelengths develop at the interface, which may cause a very fast atomization of the liquid jet with a cloud of very small droplets around it. This reasoning is further supported by numerical results of incompressible liquid round jets and planar sheets injected into a gas. For similar liquid and gas densities with reduced surface tension, a fast breakup process is induced with early small droplet formation and enhanced radial development of the two-phase mixture~\cite{jarrahbashi2014vorticity,jarrahbashi2016early,zandian2017planar,zandian2018understanding,zandian2019length}. 
Traditional experimental techniques have difficulties with visual penetration of dense sprays due to the scattering caused by the large amount of small droplets and the intrinsic difficulties of variable-density fluids. Thus, it becomes difficult to distinguish between liquid and gas phases. Some new techniques have been able to penetrate dense sprays and show clear liquid structures (i.e., ligaments, lobes, bigger droplets) emerging from the liquid core, although they have not been tested yet in supercritical environments~\cite{minniti2018ultrashort,minniti2019femtosecond}. \par 

The present work does not address the hydrodynamic instabilities that occur farther downstream as part of the atomization process. It focuses on the initial laminar mixing between the injected liquid and the gas before substantial growth of surface instabilities or transition to turbulence. Understanding how the mixing between both streams evolves is crucial to understand the beginning of high-pressure atomization and can also help create models to use as initial conditions for more complex numerical studies. \par 

Davis et al.~\cite{davis2019development} present a two-dimensional numerical study involving the non-ideal two-phase laminar mixing layer equations and provide guidelines on the validity of these relations. Results are presented for a binary mixture where a liquid stream and a gas stream come together at the end of a splitter plate at various ambient pressures, ranging from subcritical to supercritical for the pure liquid. The liquid stream is initially composed by pure \textit{n}-decane and the gas stream is pure oxygen. The mixing layer thickness grows to a few micrometers in the liquid phase and to tens of micrometers in the gas phase with downstream distance from the end of the splitter plate. At the same time, the interface reaches a near-steady solution sufficiently far away from the splitter plate. Most importantly, each individual configuration (i.e., each set of boundary conditions) shows signs of a self-similar behavior of the mixing layer evolution, even when the defined mixing layer equations involve non-ideal terms and are coupled to a non-ideal thermodynamic model involving a real-gas equation of state, thermodynamic fundamental principles and various high-pressure correlations. A self-similar behavior in this high-pressure environment is also suggested in Poblador-Ibanez and Sirignano~\cite{poblador2018transient}, although in that work a non-ideal unsteady one-dimensional model is implemented. \par 

Reducing the system of partial differential equations (i.e., mixing layer equations) to a system of ordinary differential equations in terms of a similarity variable is of special interest, not only because a simplified mathematical model describing the same phenomena is obtained, but also because of the fundamental analysis implied in the transformation. Similarity solutions have always been sought in classical fluid mechanics (e.g., the Blasius solution for flow over a flat plate or the Falker-Skan wedge flows), usually limited to incompressible flows. Some transformations are also available for compressible flows or flows with variable fluid properties, although simplifications are made to obtain a generic self-similar model (e.g., use of ideal-gas model). In general, these approaches are well documented in textbooks~\cite{white2006viscous,williams2018combustion}. When analyzing compressible or variable-density laminar shear layers, the focus is usually limited to single-phase configurations under ideal-gas assumptions with momentum and thermal mixing. However, some works also include species mixing in their analysis~\cite{kennedy1994self}. \par 

In views of the discussed numerical results~\cite{poblador2018transient,davis2019development}, the non-ideal two-phase laminar mixing layer equations are recast in terms of a similarity variable, thus forming a system of ordinary differential equations. Boundary conditions and liquid-gas interface matching relations are provided, which allow for mass, momentum and energy transfer across the interface to occur. Then, the self-similar model is coupled to a non-ideal thermodynamic model to close the system of equations. The self-similar solution is compared against the results from Davis et al.~\cite{davis2019development} with different problem configurations to assess the validity of the self-similar model. Finally, a proper non-dimensional scaling is presented, which provides a semi-collapse of the self-similar solutions for different sets of boundary conditions. This allows the development of a correlation based on freestream and interface conditions which is able to estimate mixing layer thicknesses with sufficient accuracy.

\section{Non-ideal two-phase laminar mixing layer equations}
\label{sec:mixinglayer}

The behavior of two parallel streams coming together at the edge of a splitter plate can be modeled under the boundary-layer approximation after a transitional region~\cite{white2006viscous}. For a sufficiently large Reynolds number, the flow begins to develop with the transverse velocity being much smaller than the streamwise velocity, \(v\ll u\), and variations in the \(x\)-direction become negligible compared to variations in \(y\) (i.e., \(\partial()/\partial x\ll\partial()/\partial y\)). The transverse pressure gradient, \(\partial p/\partial y\), is also negligible. Therefore, the transverse velocity solution is directly obtained from the continuity equation and not from the transverse momentum equation~\cite{he2017sharp,poblador2018transient}. For free-shear, low-Mach number flows at high pressures without confining walls, pressure can be assumed constant (\(Dp/Dt=\partial p/\partial x = \partial p/\partial y = 0\)). Furthermore, viscous dissipation and kinetic energy can be safely neglected in the energy equation. \par 

\begin{figure}[h!]
\centering
\includegraphics[width=0.8\linewidth]{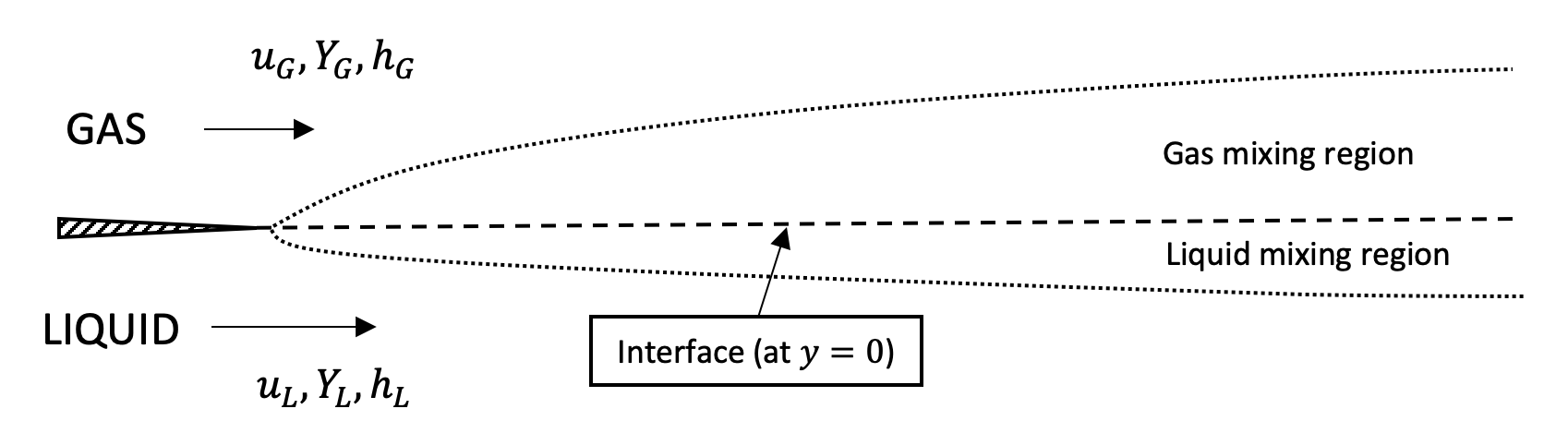}
\caption{Sketch of the mixing layer problem between a liquid stream and a gas stream. The liquid-gas interface or dividing streamline is assumed to be fixed at \(y=0\) m as discussed in \cite{davis2019development}.}
\label{fig:domain}
\end{figure}

Both fluid streams are initially pure components (e.g., a pure liquid hydrocarbon and a pure gas, such as oxygen). The governing equations are cast in non-conservative form and in terms of the mass fraction of one mixture component, \(Y_1\), where \(Y_1+Y_2=1\). Under these conditions, the steady-state governing equations describing the two-phase laminar mixing layer development of two non-ideal fluids (i.e., a liquid stream and a gas stream) are the continuity equation, Eq.~(\ref{eqn:gc}), streamwise momentum equation, Eq.~(\ref{eqn:xmom_ncons}), species continuity equation, Eq.~(\ref{eqn:sc_ncons}), and energy equation, Eq.~(\ref{eqn:ene_ncons}). For simplicity, a Fickian form is used to represent mass diffusion in Eqs.~(\ref{eqn:sc_ncons}) and~(\ref{eqn:ene_ncons}) since a binary mixture is considered. The system of equations is given as

\begin{equation}
\label{eqn:gc}
\frac{\partial}{\partial x}(\rho u)+\frac{\partial}{\partial y}(\rho v)=0
\end{equation}

\begin{equation}
\label{eqn:xmom_ncons}
\rho u \frac{\partial u}{\partial x} + \rho v \frac{\partial u}{\partial y} = \frac{\partial}{\partial y} \bigg(\mu \frac{\partial u}{\partial y} \bigg)
\end{equation}

\begin{equation}
\label{eqn:sc_ncons}
\rho u \frac{\partial Y_1}{\partial x} + \rho v \frac{\partial Y_1}{\partial y} = \frac{\partial}{\partial y} \bigg(\rho D \frac{\partial Y_1}{\partial y}\bigg)
\end{equation}

\begin{equation}
\label{eqn:ene_ncons}
\rho u \frac{\partial h}{\partial x} + \rho v \frac{\partial h}{\partial y} = \frac{\partial}{\partial y} \bigg(\frac{\lambda}{c_p} \frac{\partial h}{\partial y}\bigg) + \frac{\partial}{\partial y} \Bigg[\bigg(\rho D - \frac{\lambda}{c_p}\bigg)(h_1-h_2)\frac{\partial Y_1}{\partial y}\Bigg]
\end{equation}

where \(\rho\), \(u\) and \(v\) are the mixture density, streamwise velocity and transverse velocity, respectively. \(h\) is the mixture specific enthalpy and \(h_1\) and \(h_2\) are the partial specific enthalpies of each species of the binary mixture. Other mixture fluid properties are the mixture dynamic viscosity, \(\mu\), the mass diffusion coefficient, \(D\), the thermal conductivity, \(\lambda\), and the specific heat at constant pressure, \(c_p\). Furthermore, the energy equation is written as an enthalpy transport equation by using the relation \(\lambda\nabla T=(\lambda/c_p)\nabla h - \sum_{i=1}^{N} (\lambda/c_p)h_i\nabla Y_i\). Note that since the problem is diffusion-driven, fluid properties will vary as species and energy diffuse. \par

These steady-state equations are valid as long as the flow is laminar and surface instabilities are negligible. Davis et al.~\cite{davis2019development} provide some guidance for oxygen-hydrocarbon mixtures at various ambient pressures. The quasi-parallel laminar region may exist up to a distance from the splitter plate of the order of \(\mathcal{O}(10^{-2}\) m). The mean flow velocity is fixed at 10 m/s and the velocity difference between both streams is varied in the range of \(\Delta u = 0.3-5\) m/s to limit the maximum Reynolds number at \(\bar{L}=0.01\) m to \(Re_{\bar{L}}=10,000\). \(Re_{\bar{L}}\) is defined as~\cite{davis2019development}

\begin{equation}
\label{eqn:reynolds}
Re_{\bar{L}}=\frac{\rho_G \bar{L} \Delta u}{\mu_G}
\end{equation}

\noindent
where the subscript \(G\) refers to gas freestream values. Analogously, the subscript \(L\) will refer to liquid freestream values. \par 

With this analysis, one of the main goals is to show the existence of a distinct two-phase behavior before hydrodynamic instabilities dominate.

\section{Similarity transformation}
\label{sec:similarity}

The system of partial differential equations (PDEs) describing the problem (Eqs.~(\ref{eqn:gc}),~(\ref{eqn:xmom_ncons}),~(\ref{eqn:sc_ncons}) and~(\ref{eqn:ene_ncons})) can be transformed into a system of ordinary differential equations (ODEs) by using a proper similarity variable, \(\eta\). However, the transformation requires an intermediate step due to variable density. The mapping follows \((x,y)\rightarrow(\bar{x},z)\rightarrow\eta\). \par 

The first transformation is defined as

\begin{equation}
\bar{x} = x \quad ; \quad z = \int_{0}^{y} \rho dy'
\end{equation}

\noindent
where \(y'\) is a dummy variable for integration purposes and \(z\) is a density-weighted transverse coordinate following the so-called Howarth-Dorodnitsyn transformation~\cite{williams2018combustion}. This type of transformation was independently introduced in 1942 by Dorodnitsyn~\cite{dorodnitsyn1942boundary} and later in 1948 by Howarth~\cite{howarth1948concerning}. Other authors followed with similar work~\cite{stewartson1949correlated,lees1956laminar}. \par

As shown in Figure~\ref{fig:domain}, the interface between the two phases is placed at \(y=0\). To transform the original system of equations from \((x,y)\rightarrow(\bar{x},z)\), the following relations between partial derivatives are obtained

\begin{equation}
\label{eqn:partialx}
\frac{\partial}{\partial x}() = \frac{\partial}{\partial \bar{x}}() + \bigg(\int_{0}^{y} \frac{\partial \rho}{\partial x} dy'\bigg) \frac{\partial}{\partial z}()
\end{equation}

\begin{equation}
\label{eqn:partialy}
\frac{\partial}{\partial y}() = \rho \frac{\partial}{\partial z}()
\end{equation}

Defining the transformed transverse velocity, \(w\), as

\begin{equation}
\label{eqn:modv}
w = \rho v + u \int_{0}^{y}\frac{\partial \rho}{\partial x}dy'
\end{equation}

\noindent
the velocity components can be related to the stream function for compressible flow, \(\Psi(\bar{x},z)\), as 

\begin{equation}
\label{eqn:vel_psi}
u = \frac{\partial \Psi}{\partial z} \quad ; \quad w = -\frac{\partial \Psi}{\partial \bar{x}}
\end{equation}

Here, \(w\) does not have units of velocity and can be considered as a convenient variable in the solution process. \par 

This arrangement yields a system of PDEs in the \((\bar{x},z)\) space in terms of the compressible stream function, Eqs.~(\ref{eqn:gc_trans1})-(\ref{eqn:ene_trans1}). Note the continuity equation, Eq.~(\ref{eqn:gc_trans1}), is identically satisfied. 

\begin{equation}
\label{eqn:gc_trans1}
\frac{\partial^2 \Psi}{\partial \bar{x} \partial z} - \frac{\partial^2 \Psi}{\partial z \partial \bar{x}} = 0
\end{equation}

\begin{equation}
\label{eqn:xmom_trans1}
\frac{\partial \Psi}{\partial z} \frac{\partial^2 \Psi}{\partial \bar{x} \partial z} - \frac{\partial \Psi}{\partial \bar{x}} \frac{\partial^2 \Psi}{\partial z^2} = \frac{\partial}{\partial z} \bigg(\rho \mu \frac{\partial^2 \Psi}{\partial z^2} \bigg)
\end{equation}

\begin{equation}
\label{eqn:sc_trans1}
\frac{\partial \Psi}{\partial z} \frac{\partial Y_1}{\partial \bar{x}} - \frac{\partial \Psi}{\partial \bar{x}} \frac{\partial Y_1}{\partial z} = \frac{\partial}{\partial z} \bigg(\rho^2 D \frac{\partial Y_1}{\partial z}\bigg)
\end{equation}

\begin{equation}
\label{eqn:ene_trans1}
\frac{\partial \Psi}{\partial z} \frac{\partial h}{\partial \bar{x}} - \frac{\partial \Psi}{\partial \bar{x}} \frac{\partial h}{\partial z} = \frac{\partial}{\partial z} \bigg(\frac{\rho \lambda}{c_p} \frac{\partial h}{\partial z}\bigg) + \frac{\partial}{\partial z} \Bigg[\bigg(\rho^2 D - \frac{\rho \lambda}{c_p}\bigg)(h_1-h_2)\frac{\partial Y_1}{\partial z}\Bigg]
\end{equation}

After this intermediate step, the self-similar transformation is taken by defining the similarity variable, \(\eta\), as in the Blasius solution for a flat plate~\cite{white2006viscous}, but where \(x\) is substituted by \(\bar{x}\) and \(y\) is substituted by \(z\) (see Eq.~(\ref{eqn:simvar})).

\begin{equation}
\label{eqn:simvar}
\eta = \frac{z}{\sqrt{2\bar{x}}}
\end{equation}

The transformation from \((\bar{x},z)\rightarrow\eta\) requires the following relations between partial derivatives

\begin{equation}
\label{eqn:partialz}
\frac{\partial}{\partial z}() = \frac{1}{\sqrt{2\bar{x}}}\frac{\partial}{\partial \eta}()
\end{equation}

\begin{equation}
\frac{\partial}{\partial\bar{x}}() = \frac{\partial}{\partial \bar{\bar{x}}}() - \frac{1}{2}\frac{\eta}{\bar{x}}\frac{\partial}{\partial \eta}()
\end{equation}

\noindent
where \(\bar{\bar{x}}=\bar{x}=x\). \par 

A self-similar solution exists if \(\Psi/\sqrt{\bar{x}}\), \(Y_1\) and \(h\) depend only on \(\eta\). The following expressions are assumed, where

\begin{equation}
\Psi = \sqrt{2\bar{x}}f(\eta) \quad ; \quad Y_1 = Y(\eta) \quad ; \quad h = h(\eta)
\end{equation}

From Eq.~(\ref{eqn:vel_psi}), it reads that 

\begin{equation}
\label{eqn:vel_eta}
u=\frac{d f}{d \eta} = f'(\eta) \quad ; \quad w = \frac{1}{\sqrt{2\bar{x}}}\bigg(\eta f'(\eta)-f(\eta)\bigg)
\end{equation}

\noindent
where the \((\) \()'\) operator applied to any variable \(a\) (i.e., \(a'\)) means differentiation of \(a\) with respect to \(\eta\) or \(da/d\eta\). Second and third derivatives follow the same notation. For the sake of simplicity, single dependence on \(\eta\) is implicit in the equations to follow. \par 


Rearranging Eqs.~(\ref{eqn:xmom_trans1})-(\ref{eqn:ene_trans1}), a system of ODEs can be written as

\begin{equation}
\label{eqn:xmom_trans2}
(\rho \mu)f''' + (\rho \mu)'f'' + ff'' = 0
\end{equation}

\begin{equation}
\label{eqn:sc_trans2}
(\rho^2 D)Y'' + (\rho^2 D)'Y' + fY' = 0
\end{equation}

\begin{equation}
\label{eqn:ene_trans2}
\bigg(\frac{\rho \lambda}{c_p}\bigg) h'' + \bigg(\frac{\rho \lambda}{c_p}\bigg)'h' + \bigg(\rho^2 D - \frac{\rho \lambda}{c_p}\bigg)(h_1-h_2)Y'' + \Bigg[\bigg(\rho^2 D - \frac{\rho \lambda}{c_p}\bigg)(h_1-h_2)\Bigg]'Y' + fh' = 0
\end{equation}

\subsection{Evaluation of the transverse velocity}
\label{subsec:transvel}

The determination of the transverse velocity, \(v\), across the mixing layer from the self-similar solution in variable-density mixing layers has been given little to no attention in previous works and textbooks. Generally, the interest is only focused on solving the velocity profile in the main flow direction, together with the mixing profiles of other variables such as density and temperature. Some works addressing this issue are those of Kennedy and Gatski~\cite{kennedy1994self} and Libby and Liu~\cite{libby1968some}. However, relations for \(v\) are presented without further development. On the other hand, Pruett~\cite{pruett1993accurate} offers some insights in the process to develop such relations. Nevertheless, the analysis performed on the boundary layer of an axisymmetric body complicates the final result.  \par

In the present study, a straightforward development of an equation for \(v\)  based on \(\eta\) and downstream location is presented in a form also shown in Sirignano~\cite{sirignano2020}. The latter work develops a self-similar model for mixing in a variable-density laminar shear layer with imposed counterflow. \par 

Eqs.~(\ref{eqn:partialy}) and~(\ref{eqn:partialz}) provide a differential relation between \(y\) and \(\eta\) for a fixed downstream position or constant \(\bar{x}\) which can be integrated to relate the transverse location to the similarity variable as

\begin{equation}
\label{eqn:yvseta}
dy = \frac{\sqrt{2\bar{x}}}{\rho}d\eta \quad ; \quad y = \sqrt{2\bar{x}}\int_{0}^{\eta}\frac{d\eta'}{\rho(\eta')} = \sqrt{2\bar{x}}\tilde{I}(\eta)
\end{equation}

\noindent
where \(\eta'\) is a dummy variable for integration purposes and \(\tilde{I}(\eta)\equiv\int_{0}^{\eta}d\eta'/\rho(\eta')\). \par 

By combining expressions of the transformed transverse velocity, \(w\), in \((x,y)\) and \((\bar{x},\eta)\) coordinates (Eqs.~(\ref{eqn:modv}) and~(\ref{eqn:vel_eta})), the following relation is obtained

\begin{equation}
\label{eqn:findv1}
w\sqrt{2\bar{x}}=\rho v \sqrt{2\bar{x}} + u\sqrt{2\bar{x}}\frac{\partial z}{\partial x} = \eta f' - f
\end{equation}

\noindent
which shows that \(w\) (and thus \(v\)) are self-similar when multiplied by the square root of \(\bar{x}\) or downstream position. In the previous equation, the equality \(\int_{0}^{y}(\partial \rho/\partial x) dy'= \partial(\int_{0}^{y}\rho dy')/\partial x = \partial z/\partial x\) is used. \par 

Substitution of \(\partial z/\partial x\) in Eq.~(\ref{eqn:findv1}) is found by differentiating \(dy/dx\), \(dz/dx\) and \(dz/d\bar{x}\) at constant \(y\) using Eq.~(\ref{eqn:partialx}).

\begin{equation}
\label{eqn:findv2}
\frac{dy}{dx}=0=\frac{dy}{d\bar{x}}+\frac{1}{\rho}\frac{dz}{dx}=\tilde{I}\frac{d}{d\bar{x}}(\sqrt{2\bar{x}})+\frac{\sqrt{2\bar{x}}}{\rho}\frac{d\eta}{d\bar{x}}+\frac{1}{\rho}\frac{dz}{dx}
\end{equation}

\begin{equation}
\label{eqn:findv3}
\frac{dz}{dx} = \frac{dz}{d\bar{x}} + \frac{dz}{dx} \quad \rightarrow \quad \frac{dz}{d\bar{x}} = 0  
\end{equation}

\begin{equation}
\label{eqn:findv4}
\frac{dz}{d\bar{x}} = 0 = \sqrt{2\bar{x}}\frac{d\eta}{d\bar{x}}+\eta\frac{d}{d\bar{x}}(\sqrt{2\bar{x}})
\end{equation}

Combination of Eqs.~(\ref{eqn:findv2}) and~(\ref{eqn:findv4}) yields a relation for \(dz/dx\), Eq.~(\ref{eqn:findv5}). This relation can be substituted back into Eq.~(\ref{eqn:findv1}) to find an equation for the transverse velocity \(v\) as a function of self-similar variables and downstream position, Eq.~(\ref{eqn:findv6}).

\begin{equation}
\label{eqn:findv5}
\frac{dz}{dx}=\big(\eta-\tilde{I}\rho\big)\frac{d}{d\bar{x}}(\sqrt{2\bar{x}})
\end{equation}

\begin{equation}
\label{eqn:findv6}
\rho v \sqrt{2\bar{x}} = \tilde{I}\rho f'-f
\end{equation}

\subsection{Comments on the thermodynamic model}
\label{subsec:thermo}

Another requirement needed to achieve similarity relates to the thermodynamic model used to evaluate fluid properties and transport coefficients. For these variables to depend only on \(\eta\), the thermodynamic model must ultimately depend on pressure, temperature (i.e., enthalpy) and mixture composition. The self-similar model presented here relies on constant pressure everywhere in the domain, while temperature and mixture composition depend only on \(\eta\). \par 

The thermodynamic model implemented in this work is based on a volume-corrected Soave-Redlich-Kwong (SRK) cubic equation of state~\cite{lin2006volumetric}, which is able to represent non-ideal fluid states. The volumetric correction is needed to obtain accurate density predictions of high-density fluids (i.e., liquids) because the original SRK equation of state~\cite{soave1972equilibrium} presents liquid density errors of up to 20\% when compared to experimental values~\cite{yang2000modeling,prausnitz2004thermodynamics}. The modified SRK equation of state is expressed in terms of the compressibility factor, \(Z\), as

\begin{equation}
\label{eqn:SRKEoS}
Z^3+(3C-1)Z^2+\big[C(3C-2)+A-B-B^2\big]Z+C(C^2-C+A-B-B^2)-AB=0
\end{equation}

\noindent
with

\begin{equation}
Z = \frac{p}{\rho RT} \quad ; \quad A = \frac{a(T)p}{R_{u}^{2}T^2} \quad ; \quad B = \frac{bp}{R_uT} \quad ; \quad C = \frac{c(T)p}{R_uT}
\end{equation}

\noindent
where \(a(T)\) is a temperature-dependent cohesive energy parameter, \(b\) represents a volumetric parameter related to the space occupied by the molecules and \(c(T)\) is a temperature-dependent volume correction. \(R\) and \(R_u\) are the specific gas constant and the universal gas constant, respectively. The solution of this cubic equation provides the density of the fluid mixture, \(\rho\). \par 

The volume-corrected SRK equation of state is used, together with high-pressure correlations, to evaluate fluid properties and transport coefficients~\cite{poling2001properties,chung1988generalized,leahy2007unified} (e.g., viscosity). Details on the development and implementation of this thermodynamic model can be found in Davis et al.~\cite{davis2019development}.

\section{Boundary conditions and interface matching relations}
\label{sec:boundary_cond}

The governing equations representing the non-ideal two-phase laminar mixing layer have been reduced to a system of ODEs, Eqs.~(\ref{eqn:xmom_trans2})-(\ref{eqn:ene_trans2}), which depends only on the similarity variable, \(\eta\). To solve this system of equations, proper boundary conditions must be imposed. These can be divided into freestream conditions for each individual fluid and interface matching relations between both fluid streams. \par

The self-similar transformation is defined such that the liquid-gas interface, which is denoted by \(\Gamma\),  does not move (\(V_\Gamma = 0\)) and is located at \(y=z=\eta=0\). The justification for this approximation has been shown by Davis et al.~\cite{davis2019development}. Positive values of \(\eta\) define the gas stream, while negative values of \(\eta\) represent the liquid stream. Following this definition, freestream boundary conditions for \(f\), \(Y\) and \(h\) become

\begin{equation}
\label{eqn:BC_freeplus}
f'(+\infty) = u_G \quad ; \quad Y(+\infty) = Y_G \quad ; \quad h(+\infty) = h_G
\end{equation}

\noindent
and

\begin{equation}
\label{eqn:BC_freeminus}
f'(-\infty) = u_L \quad ; \quad Y(-\infty) = Y_L \quad ; \quad h(-\infty) = h_L
\end{equation}

At \(\pm \infty\), both streams are pure components. Thus, if \(Y_1\) is chosen to represent the pure gas species, \(Y_G=1\) and \(Y_L=0\) in Eqs.~(\ref{eqn:BC_freeplus}) and~(\ref{eqn:BC_freeminus}). \par 

The streamwise velocity component and the tangential stress are continuous across the interface. These conditions are expressed in the original space \((x,y)\) as

\begin{equation}
u_{g,\Gamma} = u_{l,\Gamma} = U_\Gamma
\end{equation}

\begin{equation}
\mu_{g,\Gamma} \Bigg(\frac{\partial u}{\partial y}\Bigg)_{g,\Gamma} = \mu_{l,\Gamma} \Bigg(\frac{\partial u}{\partial y}\Bigg)_{l,\Gamma}
\end{equation}

\noindent
which converts in the \(\eta\) space to

\begin{equation}
\label{eqn:tangentvel}
f'_g(0) = f'_l(0) = U_\Gamma
\end{equation}

\begin{equation}
\label{eqn:tangentstress}
\frac{f''_l(0)}{f''_g(0)} = \frac{(\rho \mu)_{g,\Gamma}}{(\rho \mu)_{l,\Gamma}}
\end{equation}

Therefore, \(f'\) is continuous across the interface while a jump in the second derivative, \(f''\), exists for \((\rho \mu)_{g,\Gamma}\neq(\rho \mu)_{l,\Gamma}\). Another condition for \(f\) is obtained from the mass balance across the interface. Since the interface location is assumed to remain fixed, the net mass flux, \(\dot{\omega}\), across the interface reads

\begin{equation}
\label{eqn:phasechange1}
\dot{\omega} = (\rho v)_{g,\Gamma} = (\rho v)_{l,\Gamma}
\end{equation}

Eq.~(\ref{eqn:phasechange1}) is expressed in terms of \(f(\eta)\) as 

\begin{equation}
\label{eqn:phasechange2}
\dot{\omega} = -\frac{1}{\sqrt{2\bar{x}}}f_g(0) = -\frac{1}{\sqrt{2\bar{x}}}f_l(0)
\end{equation}

\noindent
since \(\tilde{I}=0\) at the interface. This equation shows that \(f\) is also continuous across the interface (i.e., \(f_g(0)=f_l(0)=f(0)\)). From this relation, the transformed velocity evaluated at the interface corresponds to the mass flux across it, or \(w(0)=\dot{\omega}=-f(0)/\sqrt{2\bar{x}}\). \par 

Finally, expressions for the species mass balance and energy balance at the interface are needed. Eqs.~(\ref{eqn:spbalance1}) and~(\ref{eqn:enebalance1}) show these relations in the \((x,y)\) space, respectively.


\begin{equation}
\label{eqn:spbalance1}
\dot{\omega}(Y_{1,g}-Y_{1,l}) = \bigg(\rho D \frac{\partial Y_1}{\partial y}\bigg)_g - \bigg(\rho D \frac{\partial Y_1}{\partial y}\bigg)_l
\end{equation}

\begin{equation}
\label{eqn:enebalance1}
\dot{\omega}(h_g-h_l) = \bigg(\frac{\lambda}{c_p}\frac{\partial h}{\partial y}\bigg)_g - \bigg(\frac{\lambda}{c_p}\frac{\partial h}{\partial y}\bigg)_l + \Bigg[\bigg(\rho D - \frac{\lambda}{c_p}\bigg)(h_1-h_2)\frac{\partial Y_1}{\partial y}\Bigg]_g - \Bigg[\bigg(\rho D - \frac{\lambda}{c_p}\bigg)(h_1-h_2)\frac{\partial Y_1}{\partial y}\Bigg]_l
\end{equation}

Eqs.~(\ref{eqn:spbalance1}) and~(\ref{eqn:enebalance1}) are expressed in terms of \(f(\eta)\), \(Y(\eta)\) and \(h(\eta)\) as 


\begin{equation}
\label{eqn:spbalance2}
-f(0)\big( Y_g(0) - Y_l(0) \big) = \big(\rho^2 D Y'(0)\big)_g - \big(\rho^2 D Y'(0)\big)_l
\end{equation}

\begin{equation}
\label{eqn:enebalance2}
\begin{split}
-f(0)\big(h_g(0)-h_l(0)\big) &= \bigg(\frac{\rho \lambda}{c_p}h'(0)\bigg)_g - \bigg(\frac{\rho \lambda}{c_p}h'(0)\bigg)_l \\
&+ \Bigg[\bigg(\rho^2 D - \frac{\rho\lambda}{c_p}\bigg)(h_1-h_2)Y'(0)\Bigg]_g - \Bigg[\bigg(\rho^2 D - \frac{\rho\lambda}{c_p}\bigg)(h_1-h_2)Y'(0)\Bigg]_l
\end{split}
\end{equation}

Thermodynamic phase equilibrium is used to evaluate fully the interface solution of the system of ODEs. That is, an equality in chemical potential for each species on either side of the interface is imposed. This condition is expressed through an equality in fugacity for each species~\cite{soave1972equilibrium,poling2001properties}, \(\phi_i\), as

\begin{equation}
\phi_{li}(T_l,p_l,X_{li}) = \phi_{gi}(T_g,p_g,X_{gi})
\end{equation}

\noindent
which is a function of mixture temperature, pressure and composition. Under constant pressure across the interface, phase equilibrium can be expressed using the fugacity coefficient, \(\Phi_i \equiv \phi_i/pX_i\), as

\begin{equation}
\label{eqn:pheq}
X_{li}\Phi_{li}=X_{gi}\Phi_{gi}
\end{equation}

\noindent
where \(X_i\) represents the mole fraction of species \(i\). \par 

Note that liquid and gas compositions are only the same at the mixture critical point. Therefore, phase equilibrium shows that \(Y\) and \(h\) are discontinuous across the interface, as well as other fluid properties (e.g., density). \par 

Since the thickness of the interface is of the order of \(\mathcal{O}(10^{-9}\) m)~\cite{dahms2013transition,dahms2015liquid} and diffusion layers in non-ideal high-pressure conditions quickly reach thicknesses of the order of \(\mathcal{O}(10^{-6}\) m)~\cite{poblador2018transient,davis2019development}, the interface thickness is neglected and temperature is assumed to be continuous (i.e., \(T_g(0)=T_l(0)=T_\Gamma\)). The quick adjustment of the interface solution supports the thermodynamic equilibrium assumption~\cite{poblador2018transient,davis2019development}. \par 


\section{Solution method}
\label{sec:solution}

The system of ODEs, Eqs.~(\ref{eqn:xmom_trans2})-(\ref{eqn:ene_trans2}), must be solved numerically. To avoid solving a third-order differential equation, Eq.~(\ref{eqn:xmom_trans2}) is split into two equations: a first-order differential equation to solve for \(f\) and a second-order differential equation to solve for \(f'\). Therefore, Eq.~(\ref{eqn:xmom_trans2}) is rewritten as

\begin{equation}
\label{eqn:xmom1_trans3a}
(\rho \mu)g_1'' + (\rho \mu)'g_1' + g_2g_1' = 0 
\end{equation}

\begin{equation}
\label{eqn:xmom1_trans3b}
g_2' = g_1
\end{equation}

\noindent
where the new variables are \(g_1 = f'\) and \(g_2=f\). \par 

Eqs.~(\ref{eqn:sc_trans2}),~(\ref{eqn:ene_trans2}),~(\ref{eqn:xmom1_trans3a}) and~(\ref{eqn:xmom1_trans3b}) are discretized using a second-order finite difference scheme and solved using a tri-diagonal matrix solver. The asymptotic behaviors at \(+\infty\) and \(-\infty\) are well represented with a numerical range for the similarity variable given by \(-0.5\leq\eta\leq0.5\). Since the system of equations is highly coupled, some iterations are needed before reaching a converged solution. A mesh size of \(\Delta\eta=1.5625\times10^{-4}\) kg/(m\(^{5/2}\)) provides a grid-independent solution. \par 

The interface solution is also updated at every iteration. The streamwise momentum matching conditions, Eqs.~(\ref{eqn:tangentvel}) and~(\ref{eqn:tangentstress}), are combined numerically to solve for \(U_\Gamma\) or \(f'(0)\). On the other hand, Eqs.~(\ref{eqn:spbalance2}),~(\ref{eqn:enebalance2}) and~(\ref{eqn:pheq}) form a closed system that can be solved at every iteration in terms of the interface temperature. The solution of this system of equations provides the interface values for \(f(0)\), \(Y_g(0)\), \(Y_l(0)\), \(h_g(0)\) and \(h_l(0)\). More information on this interface algorithm is provided in Poblador-Ibanez and Sirignano~\cite{poblador2018transient}.

\section{Results}
\label{sec:results}

The system of ODEs has been solved for the analyzed cases from Davis et al.~\cite{davis2019development}, where numerical results for a non-ideal two-phase laminar mixing layer using the system of partial differential equations, Eqs.~(\ref{eqn:gc}),~(\ref{eqn:xmom_ncons}),~(\ref{eqn:sc_ncons}) and~(\ref{eqn:ene_ncons}), are provided. \par 

Oxygen is chosen to be the pure gas species, while the liquid consists of pure \textit{n}-decane. Mass fraction freestream conditions correspond to those discussed in Section~\ref{sec:boundary_cond}, with \(Y_1\) representing the pure gas species composition. Thus, \(Y_G=1\) and \(Y_L=0\). Temperature (i.e., enthalpy) freestream conditions are \(T(+\infty)=T_G=550\) K and \(T(-\infty)=T_L=450\) K. Using the thermodynamic model, \(h_G\) and \(h_L\) are evaluated. Reynolds number, \(Re_{\bar{L}}\), is based on gas properties and the relative velocity between the two free streams. To keep the boundary-layer approximation valid, freestream conditions for the streamwise velocity vary among the different analyzed pressures. Their values are summarized in Table~\ref{tab:scenarios}.

\begin{table}[!h]
\centering
\begin{tabular}{ c c c c c }
\hline
 & \(p=10\) bar & \(p=50\) bar & \(p=100\) bar & \(p=150\) bar \\
 \hline
\(u_G\) (m/s) & 7.673 & 9.525 & 9.755 & 9.830 \\
\(u_L\) (m/s) & 12.327 & 10.475 & 10.246 & 10.170 \\
\hline
\end{tabular}
\caption{Freestream conditions for the streamwise component of the velocity field, \(u\), at different ambient pressures. These values satisfy \(Re_{\bar{L}}=10,000\) at \(\bar{L}=0.01\) m for a fixed mean velocity of \(10\) m/s as the freestream gas density and viscosity change with pressure~\cite{davis2019development}.}
\label{tab:scenarios}
\end{table}

In Section~\ref{subsec:profiles}, the mixing layer solution from the system of ODEs is given for the four pressure cases from Table~\ref{tab:scenarios} and compared with the solution from the system of PDEs obtained in Davis et al.~\cite{davis2019development} for the same configurations. Then, in Section~\ref{subsec:interface}, the interface equilibrium solution is obtained with both approaches. Lastly, in Section~\ref{subsec:thickness}, the evolution of the mass, momentum and thermal mixing layer thicknesses is determined and a generalization of their evaluation is made to avoid solving either the system of partial differential equations or the self-similar system.

\subsection{Mixing layer profiles}
\label{subsec:profiles}

For visualization purposes, the results are non-dimensionalized by the liquid freestream conditions to provide a consistent comparison between different cases. The non-dimensional similarity variable and \(f\) function are, respectively,

\begin{equation}
\label{eqn:nondim_f}
\eta^* = \frac{\eta}{\sqrt{\frac{\rho_L \mu_L}{u_L}}} \quad ; \quad f^* = \frac{f}{\sqrt{\rho_L \mu_L u_L}}
\end{equation}

%

The non-dimensional second derivative of \(f\) (or velocity gradient) follows that

\begin{equation}
f'' = \frac{d^2f}{d\eta^2} = \frac{\sqrt{\rho_L \mu_L u_L}}{\frac{\rho_L\mu_L}{u_L}}\frac{d^2f^*}{d{\eta^*}^2} \quad ; \quad f''^* = \frac{f''}{\sqrt{\frac{u_{L}^{3}}{\rho_L \mu_L}}}
\end{equation}

Any other variable, \(\phi\), is made dimensionless by dividing by the liquid freestream value, \(\phi_L\) (e.g., \(h^* = h/h_L\) or \(f'^* = f'/u_L\)). Note that \(Y\) is already non-dimensional. Furthermore, for continuous variables across the interface, such as \(f'\) and \(T\), the following non-dimensional definitions are made to obtain distributions ranging from 0 to 1, being 1 the largest value of the variable,

\begin{equation}
\theta_T(\eta) = \frac{T(\eta)-T_L}{T_G-T_L} \quad ; \quad \theta_u(\eta) = \frac{f'(\eta)-u_G}{u_L-u_G}
\end{equation}

The transformed transverse velocity, \(w\), is not self-similar since it depends on \(\bar{x}\) and \(\eta\) (see Eq.~(\ref{eqn:vel_eta})). However, as stated in Section~\ref{subsec:transvel}, \(w\) becomes a function of \(\eta\) only when multiplied by the square root of \(\bar{x}\). Thus, \(\hat{w}=w\sqrt{2\bar{x}} = \eta f' - f\) is self-similar. To obtain a non-dimensional \(\hat{w}\) for representation purposes, the results are scaled as \(\hat{w}^* = \hat{w}/\sqrt{\rho_L \mu_L u_L}\). \par

\begin{figure}[h!]
\centering
\begin{subfigure}{.5\textwidth}
  \centering
  \includegraphics[width=1.0\linewidth]{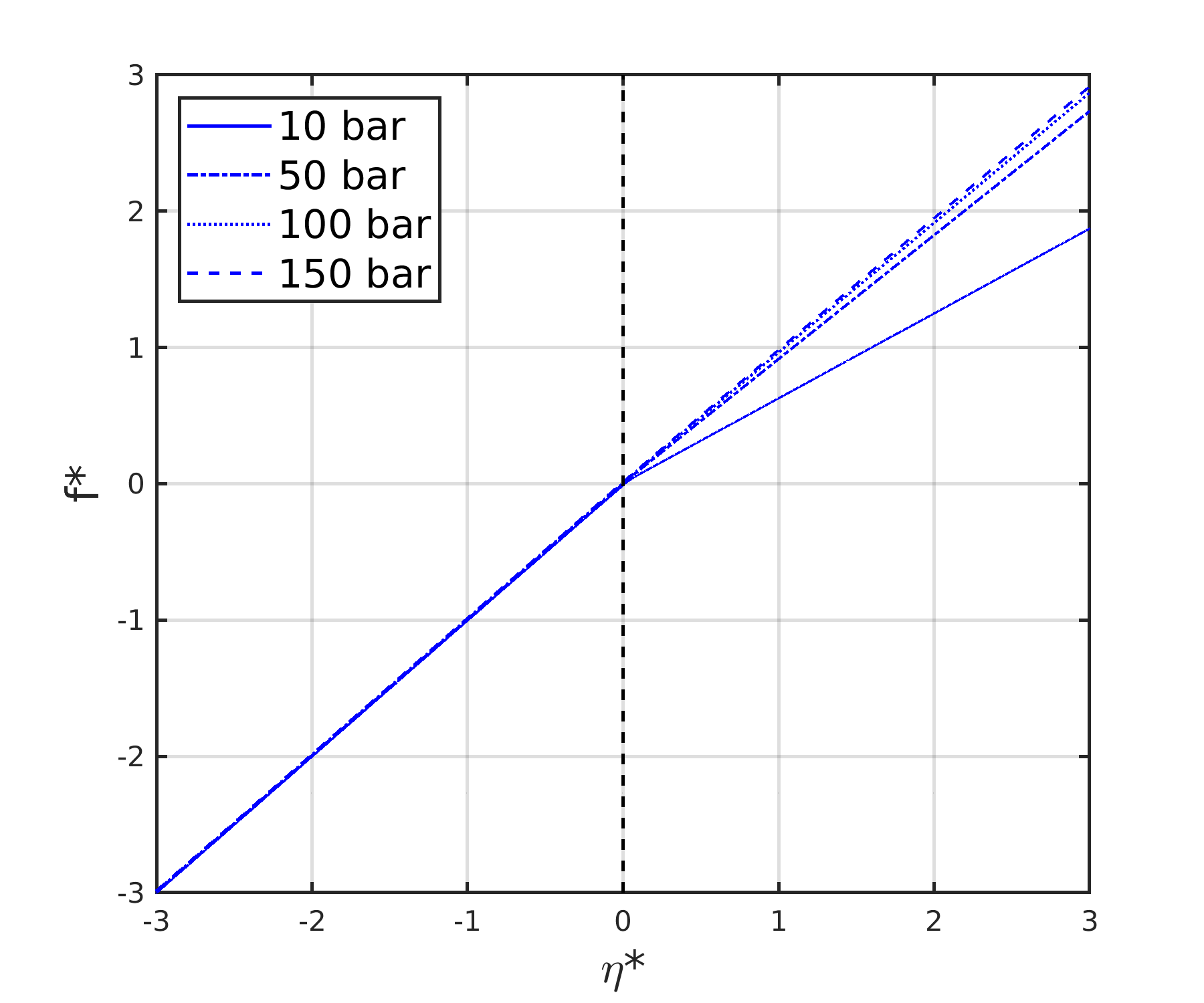}
  \caption{}
\end{subfigure}%
\begin{subfigure}{.5\textwidth}
  \centering
  \includegraphics[width=1.0\linewidth]{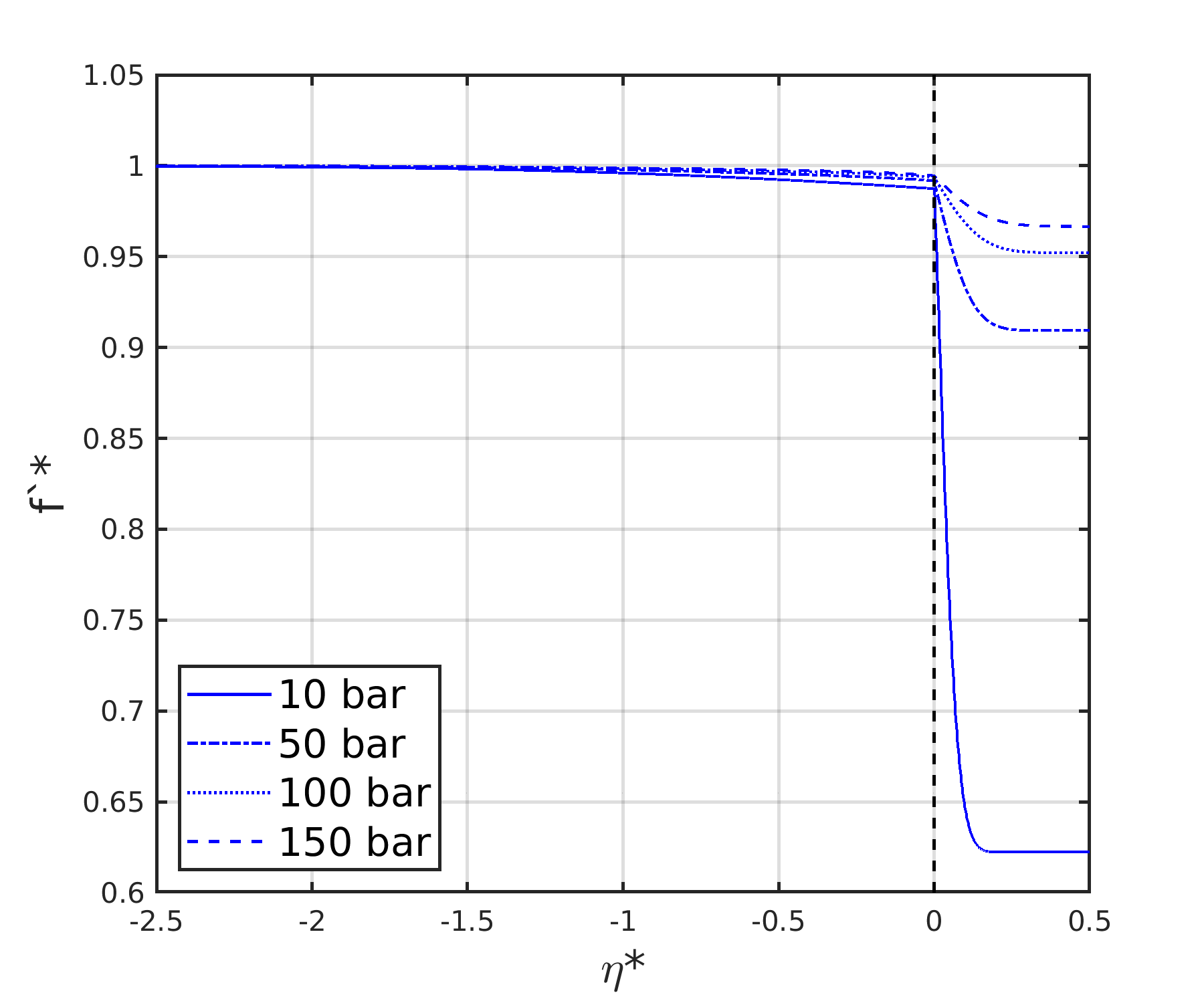}
  \caption{}
  \label{subfig:fp}
\end{subfigure}%
\\
\begin{subfigure}{.5\textwidth}
  \centering
  \includegraphics[width=1.0\linewidth]{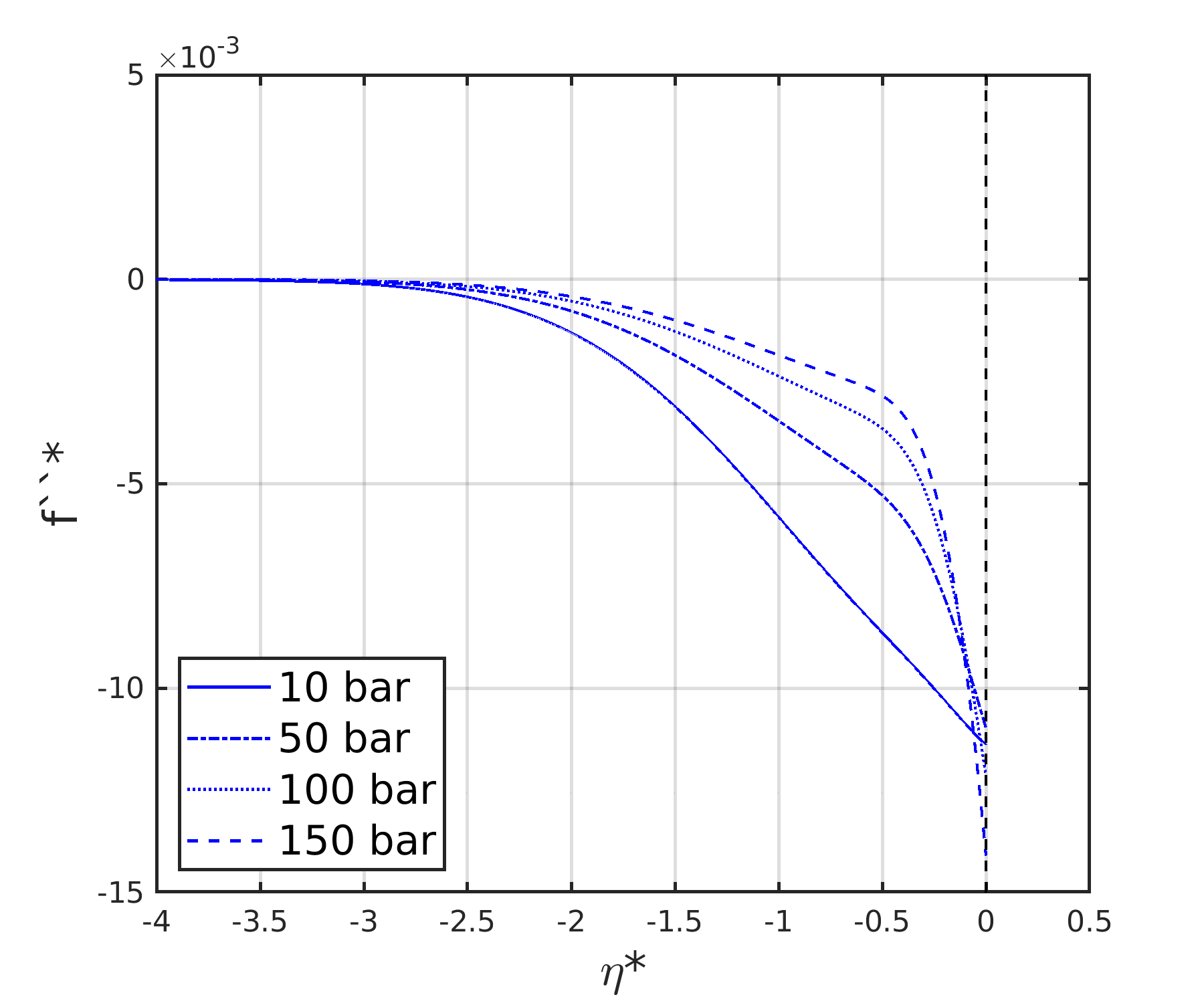}
  \caption{}
  \label{subfig:velgradL}
\end{subfigure}%
\begin{subfigure}{.5\textwidth}
  \centering
  \includegraphics[width=1.0\linewidth]{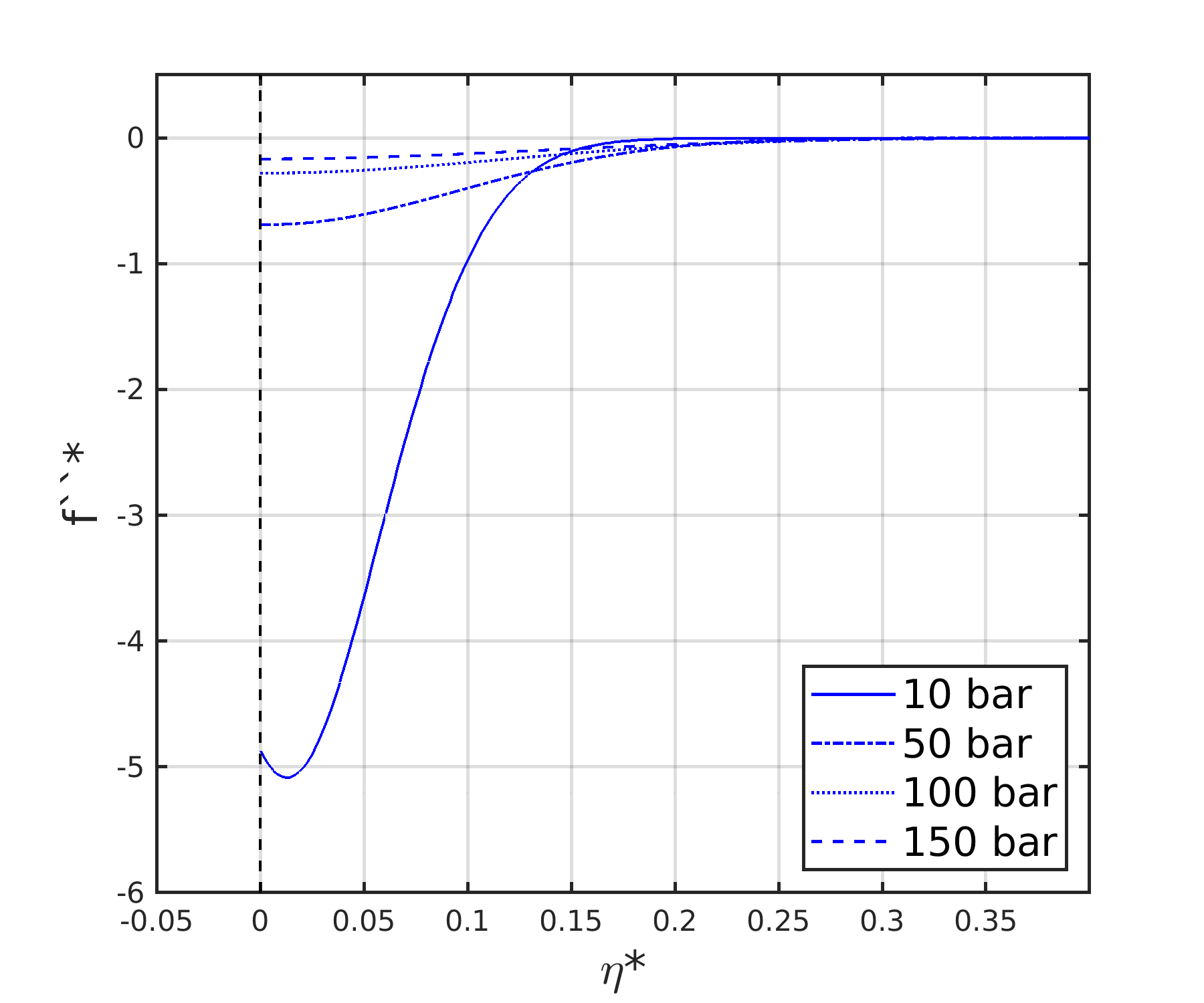}
  \caption{}
\end{subfigure}%
\caption{Solution of the self-similar system of ordinary differential equations. (a) \(f^*\); (b) \(f'^*\); (c) \(f''^*\) in the liquid phase; (d) \(f''^*\) in the gas phase.}
\label{fig:profiles1}
\end{figure}

\begin{figure}[h!]
\centering
\begin{subfigure}{.5\textwidth}
  \centering
  \includegraphics[width=1.0\linewidth]{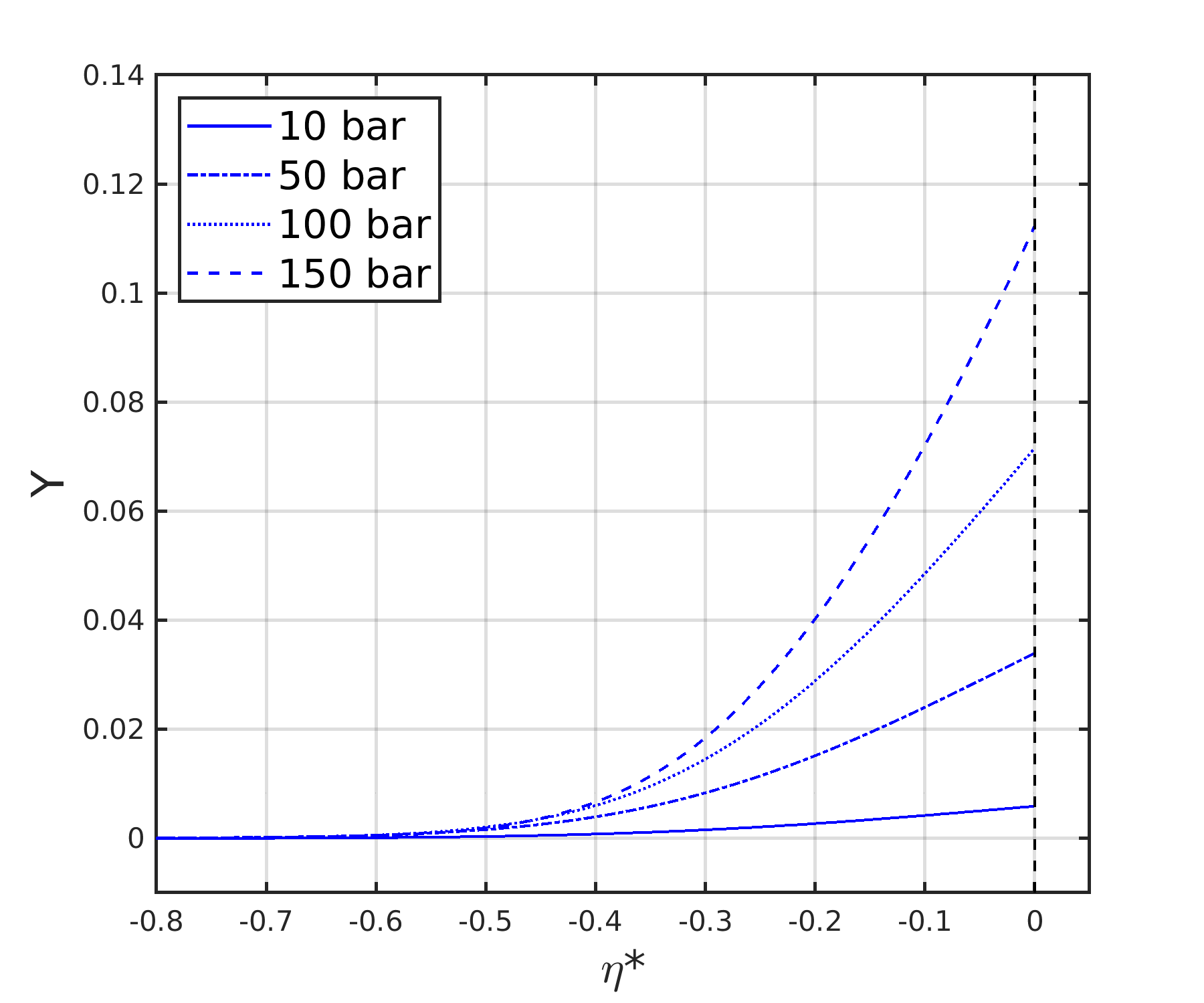}
  \caption{}
  \label{subfig:YL}
\end{subfigure}%
\begin{subfigure}{.5\textwidth}
  \centering
  \includegraphics[width=1.0\linewidth]{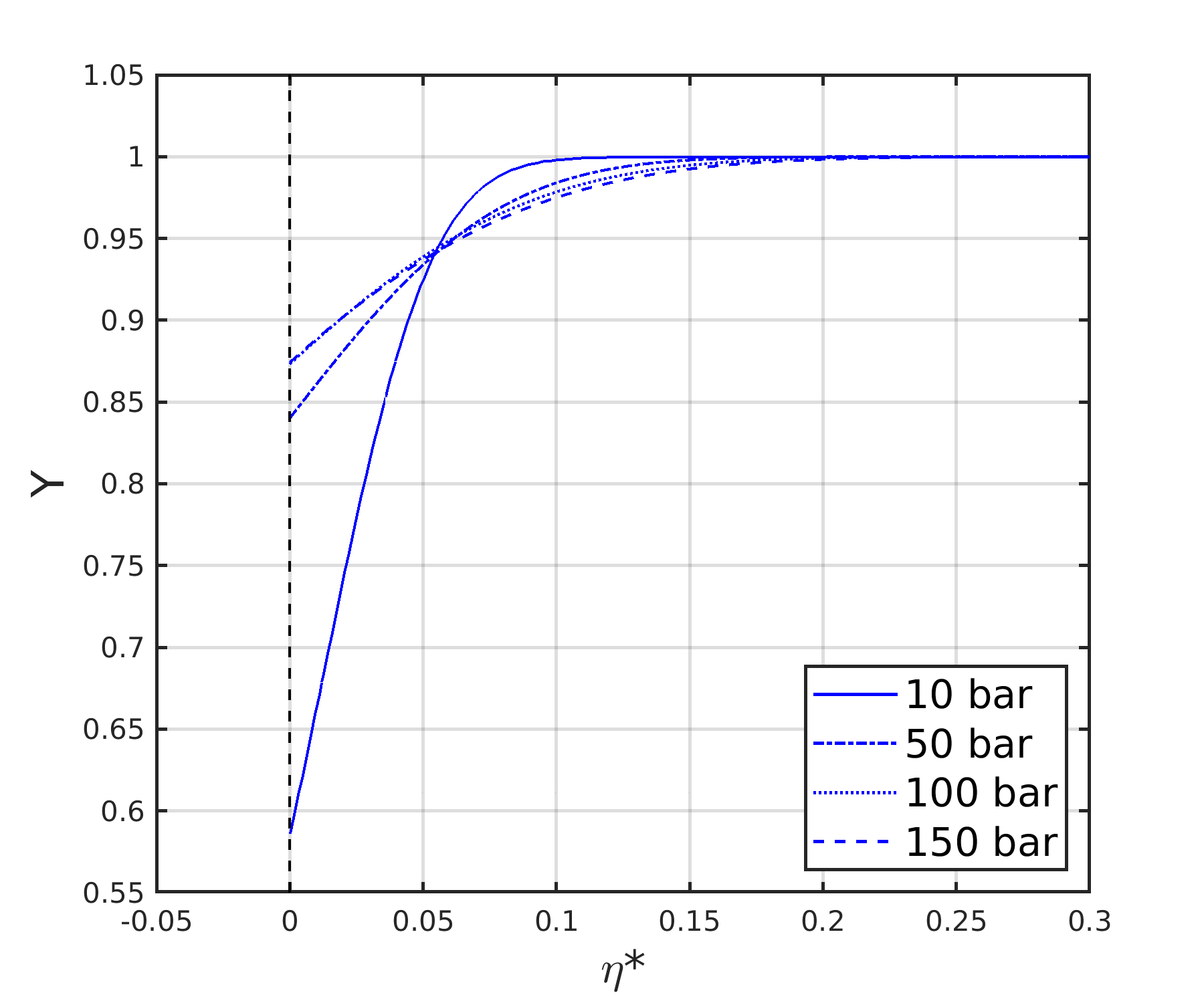}
  \caption{}
\end{subfigure}%
\caption{Solution of the self-similar system of ordinary differential equations. (a) \(Y\) in the liquid phase; (b) \(Y\) in the gas phase.}
\label{fig:profiles2}
\end{figure}

\begin{figure}[h!]
\centering
\begin{subfigure}{.5\textwidth}
  \centering
  \includegraphics[width=1.0\linewidth]{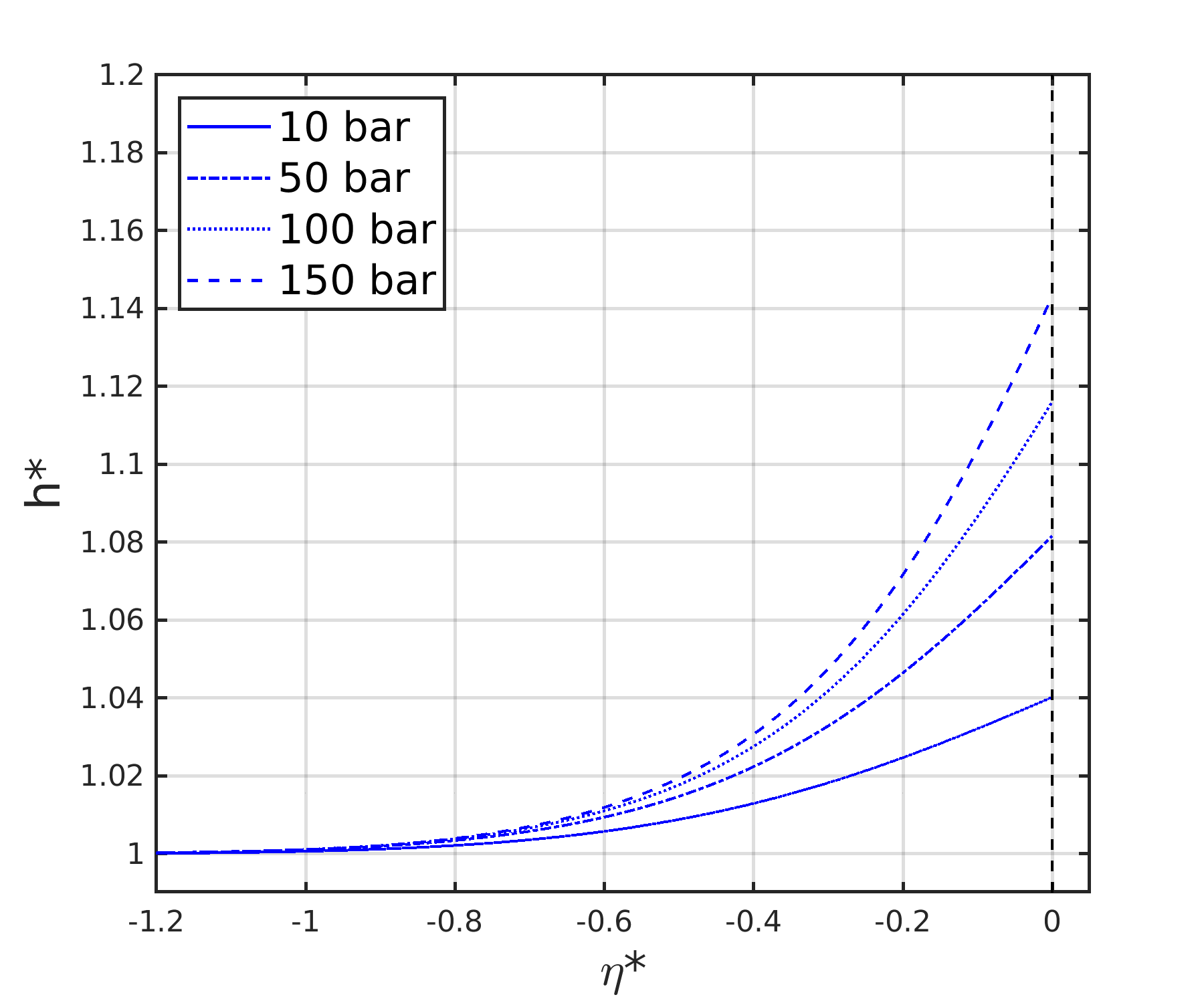}
  \caption{}
\end{subfigure}%
\begin{subfigure}{.5\textwidth}
  \centering
  \includegraphics[width=1.0\linewidth]{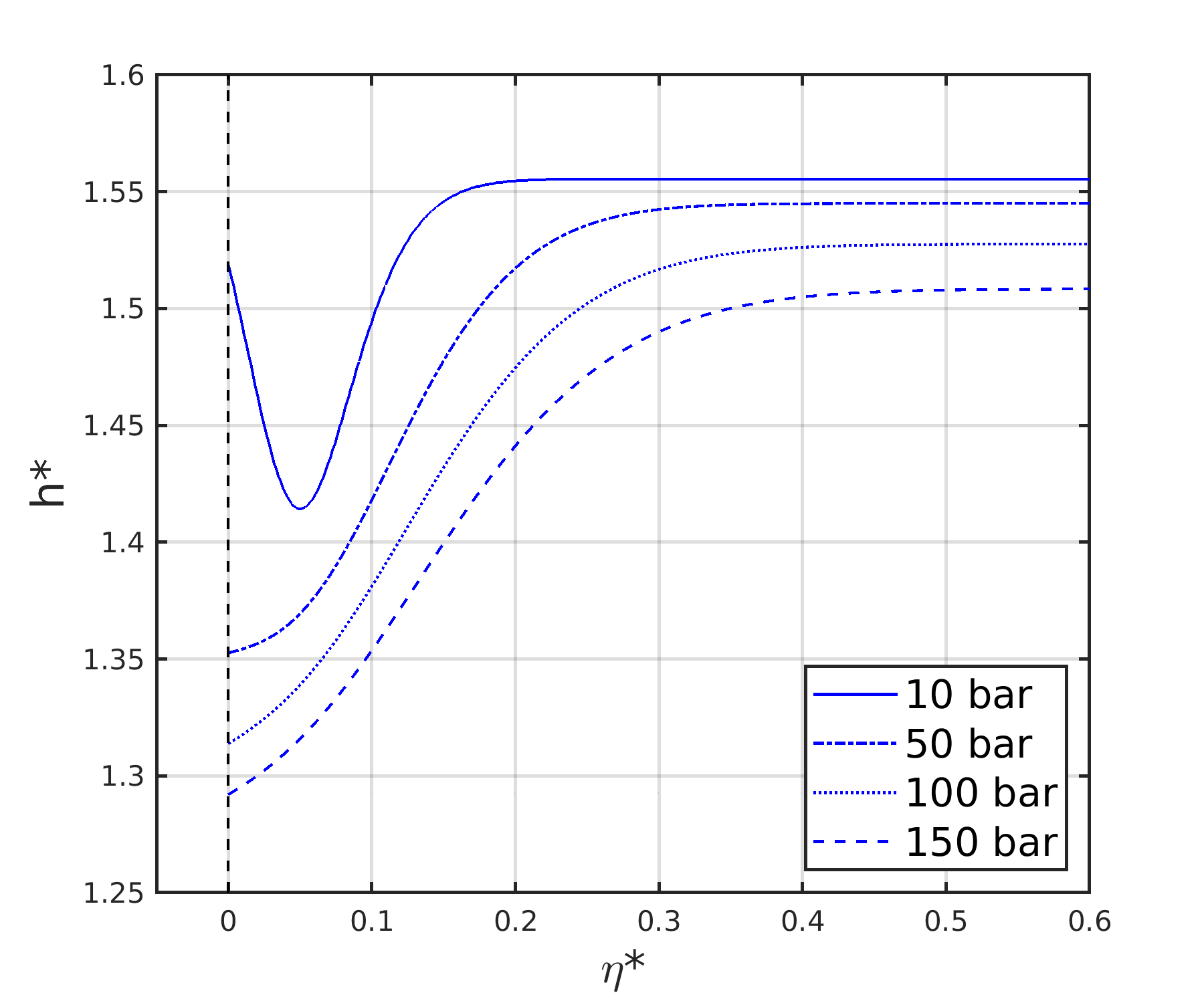}
  \caption{}
\end{subfigure}%
\caption{Solution of the self-similar system of ordinary differential equations. (a) \(h^*\) in the liquid phase; (b) \(h^*\) in the gas phase.}
\label{fig:profiles3}
\end{figure}

Figure~\ref{fig:profiles1} presents the non-dimensional solution of \(f^*\), \(f'^*\) and \(f''^*\). Figure~\ref{subfig:fp} shows how the interface velocity tends to be very close to the freestream liquid velocity. That is, the liquid phase is much more dense and viscous than the gas phase and it becomes hard for the slower gas stream to slow the liquid stream down. The solutions of \(Y\) and \(h^*\) are plotted in Figures~\ref{fig:profiles2} and~\ref{fig:profiles3}. \par 

\begin{figure}[h!]
\centering
\begin{subfigure}{.5\textwidth}
  \centering
  \includegraphics[width=1.0\linewidth]{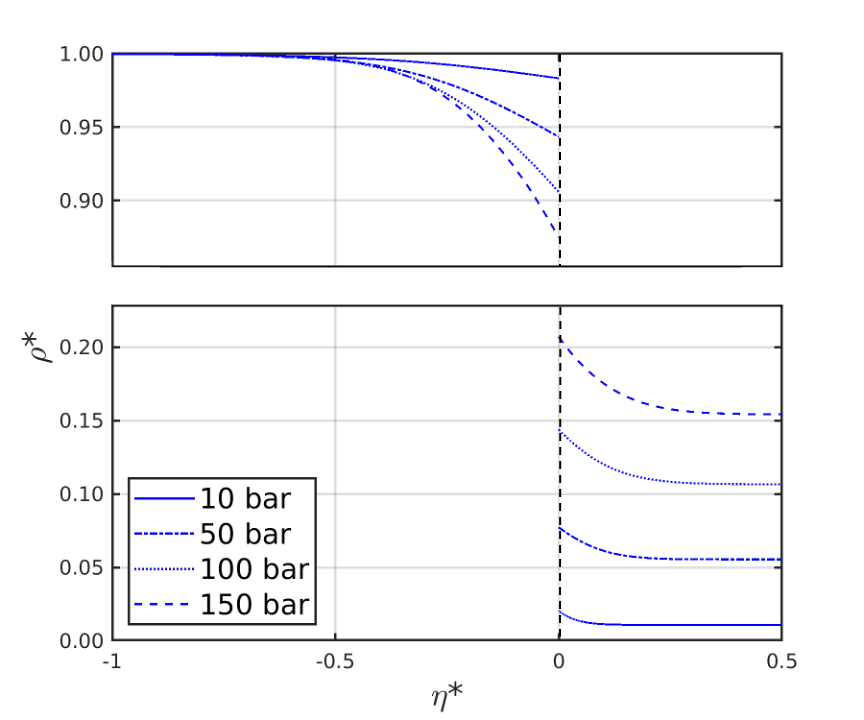}
  \caption{}
\end{subfigure}%
\begin{subfigure}{.5\textwidth}
  \centering
  \includegraphics[width=1.0\linewidth]{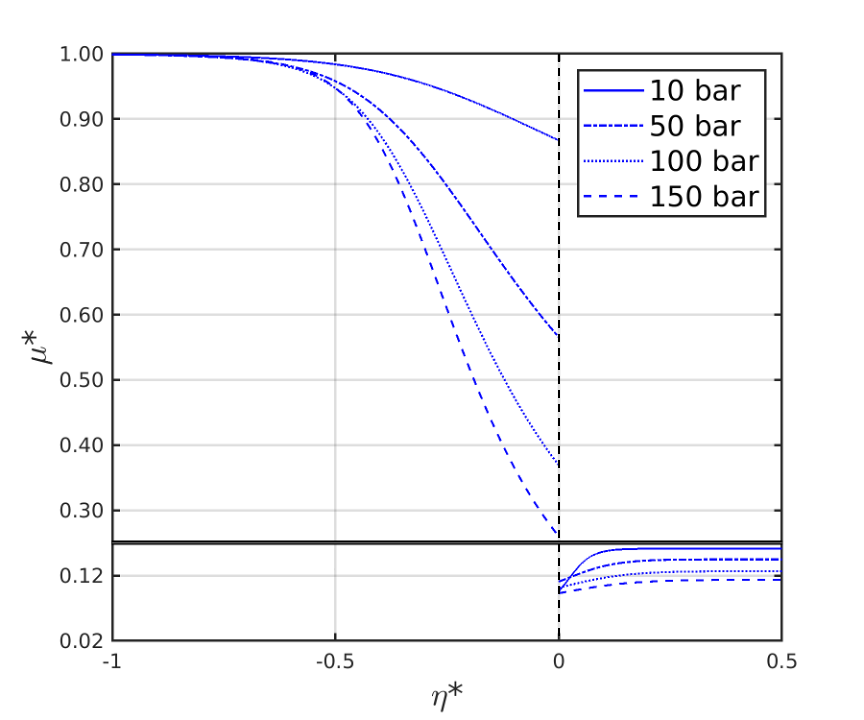}
  \caption{}
\end{subfigure}%
\caption{Solution of the self-similar system of ordinary differential equations. (a) \(\rho^*\); (b) \(\mu^*\).}
\label{fig:profiles4}
\end{figure}

As pressure increases well above the liquid critical pressure, the dissolution of the lighter gas species into the liquid phase is enhanced~\cite{juanos2015thermodynamic,poblador2018transient} (see Figure~\ref{subfig:YL}). This generates sharper variations in the fluid properties within the liquid phase, which can be responsible for the change in behavior seen in the liquid velocity gradient (see Figure~\ref{fig:profiles4}). As seen in Figure~\ref{subfig:velgradL}, the velocity gradient behaves almost linearly within the liquid momentum mixing layer at subcritical pressures (i.e., 10 bar) where liquid density and viscosity remain fairly constant. However, as pressure increases well above the critical pressure of \textit{n}-decane (\(p_c = 21.03\) bar), the velocity gradient shows an inflection point where its rate of change experiences a transition from a stronger deceleration of the liquid near the interface to a slower deceleration in a larger portion of the mixing layer. These two differentiated regions match the region where density and viscosity are experiencing the largest drop towards interface values. That is, the less dense and viscous liquid phase is decelerated much more easily by the slower gas. \par

The specific mixture enthalpy distribution is consistent with the increase of temperature in the liquid phase and the decrease of temperature in the gas phase. However, a wavy distribution is observed at 10 bar. A discussion based on a thermodynamic analysis is provided in Poblador-Ibanez and Sirignano~\cite{poblador2018transient}. Considering a mass element containing the whole mixing layer (liquid and gas), the mixing process is just an internal phenomenon. Negligible net heat flux, \(Q\), crosses the boundaries of the mass element. Therefore for a constant pressure process, \(\Delta H = Q|_p = 0\). To compensate for the increase in enthalpy, \(H\), in the liquid phase, the gas phase enthalpy must decrease accordingly. At 10 bar, this requirement cannot be solely satisfied with a monotonic enthalpy distribution.

\begin{figure}[h!]
\centering
\begin{subfigure}{.5\textwidth}
  \centering
  \includegraphics[width=1.0\linewidth]{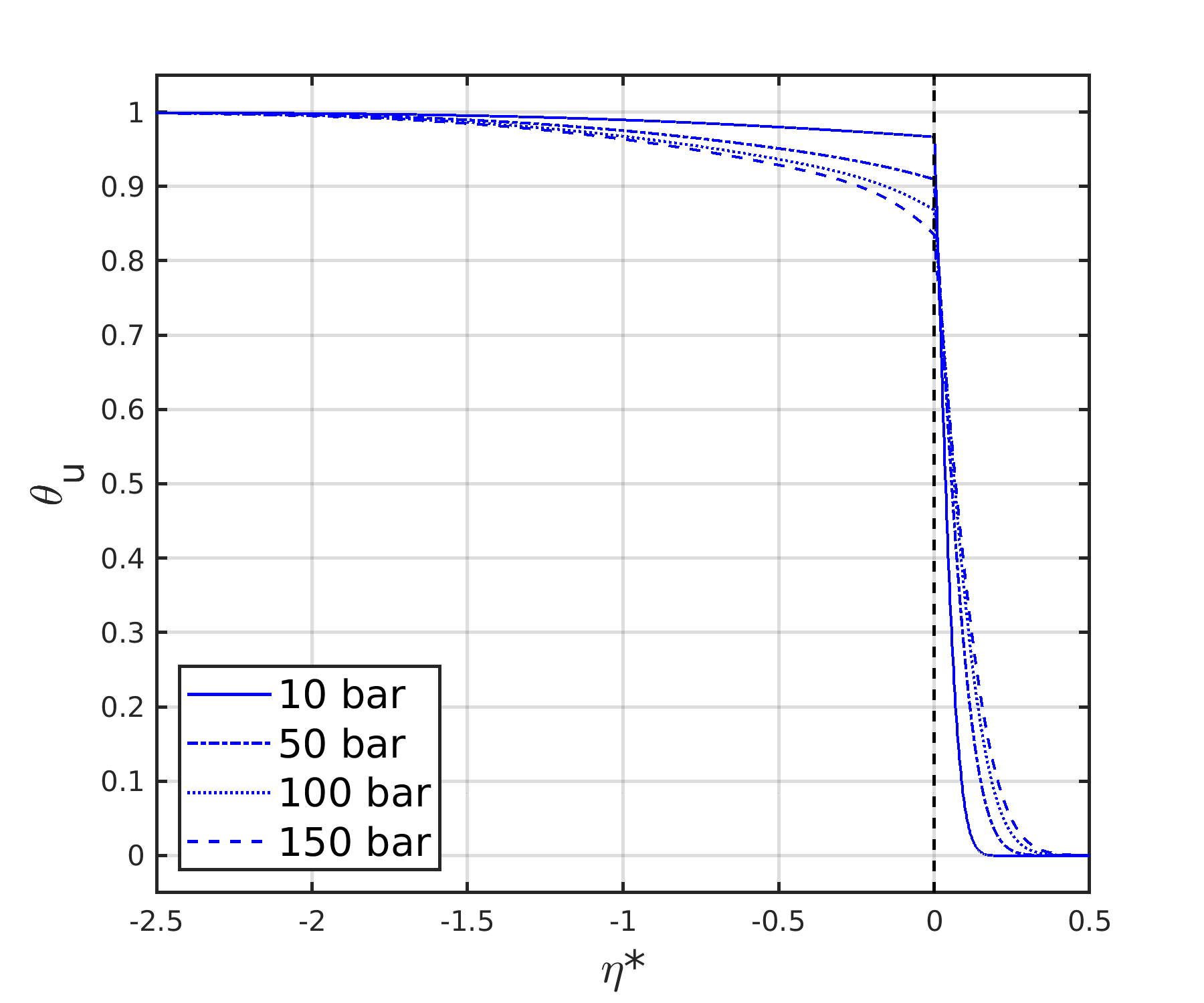}
  \caption{}
\end{subfigure}%
\begin{subfigure}{.5\textwidth}
  \centering
  \includegraphics[width=1.0\linewidth]{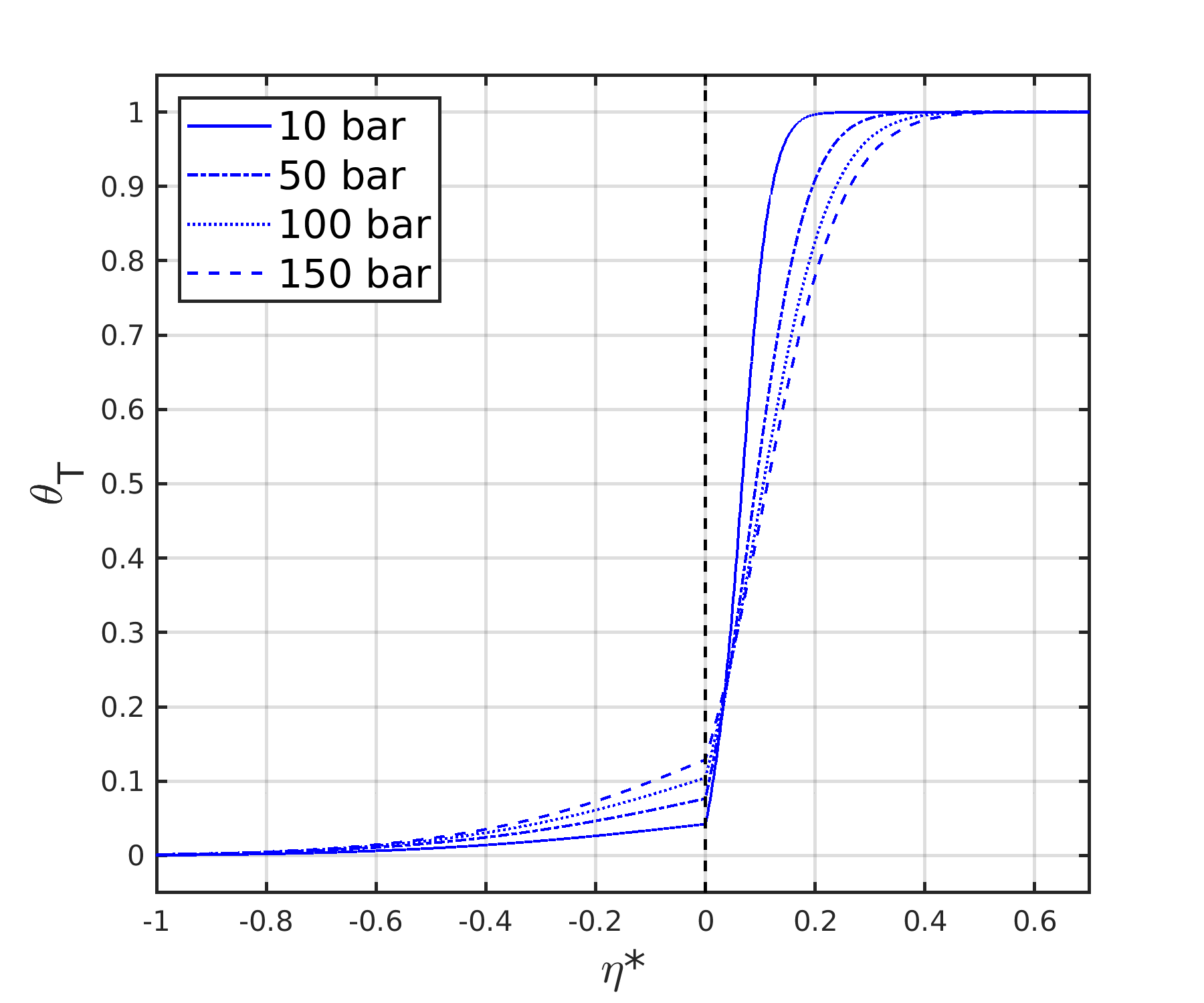}
  \caption{}
\end{subfigure}%
\caption{Solution of the self-similar system of ordinary differential equations. (a) \(\theta_u\); (b) \(\theta_T\).}
\label{fig:profiles5}
\end{figure}

Figure~\ref{fig:profiles4} presents distributions for the non-dimensional density and viscosity for both phases. As previously mentioned, these fluid properties present sharper variations as pressure increases. As the mixture critical point is approached, both liquid and gas look more alike. A discussion on the behavior of the viscosity profiles is provided in Davis et al.~\cite{davis2019development}. Figure~\ref{fig:profiles5} shows the non-dimensional continuous distributions of streamwise velocity, \(\theta_u\), and temperature, \(\theta_T\). As pressure increases and the gas dissolves more easily into the liquid phase, the interface velocity and temperature deviate further from the freestream liquid values. That is, as the liquid and gas phases come closer together, the liquid loses inertia and the gas stream can slow it down more. Furthermore, the enhanced dissolution of the gas into the liquid increases the heat flux by diffusion into the liquid, raising its temperature.

\begin{figure}[h!]
\centering
\includegraphics[width=0.5\linewidth]{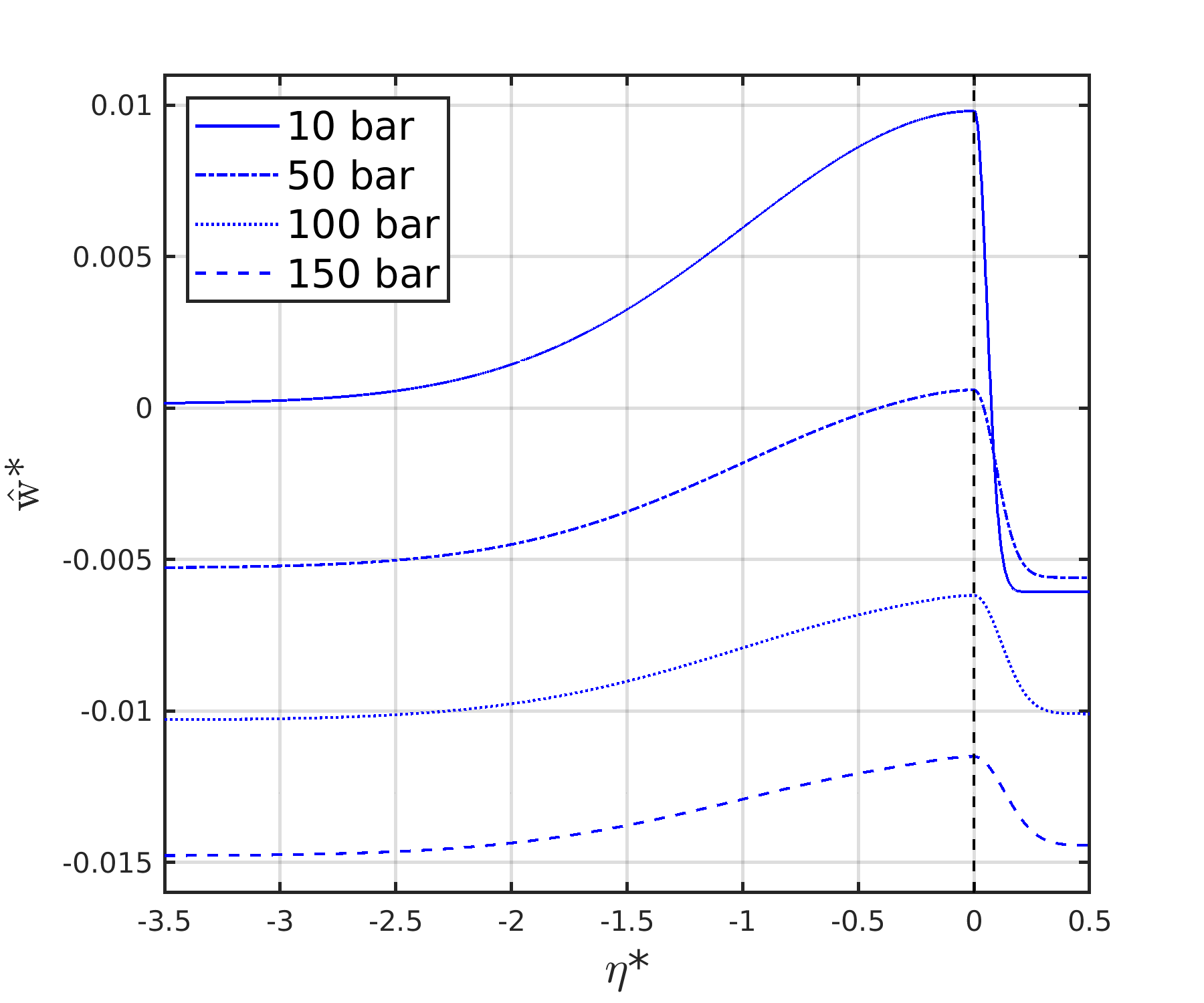}
\caption{\(\hat{w}^*\) profiles from the solution of the self-similar system of ordinary differential equations.}
\label{fig:profiles6}
\end{figure}

Distributions of the non-dimensional modified transverse velocity, \(\hat{w}^*\), are plotted in Figure~\ref{fig:profiles6}. Information on the phase change behavior at the interface can be extracted from these results. At the interface, \(w=\rho v\) from Eq.~(\ref{eqn:modv}). Therefore, \(w>0\) or \(\hat{w}>0\) at the interface implies that the transverse velocity is also positive, \(v>0\). That is, net vaporization at the interface is occurring under the assumption that \(V_\Gamma=0\). On the other hand, \(w<0\) or \(\hat{w}<0\) implies \(v<0\) or net condensation. As inferred from Eq.~(\ref{eqn:vel_eta}), the strength of the mass flux across the interface decreases with downstream distance as the \(1/\sqrt{\bar{x}}\) or \(1/\sqrt{x}\). At 10 and 50 bar, net vaporization is occurring, while condensation dominates at higher pressures. This interface behavior and its implications on the liquid jet breakup are further discussed in~\cite{poblador2018transient,poblador2019axisymmetric}.

\begin{figure}[h!]
\centering
\begin{subfigure}{.5\textwidth}
  \centering
  \includegraphics[width=1.0\linewidth]{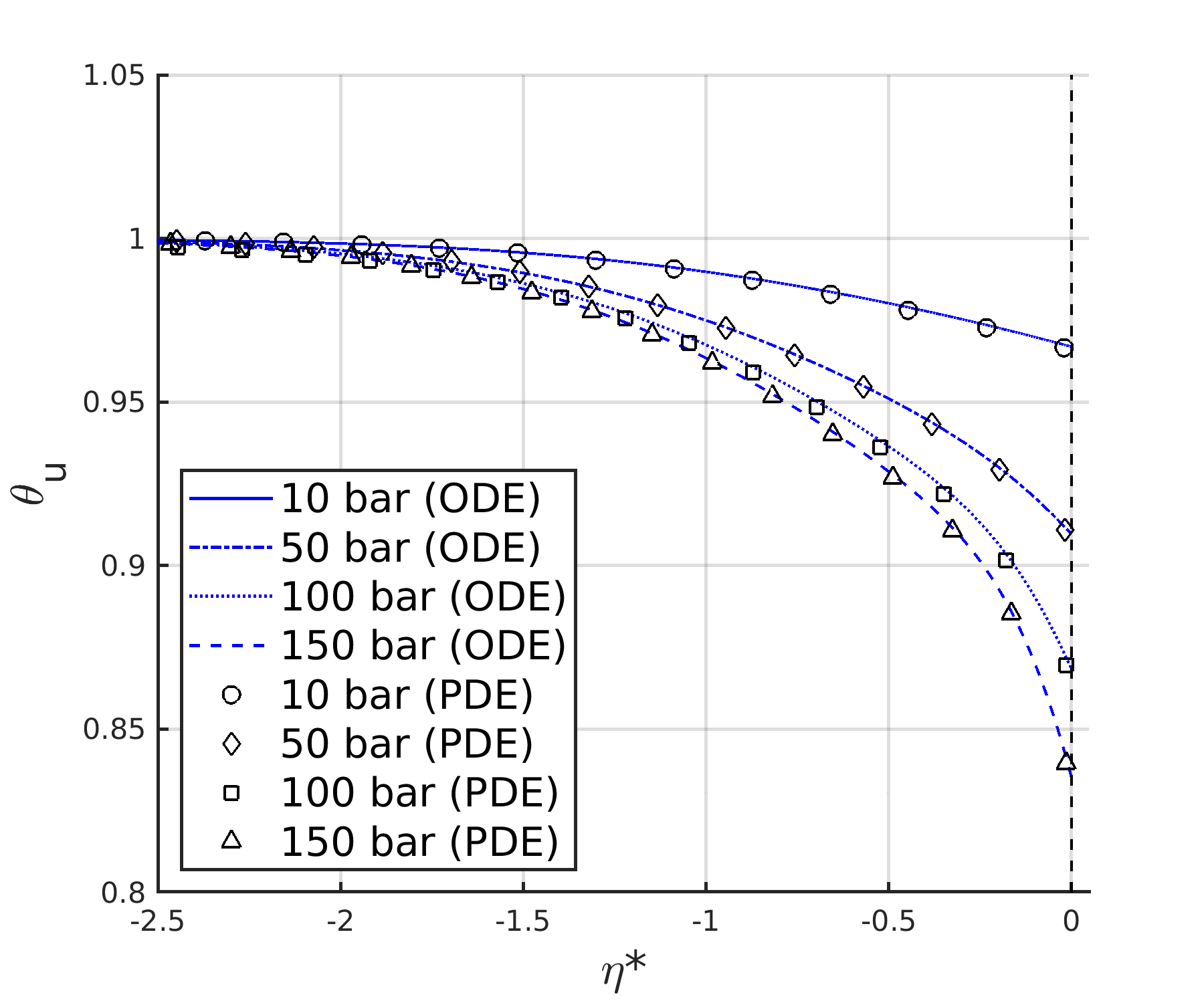}
  \caption{}
\end{subfigure}%
\begin{subfigure}{.5\textwidth}
  \centering
  \includegraphics[width=1.0\linewidth]{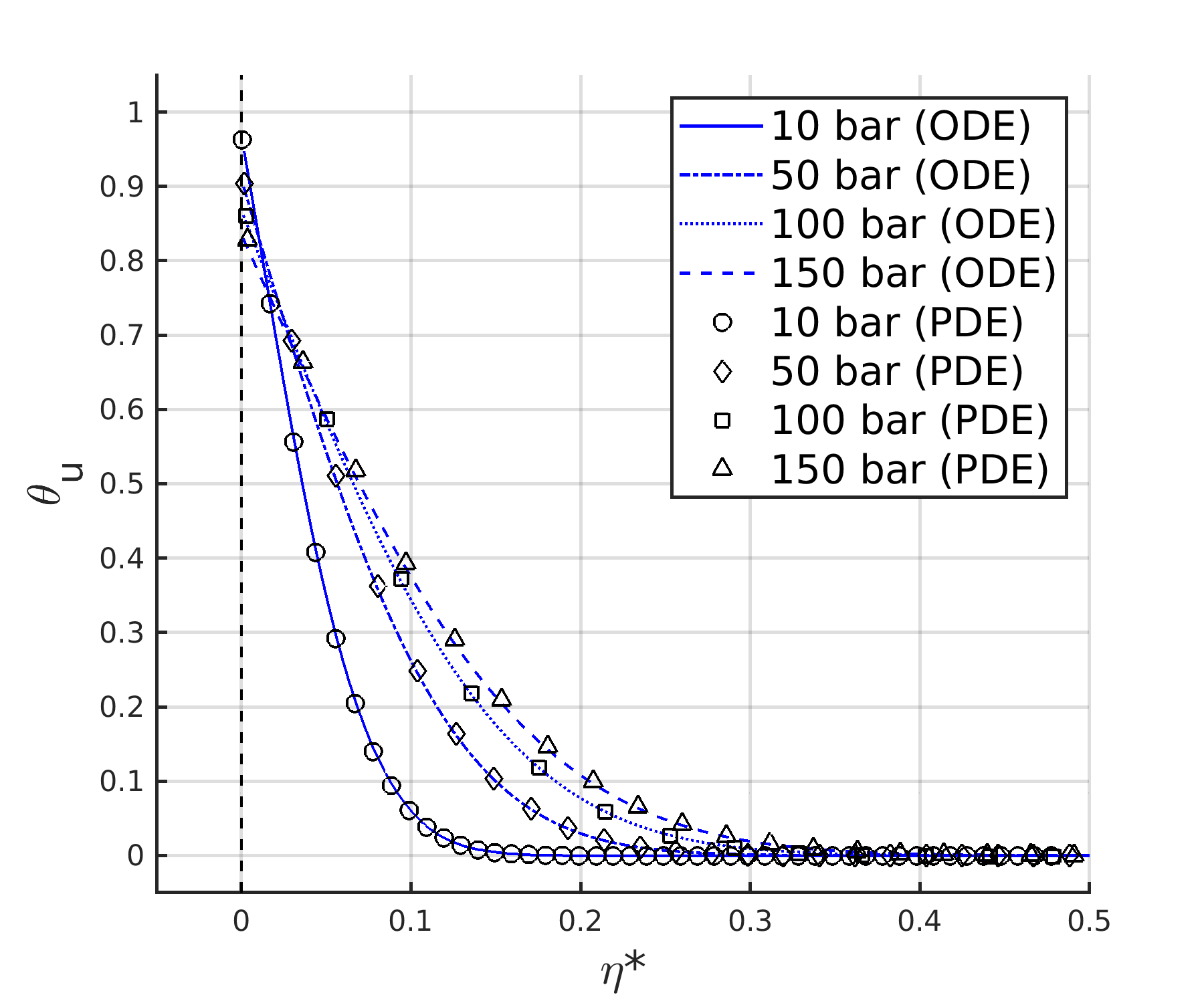}
  \caption{}
\end{subfigure}%
\caption{Comparison between the solution of the system of ordinary differential equations (ODE) and the solution of the system of partial differential equations (PDE) at \(x=0.01\) m mapped from (\(x\),\(y\)) to \(\eta\). (a) \(\theta_u\) in the liquid phase; (b) \(\theta_u\) in the gas phase.}
\label{fig:comp1}
\end{figure}

\begin{figure}[h!]
\centering
\begin{subfigure}{.5\textwidth}
  \centering
  \includegraphics[width=1.0\linewidth]{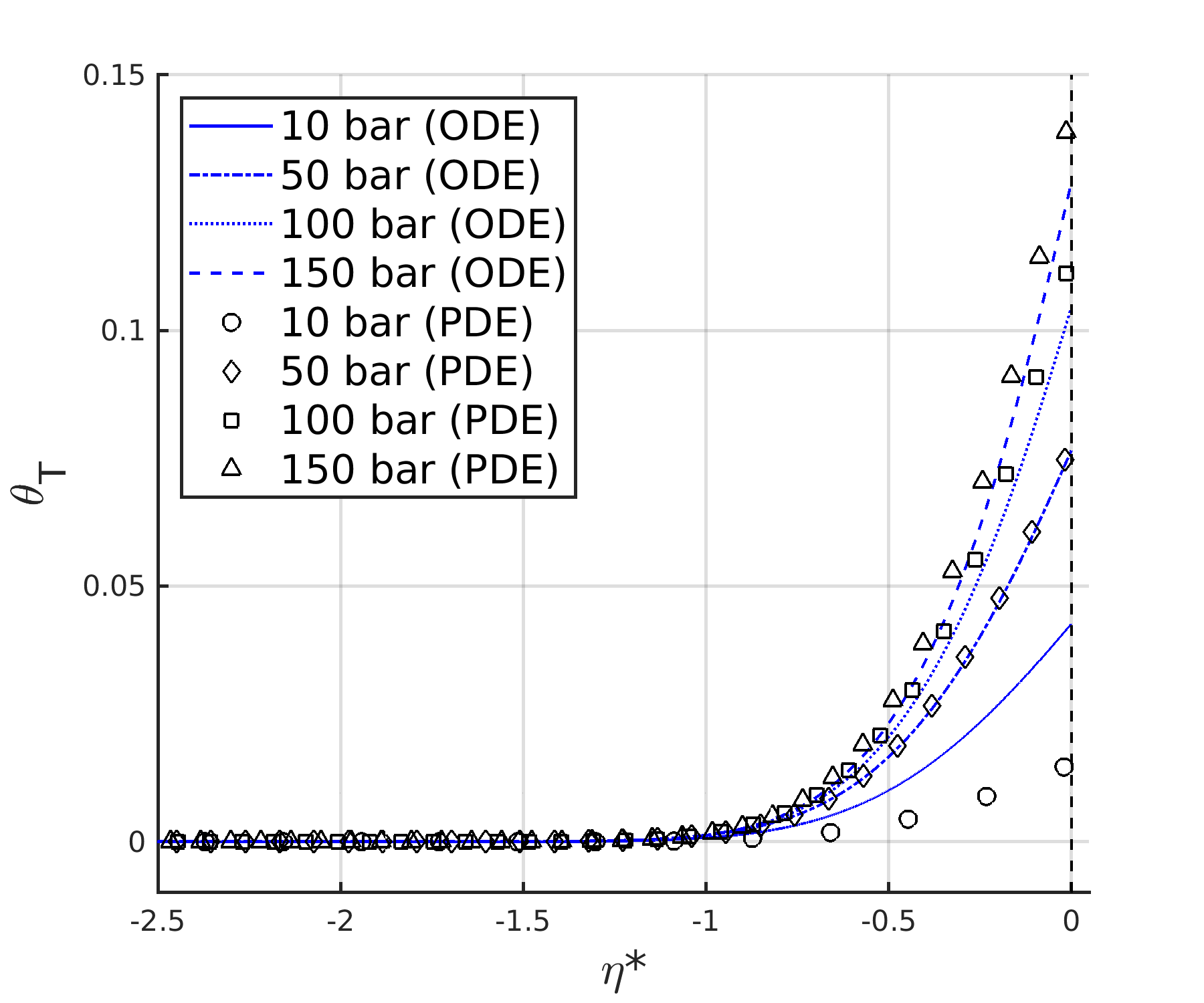}
  \caption{}
\end{subfigure}%
\begin{subfigure}{.5\textwidth}
  \centering
  \includegraphics[width=1.0\linewidth]{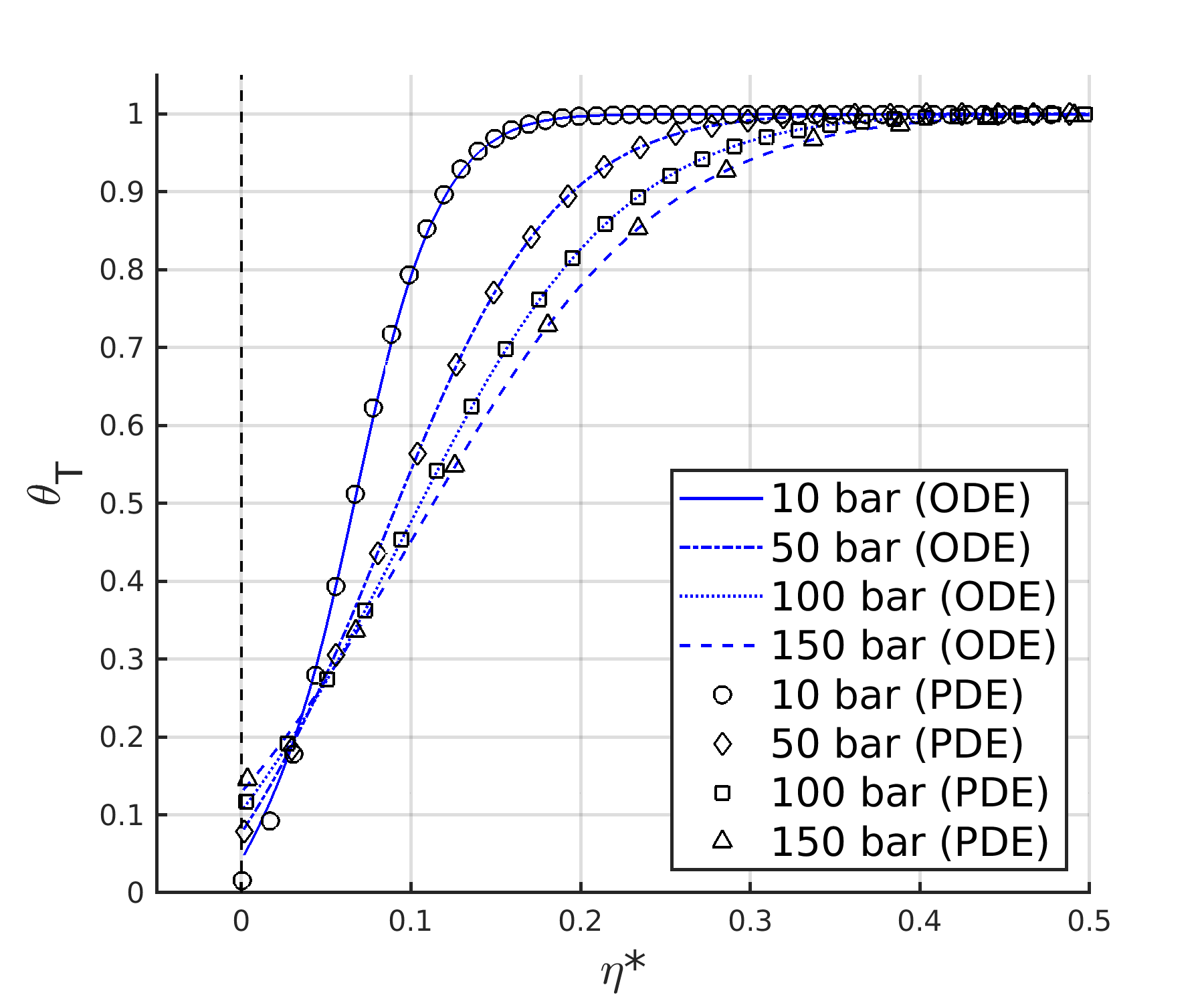}
  \caption{}
\end{subfigure}%
\caption{Comparison between the solution of the system of ordinary differential equations (ODE) and the solution of the system of partial differential equations (PDE) at \(x=0.01\) m mapped from (\(x\),\(y\)) to \(\eta\). (a) \(\theta_T\) in the liquid phase; (b) \(\theta_T\) in the gas phase.}
\label{fig:comp2}
\end{figure}

The agreement between the solution of the self-similar system of ODEs and the system of PDEs is initially tested by comparing various profiles of different variables of interest. First, the PDE solution from Davis et al.~\cite{davis2019development} at a downstream position of \(x=0.01\) m has been mapped from the (\(x\),\(y\)) space to the \(\eta\) space for all four pressure cases. Figures~\ref{fig:comp1} and~\ref{fig:comp2} present these results for the non-dimensional distribution of continuous variables (i.e., \(\theta_u\) and \(\theta_T\)), where both methods are seen to concur for 50, 100 and 150 bar. At 10 bar, larger discrepancies are observed in the liquid temperature distribution. Possible reasons for the interface temperature divergence seen at 10 bar are discussed later in Section~\ref{subsec:interface}, where a comparison of the interface matching solution obtained with each approach is presented.

\begin{figure}[h!]
\centering
\includegraphics[width=0.5\linewidth]{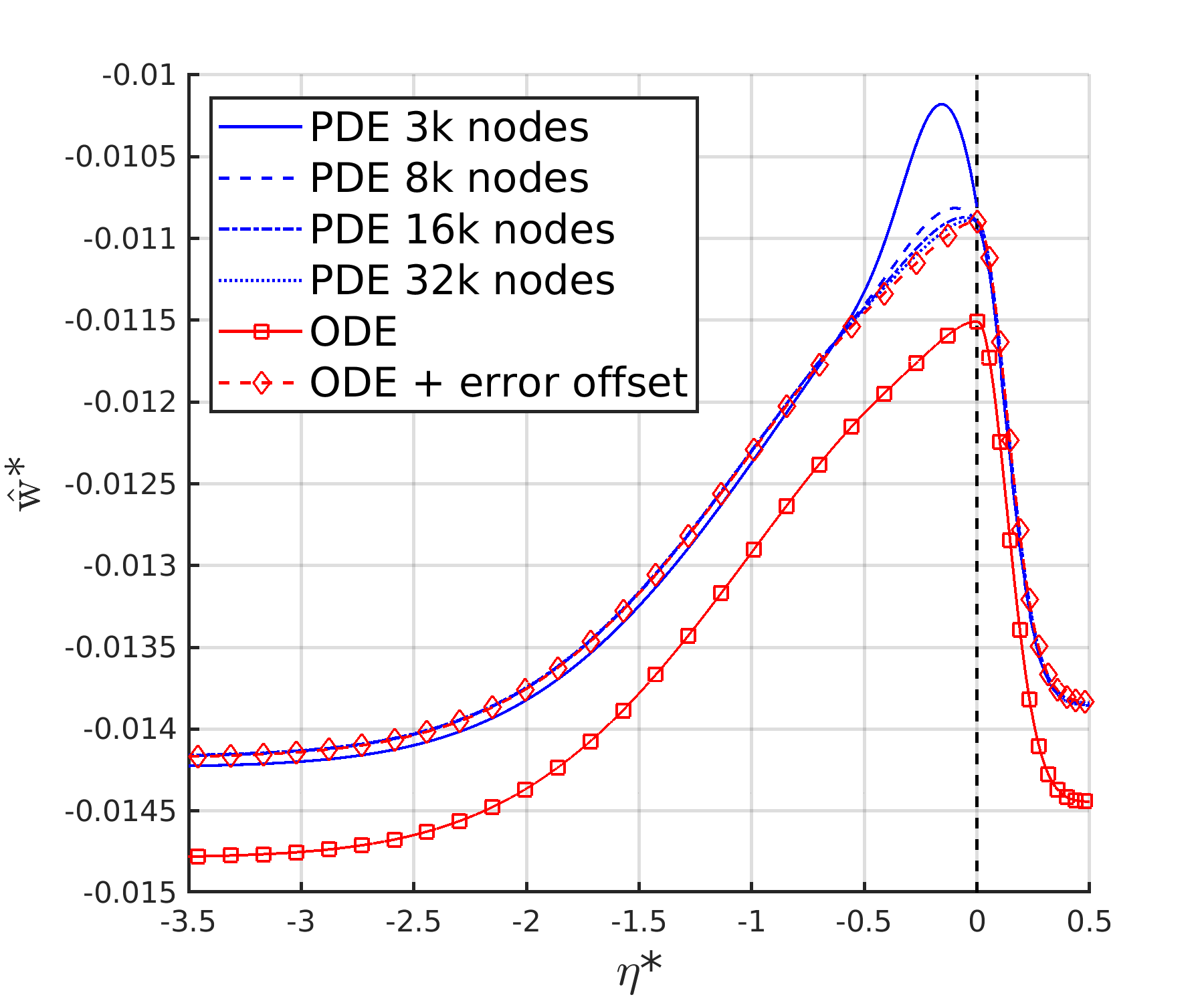}
\caption{Comparison of \(\hat{w}^*\) at 150 bar between the solution of the system of ordinary differential equations (ODE) and the solution of the system of partial differential equations (PDE) at \(x=0.01\) m mapped from (\(x\),\(y\)) to \(\eta\).}
\label{fig:comp8}
\end{figure}

Moreover, Figure~\ref{fig:comp8} presents a comparison of \(\hat{w}^*\) at 150 bar between the ODE solution and the PDE solution at \(x=0.01\) m. Two important issues can be observed in this figure. First, an offset exists between the ODE and the PDE profiles, which is related to slightly different interface equilibrium solutions, as shown in Section~\ref{subsec:interface}. Both the transverse velocity, \(v\), and the modified transverse velocity, \(w\), are variables with very small magnitudes compared to other flow parameters (e.g., temperature and density). Thus, relatively small differences of these variables at the interface can cause substantial relative errors in \(v\) and \(w\). The evolution of \(v\) or \(w\) ultimately depend on a first-order ODE, with the interface value acting as the only needed boundary condition. When introducing a numerical offset as the difference between \(w_{\Gamma,\text{ODE}}\) and \(w_{\Gamma,\text{PDE}}\) (or being the same the difference between mass flux predictions at the interface, \(\dot{\omega}_\text{ODE}\) and \(\dot{\omega}_\text{PDE}\)), the profiles are seen to collapse onto each other. Second, the PDE solution has some problems predicting the correct evolution of \(w\) in the liquid phase. From the ODE, \(w\) should be monotonically decreasing as \(\eta^*\rightarrow -\infty\). However, the PDE solution predicts a bump before decreasing. Refining the mesh used in Davis et al.~\cite{davis2019development} mitigates the problem and the PDE solution tends asymptotically to the ODE solution. This issue has been identified as a numerical problem, but proper analysis of the error is out of scope of the present work. \par 

\begin{figure}[h!]
\centering
\begin{subfigure}{.5\textwidth}
  \centering
  \includegraphics[width=1.0\linewidth]{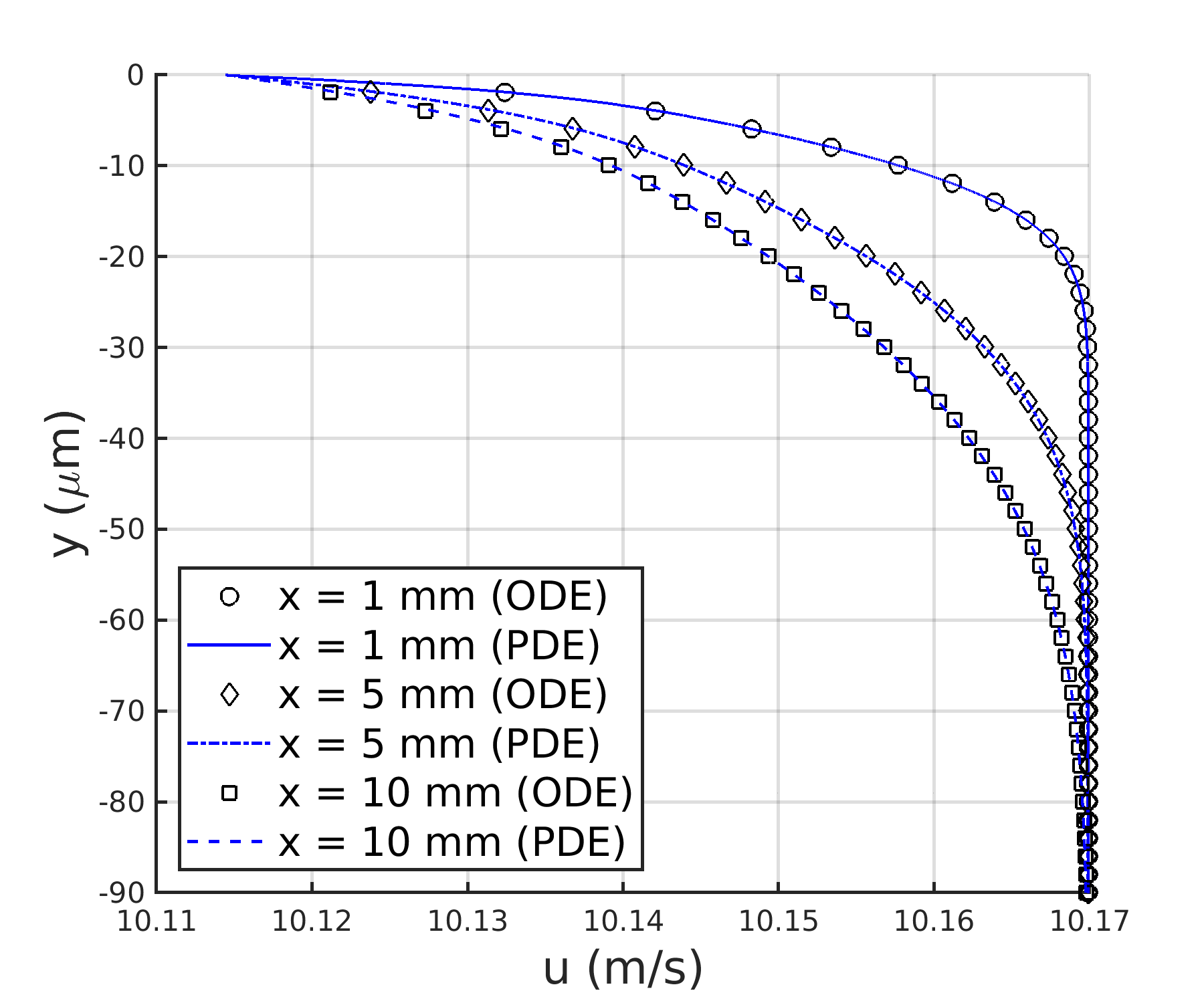}
  \caption{}
\end{subfigure}%
\begin{subfigure}{.5\textwidth}
  \centering
  \includegraphics[width=1.0\linewidth]{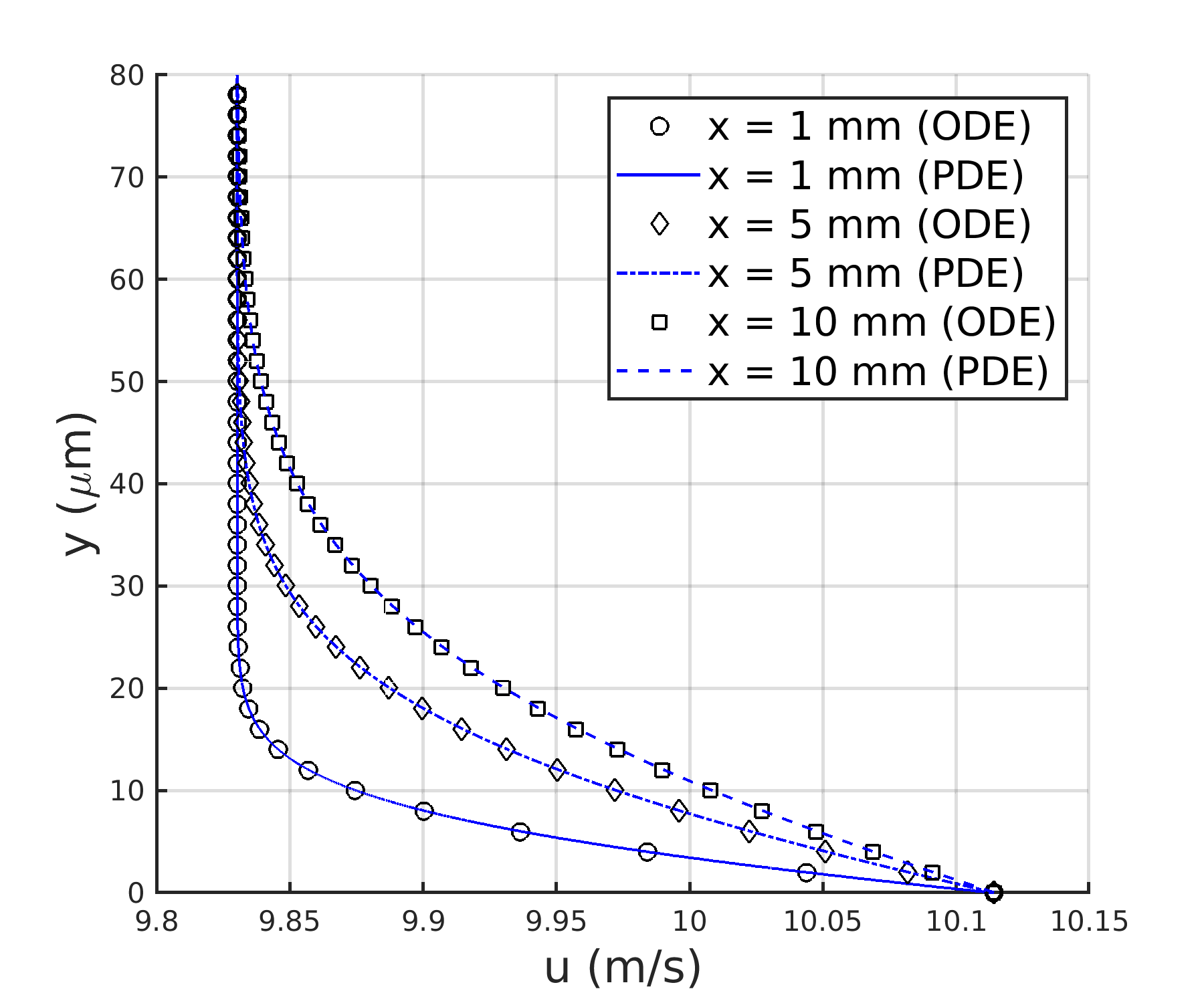}
  \caption{}
\end{subfigure}%
\caption{Comparison between the solution at 150 bar of the system of ordinary differential equations (ODE) and the solution of the system of partial differential equations (PDE) mapped from \(\eta\) to (\(x\),\(y\)). (a) \(u\) in the liquid phase; (b) \(u\) in the gas phase.}
\label{fig:comp3}
\end{figure}

\begin{figure}[h!]
\centering
\begin{subfigure}{.5\textwidth}
  \centering
  \includegraphics[width=1.0\linewidth]{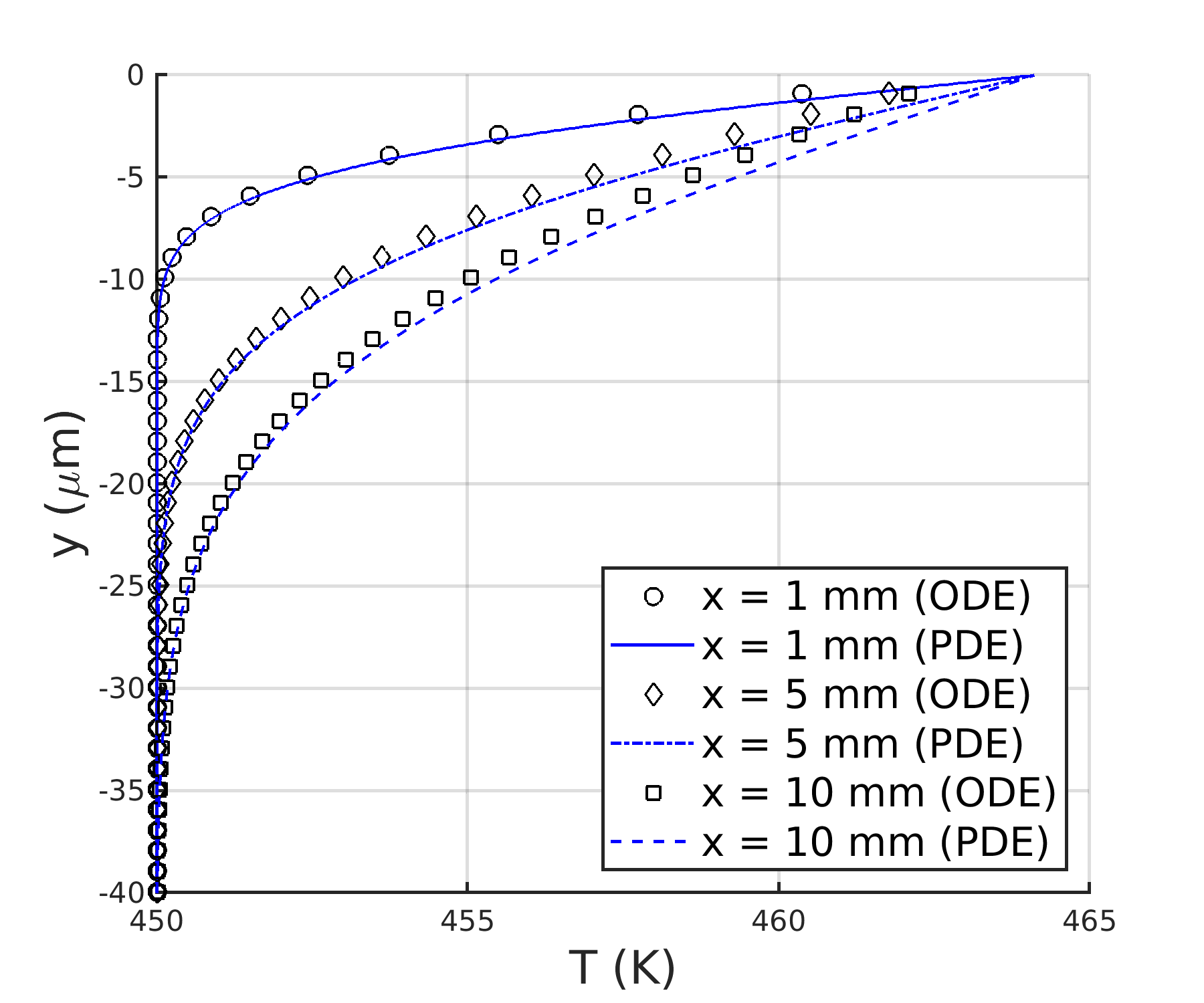}
  \caption{}
\end{subfigure}%
\begin{subfigure}{.5\textwidth}
  \centering
  \includegraphics[width=1.0\linewidth]{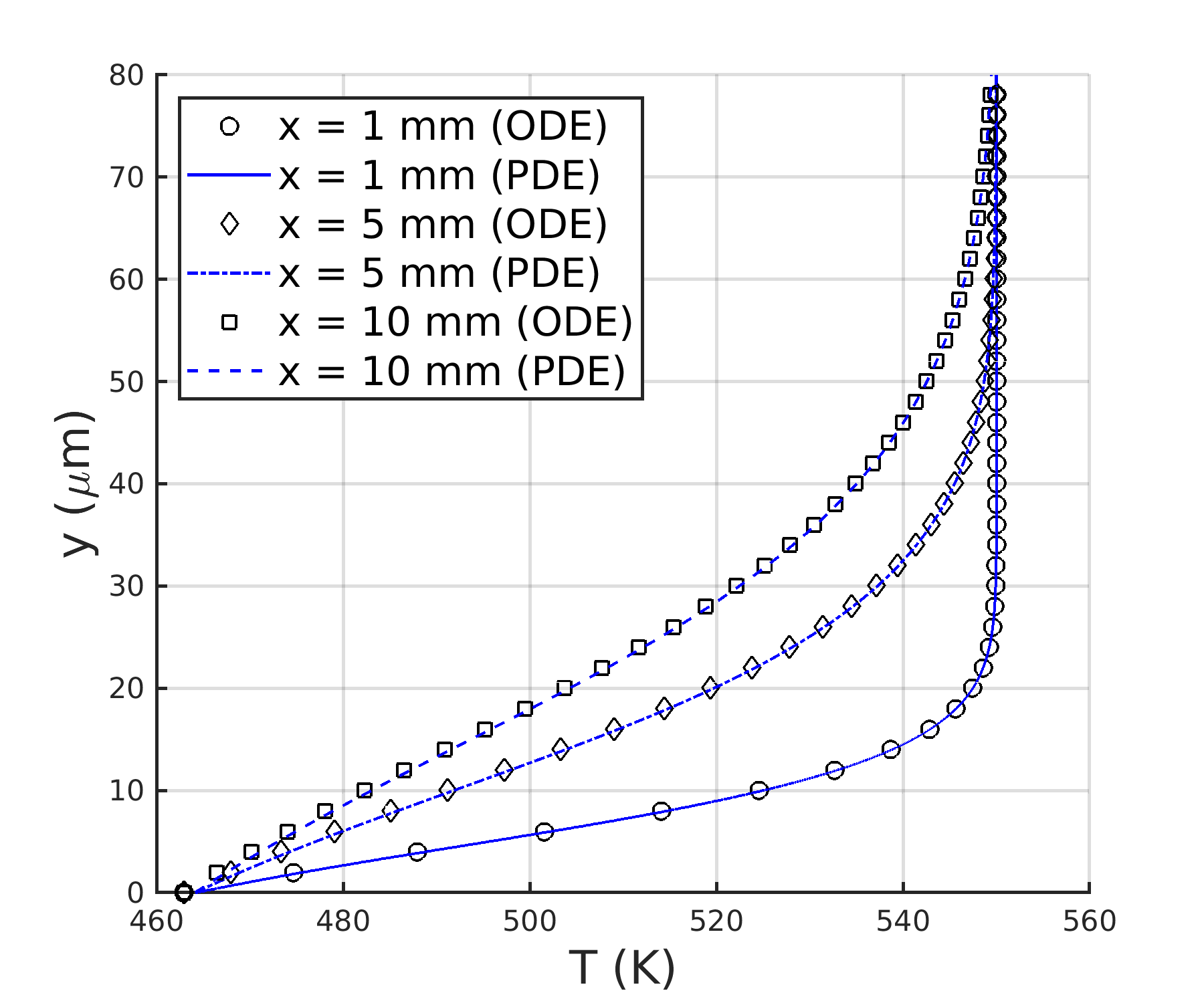}
  \caption{}
\end{subfigure}%
\caption{Comparison between the solution at 150 bar of the system of ordinary differential equations (ODE) and the solution of the system of partial differential equations (PDE) mapped from \(\eta\) to (\(x\),\(y\)). (a) \(T\) in the liquid phase; (b) \(T\) in the gas phase.}
\label{fig:comp4}
\end{figure}

\begin{figure}[h!]
\centering
\begin{subfigure}{.5\textwidth}
  \centering
  \includegraphics[width=1.0\linewidth]{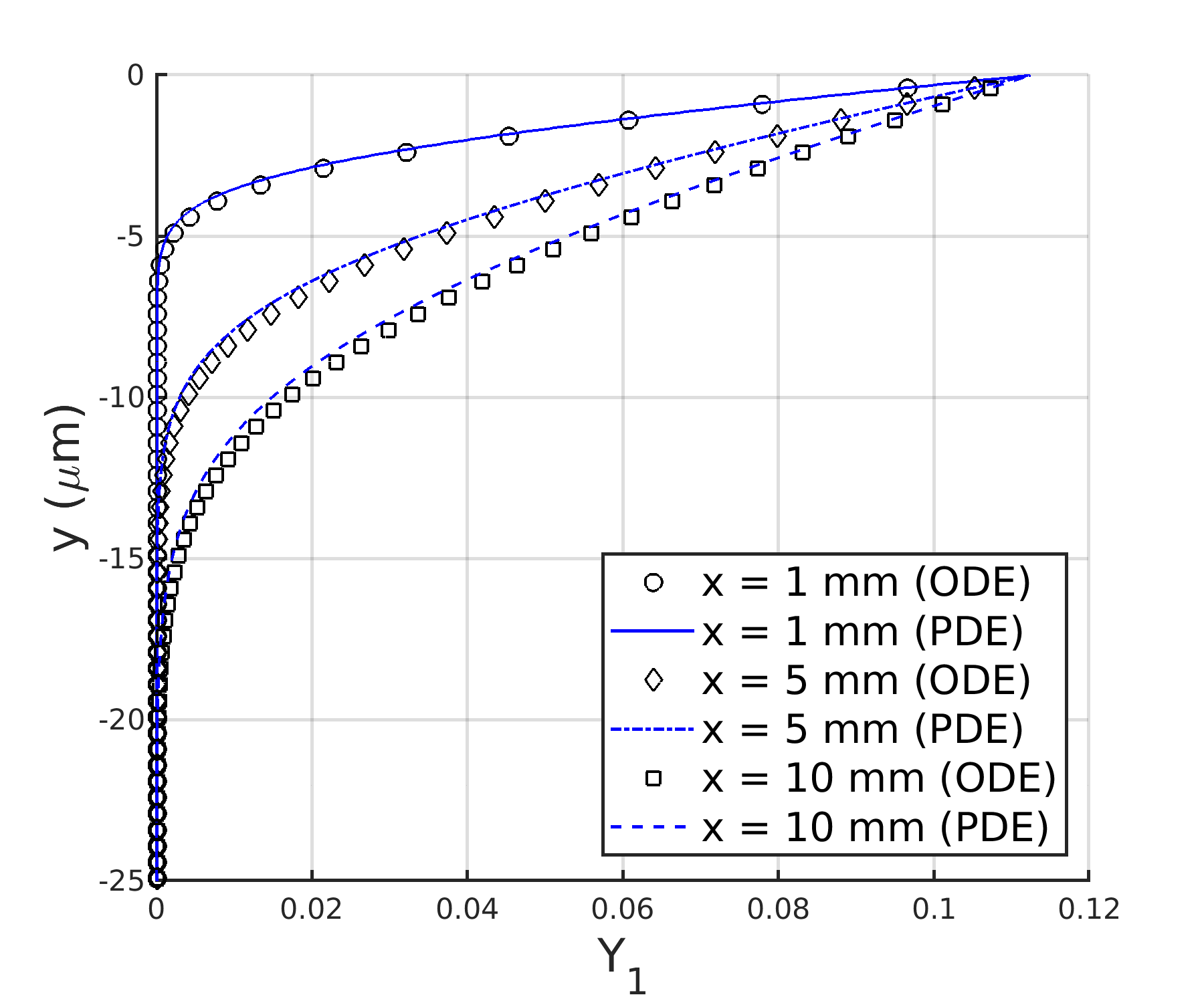}
  \caption{}
\end{subfigure}%
\begin{subfigure}{.5\textwidth}
  \centering
  \includegraphics[width=1.0\linewidth]{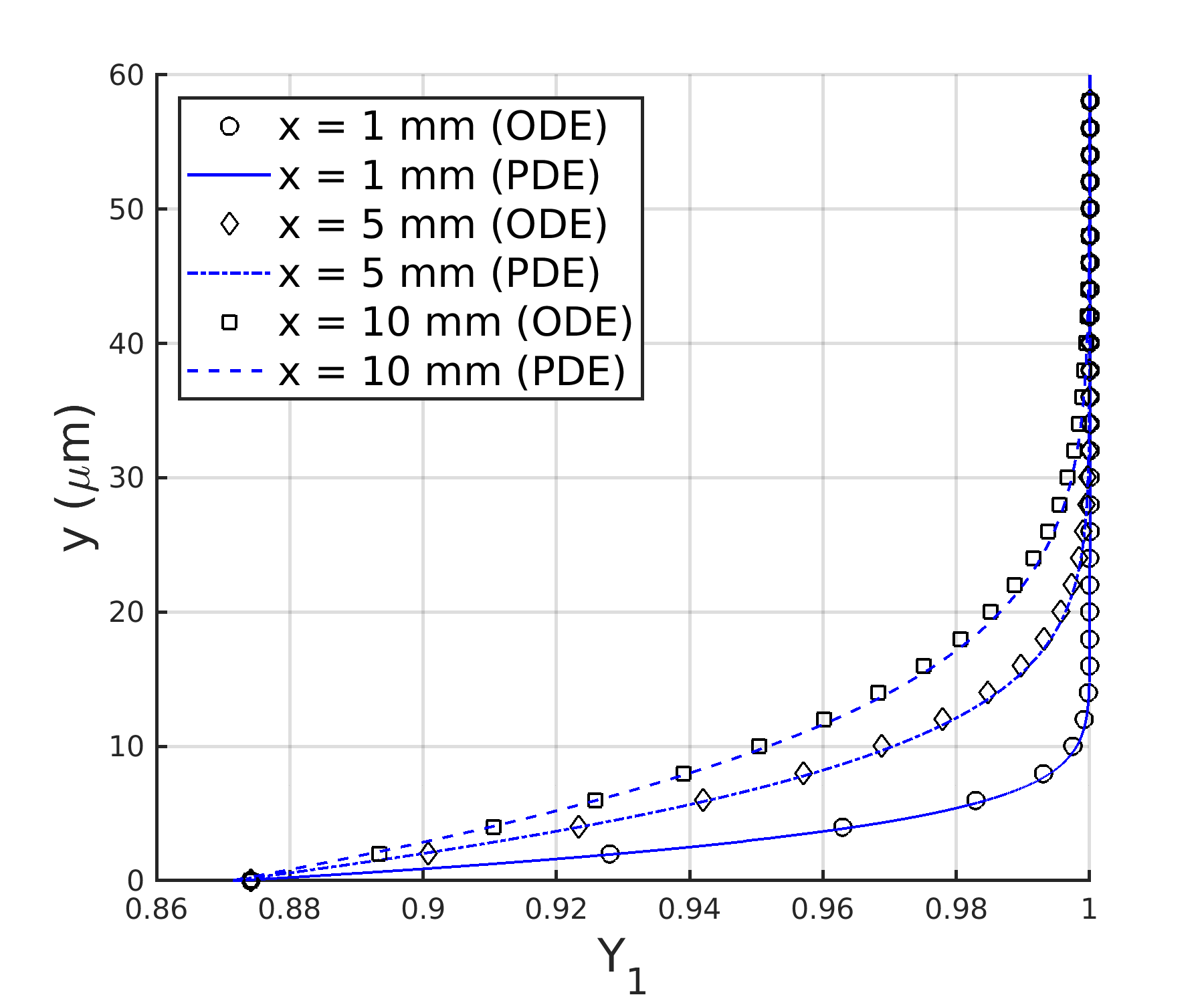}
  \caption{}
\end{subfigure}%
\caption{Comparison between the solution at 150 bar of the system of ordinary differential equations (ODE) and the solution of the system of partial differential equations (PDE) mapped from \(\eta\) to (\(x\),\(y\)). (a) \(Y_1\) in the liquid phase; (b) \(Y_1\) in the gas phase.}
\label{fig:comp5}
\end{figure}

\begin{figure}[h!]
\centering
\begin{subfigure}{.5\textwidth}
  \centering
  \includegraphics[width=1.0\linewidth]{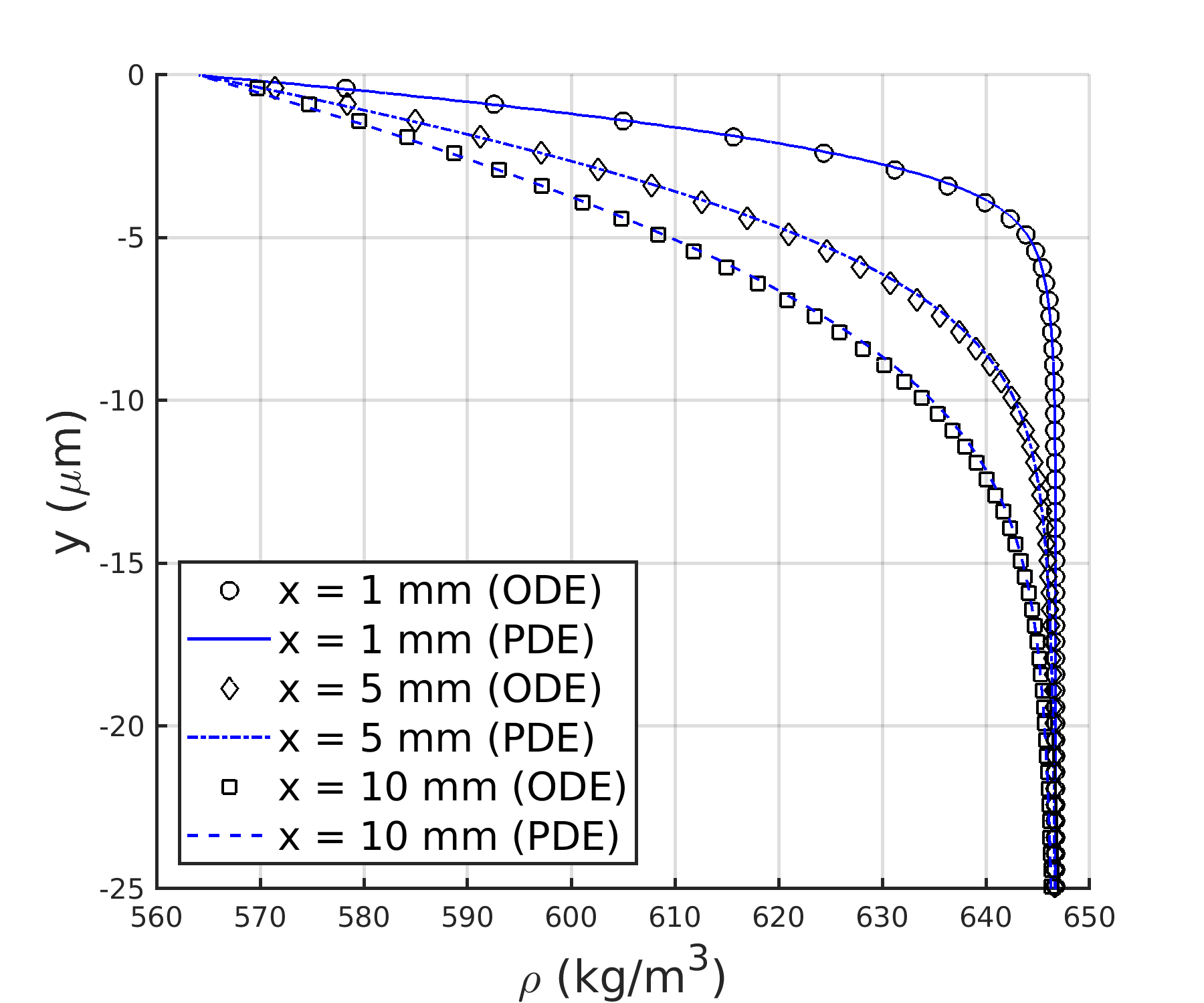}
  \caption{}
\end{subfigure}%
\begin{subfigure}{.5\textwidth}
  \centering
  \includegraphics[width=1.0\linewidth]{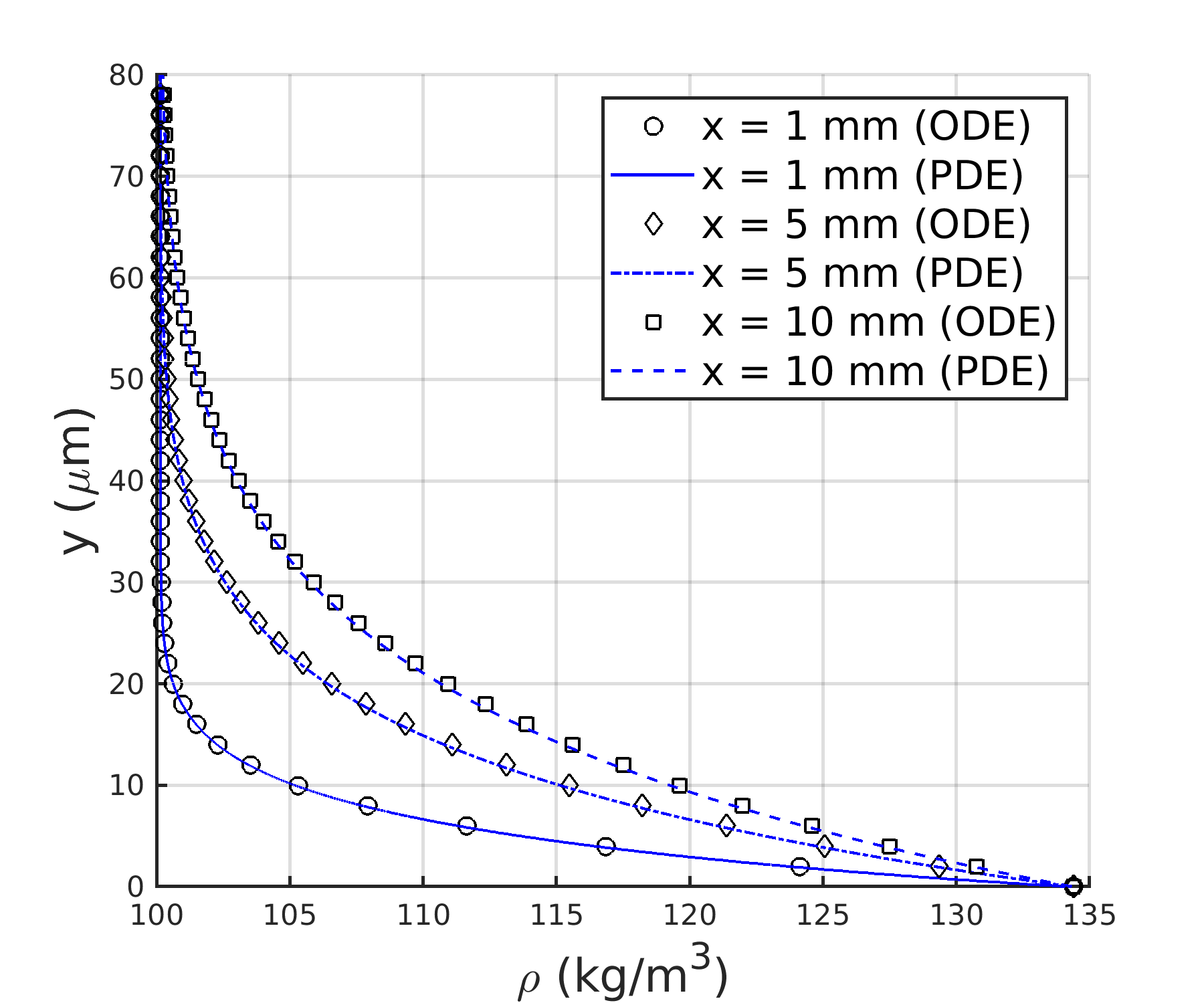}
  \caption{}
\end{subfigure}%
\caption{Comparison between the solution at 150 bar of the system of ordinary differential equations (ODE) and the solution of the system of partial differential equations (PDE) mapped from \(\eta\) to (\(x\),\(y\)). (a) \(\rho\) in the liquid phase; (b) \(\rho\) in the gas phase.}
\label{fig:comp6}
\end{figure}

\begin{figure}[h!]
\centering
\begin{subfigure}{.5\textwidth}
  \centering
  \includegraphics[width=1.0\linewidth]{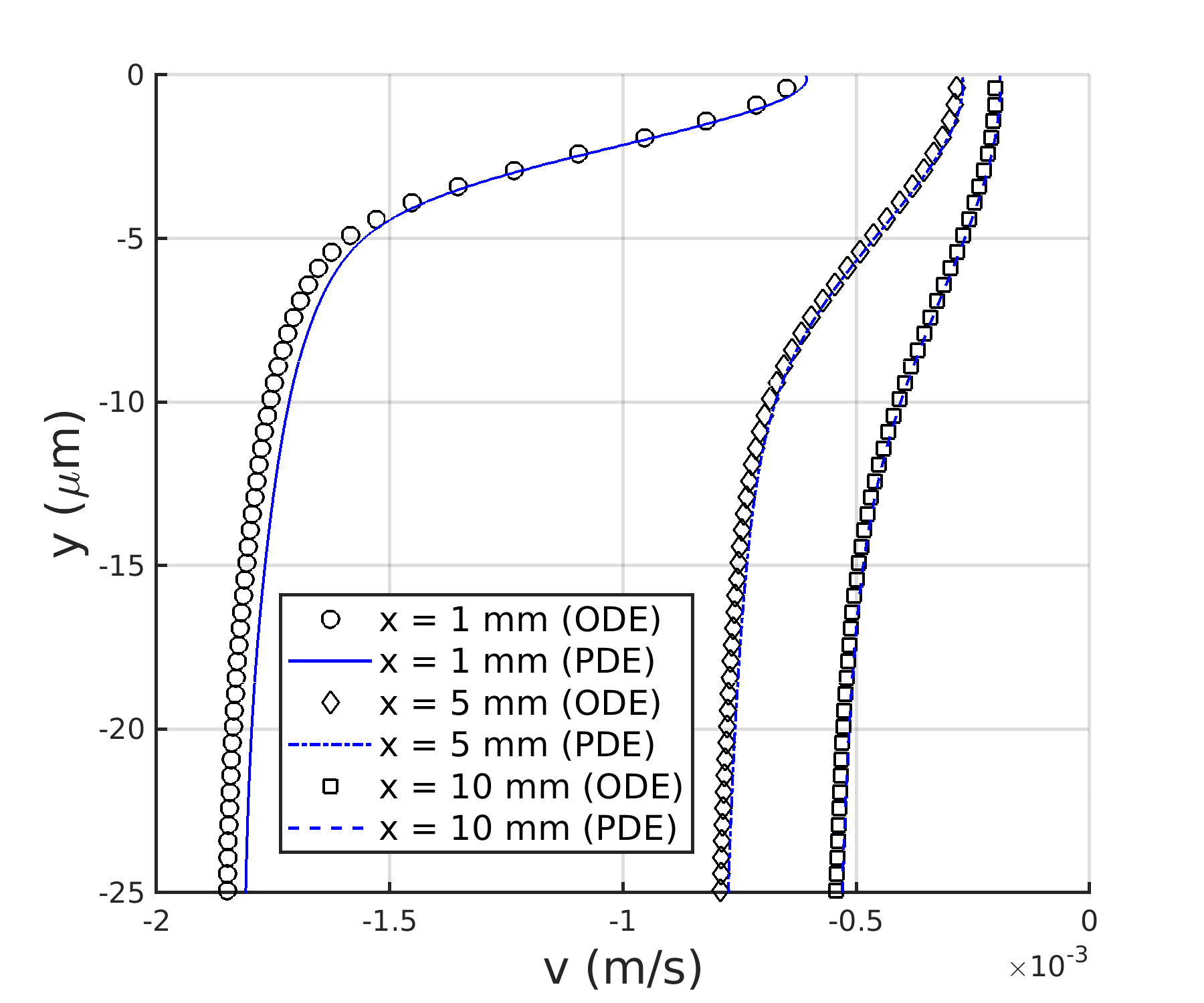}
  \caption{}
\end{subfigure}%
\begin{subfigure}{.5\textwidth}
  \centering
  \includegraphics[width=1.0\linewidth]{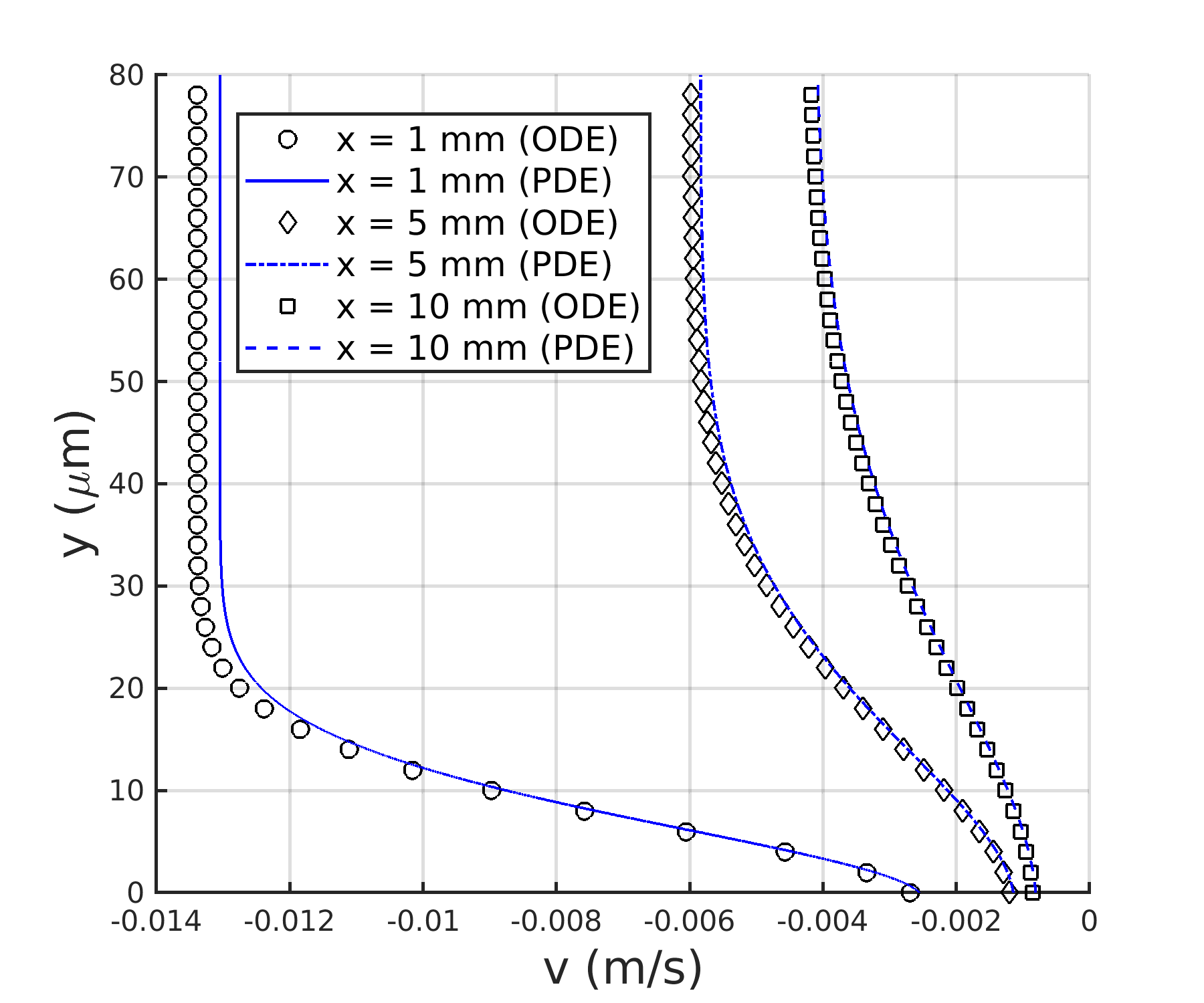}
  \caption{}
\end{subfigure}%
\caption{Comparison between the solution at 150 bar of the system of ordinary differential equations (ODE) and the solution of the system of partial differential equations (PDE) mapped from \(\eta\) to (\(x\),\(y\)). (a) \(v\) in the liquid phase; (b) \(v\) in the gas phase.}
\label{fig:comp7}
\end{figure}

Furthermore, the opposite mapping direction has been tested as well. That is, the ODE solution is mapped from the \(\eta\) space to the (\(x\),\(y\)) space. Again, the main goal is to assess the validity of both approaches and identify numerical errors that might have arisen from the PDE-to-ODE domain mapping. Figures~\ref{fig:comp3}-\ref{fig:comp7} show the transverse profiles of different variables of interest (\(u\), \(T\), \(Y_1\), \(\rho\) and \(v\)) at 150 bar at various downstream locations well into the self-similar region of the mixing layer. According to Davis et al.~\cite{davis2019development}, the mixing layer approximation with the boundary layer equations becomes valid for the analyzed configuration at \(Re_x \approx 60\) in the gas mixing layer and \(Re_x \approx 240\) in the liquid mixing layer, as defined by Eq.~(\ref{eqn:reynolds}). This corresponds to a streamwise distance from the splitter plate of 0.06 mm and 0.24 mm, respectively. Following this restriction, the analyzed locations are \(x=1\) mm, \(x=5\) mm and \(x=10\) mm. \par

The profiles shown in Figures~\ref{fig:comp3}-\ref{fig:comp7} tend to collapse onto each other although some small deviations are seen depending on the variable and the fluid phase analyzed. The streamwise velocity profiles agree almost perfectly in both phases, but the mass fraction and density profiles deviate slightly. The greatest errors appear when comparing the temperature profile in the liquid phase, although the gas phase seems to match perfectly. The temperature range across the liquid thermal mixing layer is very small compared to the temperature difference between the two pure fluids, which can magnify errors on the liquid side plot while hiding them on the gas side plot. \par

Figure~\ref{fig:comp7} presents the transverse velocity profiles evaluated using Eq.~(\ref{eqn:findv6}). A mismatch between the ODE and the PDE solutions exist, which is easily observed in the profiles at \(x=1\) mm due to the larger magnitude of \(v\) at this location. Following what has been commented in previous paragraphs, small differences in the interface solution between the ODE and the PDE models are responsible for this mismatch. However, the profiles from both solutions evolve in a similar manner. \par 

The results shown in Figures~\ref{fig:comp1}-\ref{fig:comp7} suggest that the non-ideal two-phase laminar mixing layer equations can be represented with a self-similar system of equations even though they are coupled to a complex thermodynamic model involving a cubic equation of state, high-pressure correlations used to evaluate transport properties, and non-ideal thermodynamic principles. To further compare the two approaches, Section~\ref{subsec:interface} examines the interface solution obtained with the system of ODEs and the steady interface solution obtained far downstream of the splitter plate (i.e., \(x=0.01\) m) with the system of PDEs.

\subsection{Interface equilibrium solution}
\label{subsec:interface}

This section compares the interface equilibrium solution obtained with the self-similar model with the results shown in Davis et al.~\cite{davis2019development} obtained with the mixing layer equations. To assess the performance of the self-similar system of equations, the error of a given variable \(\Theta\) between the solution of the ODE system and the PDE system is defined as

\begin{equation}
\label{eqn:error_comparison}
E_\Theta = \frac{|\Theta_\text{ODE}-\Theta_\text{PDE}|}{|\Theta_\text{PDE}|}
\end{equation}

\noindent
representing a relative deviation error from the PDE solution. \par 

The interface values and the relative error of different variables of interest (\(U_\Gamma\), \(T_\Gamma\), interface composition, etc.) obtained from the ODE and PDE solutions are shown in Tables~\ref{tab:comparison1}-\ref{tab:comparison5}. A very good agreement is seen, especially for the high pressure cases (i.e., 50, 100 and 150 bar). At 10 bar, errors are larger, which could be related to a slightly worse resolution of the liquid mixing layer in the PDE solution for a fixed mesh size. That is, diffusion is almost negligible in the liquid phase at very low pressures and it could affect the evaluation of enthalpy and concentration gradients needed to solve the interface matching conditions. \par 

Larger errors are observed when comparing the mass flux across the interface at a given location (see Table~\ref{tab:comparison5}). For 50, 100 and 150 bar, the errors are about 5\%; but again the 10 bar case presents a larger error, in line with the comparison of other interface properties. Nevertheless, the transition from net vaporization to net condensation is still well captured. \par 

Sensitivity issues might arise when evaluating \(\dot{\omega}\). Even though the relative error in interface temperature, composition, etc. is well below 1\%, the magnitude of \(\dot{\omega}\) itself is very small and easily perturbed by a small change in the other variables. This issue is then reflected in the transverse velocity profiles, as discussed in Section~\ref{subsec:profiles} and seen in Figures~\ref{fig:comp8} and~\ref{fig:comp7}, and might be one of the main causes of the small mismatch between the ODE and the PDE results when comparing mixing profiles across the mixing layer. \par 

\begin{table}[!h]
\centering
\begin{tabular}{ c c c c c c c}
\hline
\(p\) (bar) & \(U_{\Gamma,\text{ODE}}\) (m/s) & \(U_{\Gamma,\text{PDE}}\) (m/s) & \(E_{U_\Gamma}\) (\%) & \(T_{\Gamma,\text{ODE}}\) (K) & \(T_{\Gamma,\text{PDE}}\) (K) & \(E_{T_\Gamma}\) (\%) \\
\hline
10 & 12.174 & 12.171 & 0.025 & 454.261 & 451.497 & 0.612 \\
50 & 10.389 & 10.389 & \(\sim 0\) & 457.653 & 457.610 & 0.009 \\
100 & 10.181 & 10.181 & \(\sim 0\) & 460.452 & 461.326 & 0.189 \\
150 & 10.114 & 10.114 & \(\sim 0\) & 462.890 & 464.121 & 0.265 \\
\hline
\end{tabular}
\caption{Comparison of the interface velocity, \(U_\Gamma\), and equilibrium temperature, \(T_\Gamma\), between the solutions obtained using the system of ordinary differential equations (ODE) and the system of partial differential equations (PDE)~\cite{davis2019development}.}
\label{tab:comparison1}
\end{table}

\begin{table}[!h]
\centering
\begin{tabular}{ c c c c c c c c c c}
\hline
\(p\) (bar) & \(Y_{g,O_2,\text{ODE}}\) & \(Y_{g,O_2,\text{PDE}}\) & \(E_{Y_{g,O_2}}\) (\%) & \(Y_{l,O_2,\text{ODE}}\) & \(Y_{l,O_2,\text{PDE}}\) & \(E_{Y_{l,O_2}}\) (\%) \\
\hline
10 & 0.5860 & 0.6052 & 3.173 & 0.0059 & 0.0059 & \(\sim 0\) \\
50 & 0.8401 & 0.8403 & 0.024 & 0.0339 & 0.0339 & \(\sim 0\) \\
100 & 0.8732 & 0.8711 & 0.241 & 0.0715 & 0.0717 & 0.279 \\
150 & 0.8741 & 0.8713 & 0.321 & 0.1123 & 0.1126 & 0.266 \\
\hline
\end{tabular}
\caption{Comparison of the interface equilibrium composition on each phase, \(Y_{g,O_2}\) and \(Y_{l,O_2}\), between the solutions obtained using the system of ordinary differential equations (ODE) and the system of partial differential equations (PDE)~\cite{davis2019development}.}
\label{tab:comparison2}
\end{table}

\begin{table}[!h]
\centering
\begin{tabular}{ c c c c c c c c c c}
\hline
\(p\) (bar) & \(\rho_{g,\text{ODE}}\) (kg/m\(^3\)) & \(\rho_{g,\text{PDE}}\) (kg/m\(^3\)) & \(E_{\rho_{g}}\) (\%) & \(\rho_{l,\text{ODE}}\) (kg/m\(^3\)) & \(\rho_{l,\text{PDE}}\) (kg/m\(^3\)) & \(E_{\rho_{l}}\) (\%) \\
\hline
10 & 12.622 & 12.413 & 1.684 & 590.485 & 593.529 & 0.513 \\
50 & 47.934 & 47.932 & 0.004 & 580.350 & 580.403 & 0.009 \\
100 & 91.435 & 91.434 & 0.001 & 572.964 & 571.833 & 0.198 \\
150 & 134.396 & 134.362 & 0.025 & 565.565 & 563.888 & 0.297 \\
\hline
\end{tabular}
\caption{Comparison of the interface equilibrium mixture densities on each phase, \(\rho_{g}\) and \(\rho_{l}\), between the solutions obtained using the system of ordinary differential equations (ODE) and the system of partial differential equations (PDE)~\cite{davis2019development}.}
\label{tab:comparison3}
\end{table}

\begin{table}[!h]
\centering
\begin{tabular}{ c c c c c c c c c c}
\hline
\(p\) (bar) & \(h_{g,\text{ODE}}\) (kJ/kg) & \(h_{g,\text{PDE}}\) (kJ/kg) & \(E_{h_{g}}\) (\%) & \(h_{l,\text{ODE}}\) (kJ/kg) & \(h_{l,\text{PDE}}\) (kJ/kg) & \(E_{h_{l}}\) (\%) \\
\hline
10 & 496.990 & 489.153 & 1.602 & 340.530 & 332.696 & 2.355 \\
50 & 444.700 & 444.624 & 0.017 & 355.620 & 355.498 & 0.034 \\
100 & 435.700 & 437.093 & 0.319 & 370.160 & 372.589 & 0.652 \\
150 & 433.110 & 435.013 & 0.438 & 383.250 & 386.624 & 0.873 \\
\hline
\end{tabular}
\caption{Comparison of the interface equilibrium mixture enthalpies on each phase, \(h_{g}\) and \(h_{l}\), between the solutions obtained using the system of ordinary differential equations (ODE) and the system of partial differential equations (PDE)~\cite{davis2019development}.}
\label{tab:comparison4}
\end{table}

\begin{table}[!h]
\centering
\begin{tabular}{ c c c c }
\hline
\(p\) (bar) & \(\dot{\omega}_\text{ODE}\) (kg/m\(\text{s}^2\)) & \(\dot{\omega}_\text{PDE}\) (kg/m\(\text{s}^2\)) & \(E_{\dot{\omega}}\) (\%) \\
\hline
10 & 0.0849 & 0.0760 & 11.711 \\
50 & 0.00511 & 0.00488 & 4.713 \\
100 & -0.0573 & -0.0539 & 6.308 \\
150 & -0.115 & -0.109 & 5.505 \\
\hline
\end{tabular}
\caption{Comparison of the net mass flux across the interface, \(\dot{\omega}\), at the downstream location \(x=\bar{x}=0.01\) m between the solutions obtained using the system of ordinary differential equations (ODE) and the system of partial differential equations (PDE)~\cite{davis2019development}.}
\label{tab:comparison5}
\end{table}

\subsection{Mixing layer thickness}
\label{subsec:thickness}

Discussion in prior subsections addressed the similarity of solutions at differing downstream positions for a given problem configuration. The similarity between cases with differing constraints is discussed in this section. \par 

The self-similar solution can provide information on the exact evolution of the mixing layer thickness. This is true for simpler analyses, such as the self-similar solution of incompressible flow over a flat plate (a.k.a., the Blasius equation)~\cite{white2006viscous}. However, semi-empirical formulas are usually provided to estimate boundary layer or mixing layer thicknesses for more complex flows, such as compressible flows, with different boundary conditions~\cite{white2006viscous}. These equations rely on many flow parameters and are not as accurate as obtaining the mixing layer thickness directly from the similarity solution of the specific constraints. \par

A more accurate approach is presented in this paper, which covers variable density flows under non-ideal conditions. It relies on solving the self-similar system of equations coupled with a non-ideal thermodynamic model. It is not as simple to implement as other semi-empirical models, but as discussed later, some generalization is possible. To evaluate the layer thickness on each side of the interface for mass, momentum and energy mixing regions, the following non-dimensional parameters are defined

\begin{equation}
\label{eqn:thick_Y}
\theta_{Y,l}^*(\eta) = \frac{Y_l(0)-Y(\eta)}{Y_l(0)-Y_L} \quad ; \quad \theta_{Y,g}^*(\eta) = \frac{Y(\eta)-Y_g(0)}{Y_G-Y_g(0)}
\end{equation}

\begin{equation}
\label{eqn:thick_u}
\theta_{u,l}^*(\eta) = \frac{U_\Gamma-u(\eta)}{U_\Gamma-U_L} \quad ; \quad \theta_{u,g}^*(\eta) = \frac{u(\eta)-U_\Gamma}{U_G-U_\Gamma}
\end{equation}

\begin{equation}
\label{eqn:thick_T}
\theta_{T,l}^*(\eta) = \frac{T_\Gamma-T(\eta)}{T_\Gamma-T_L} \quad ; \quad \theta_{T,g}^*(\eta) = \frac{T(\eta)-T_\Gamma}{T_G-T_\Gamma}
\end{equation}

\noindent
where all profiles vary from 0 to 1 going from the interface to the freestream condition. The mixture composition is used to analyze the mass mixing layer, while the streamwise velocity represents the momentum mixing layer. For the energy mixing layer, temperature has been used since it is a more representative variable than enthalpy. That is, the thermal mixing layer is represented, separately from composition influence on the enthalpy. \par 

\begin{table}[!h]
\centering
\begin{tabular}{ c c c c c c c }
\hline
Case & Mixture & \(p\) (bar) & \(u_G\) (m/s) & \(u_L\) (m/s) & \(T_G\) (K) & \(T_L\) (K) \\
\hline
A & \(O_2/C_{10}H_{22}\) & 10  & 7.673 & 12.327 & 550 & 450 \\
B & \(O_2/C_{10}H_{22}\) & 50  & 9.525 & 10.475 & 550 & 450 \\
C & \(O_2/C_{10}H_{22}\) & 100 & 9.755 & 10.246 & 550 & 450 \\
D & \(O_2/C_{10}H_{22}\) & 150 & 9.830 & 10.170 & 550 & 450 \\
E & \(O_2/C_{10}H_{22}\) & 100 & 9.755 & 10.246 & 590 & 490 \\
F & \(O_2/C_{10}H_{22}\) & 150 & 9.830 & 10.170 & 510 & 410 \\
G & \(O_2/C_{8}H_{18}\)  & 100 & 9.755 & 10.246 & 550 & 450 \\
\hline
\end{tabular}
\caption{Analyzed cases for the evaluation of the mixing layer thickness model. Cases A-D correspond to those analyzed in Davis et al.~\cite{davis2019development}.}
\label{tab:thick_case}
\end{table}

\begin{figure}[h!]
\centering
\begin{subfigure}{.5\textwidth}
  \centering
  \includegraphics[width=1.0\linewidth]{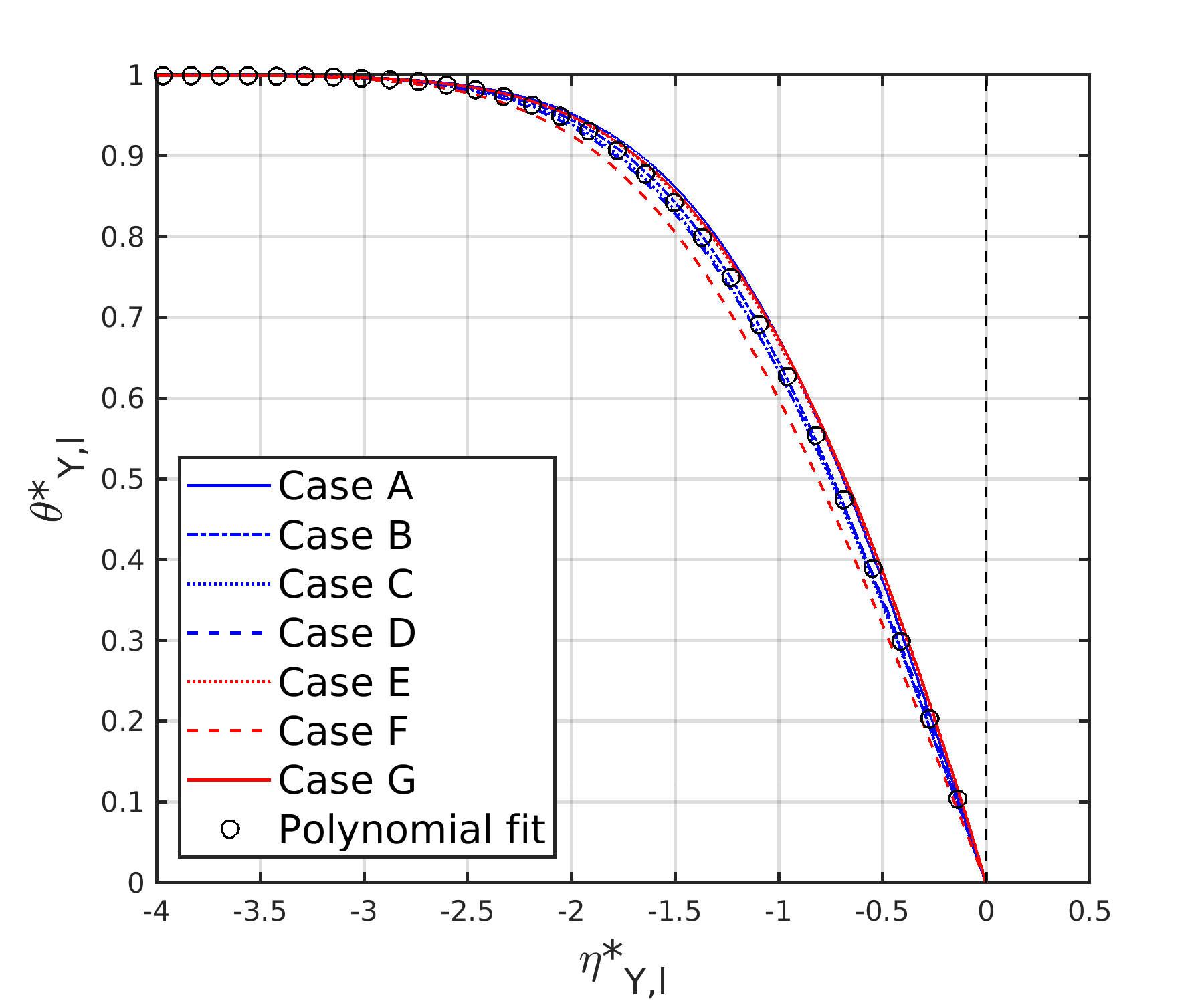}
  \caption{}
\end{subfigure}%
\begin{subfigure}{.5\textwidth}
  \centering
  \includegraphics[width=1.0\linewidth]{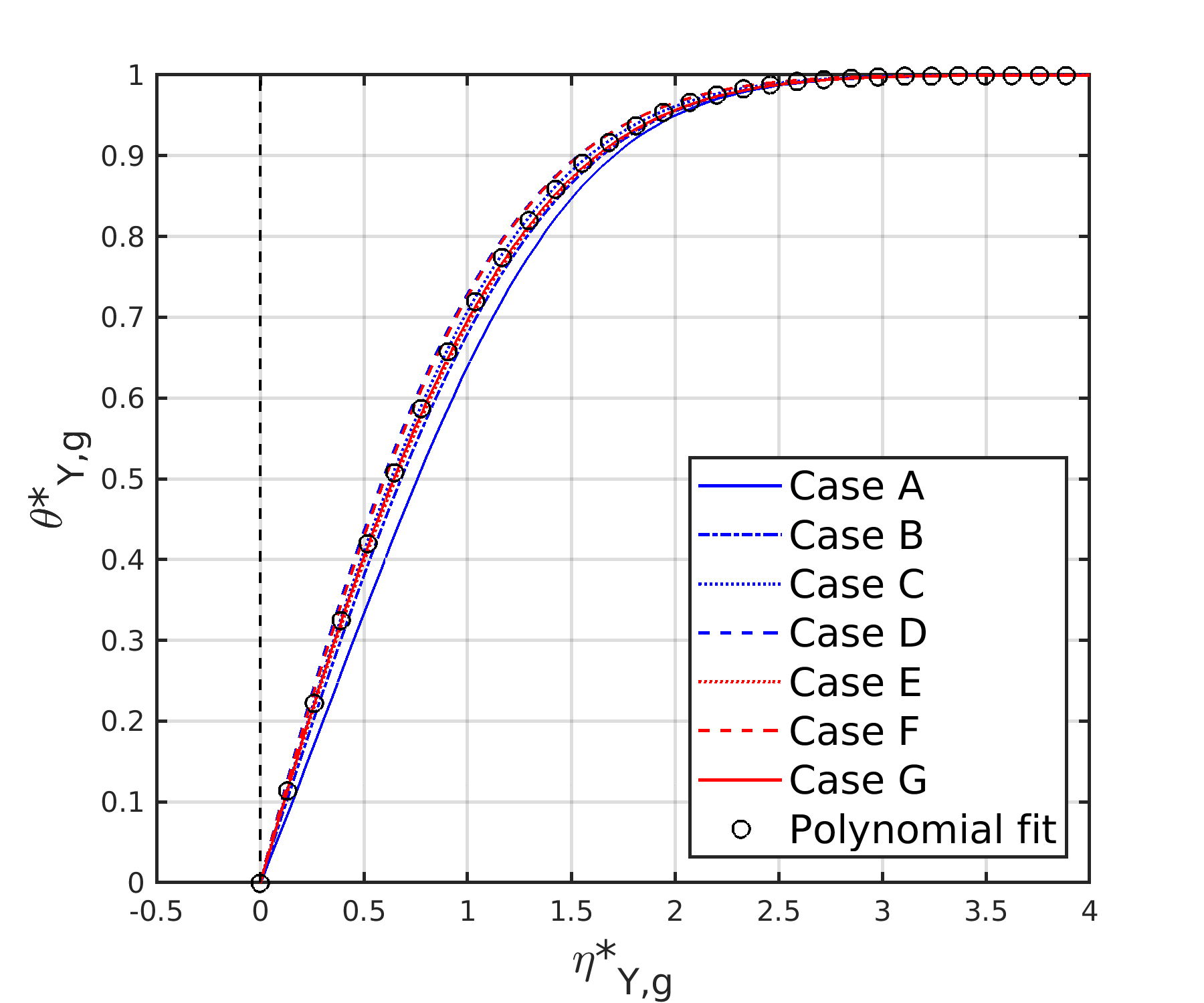}
  \caption{}
\end{subfigure}%
\caption{Mass mixing layer thickness for different pressures, freestream velocities, freestream temperatures and mixture components. See Table~\ref{tab:thick_case} for reference. (a) liquid phase; (b) gas phase.}
\label{fig:thick_Y}
\end{figure}

\begin{figure}[h!]
\centering
\begin{subfigure}{.5\textwidth}
  \centering
  \includegraphics[width=1.0\linewidth]{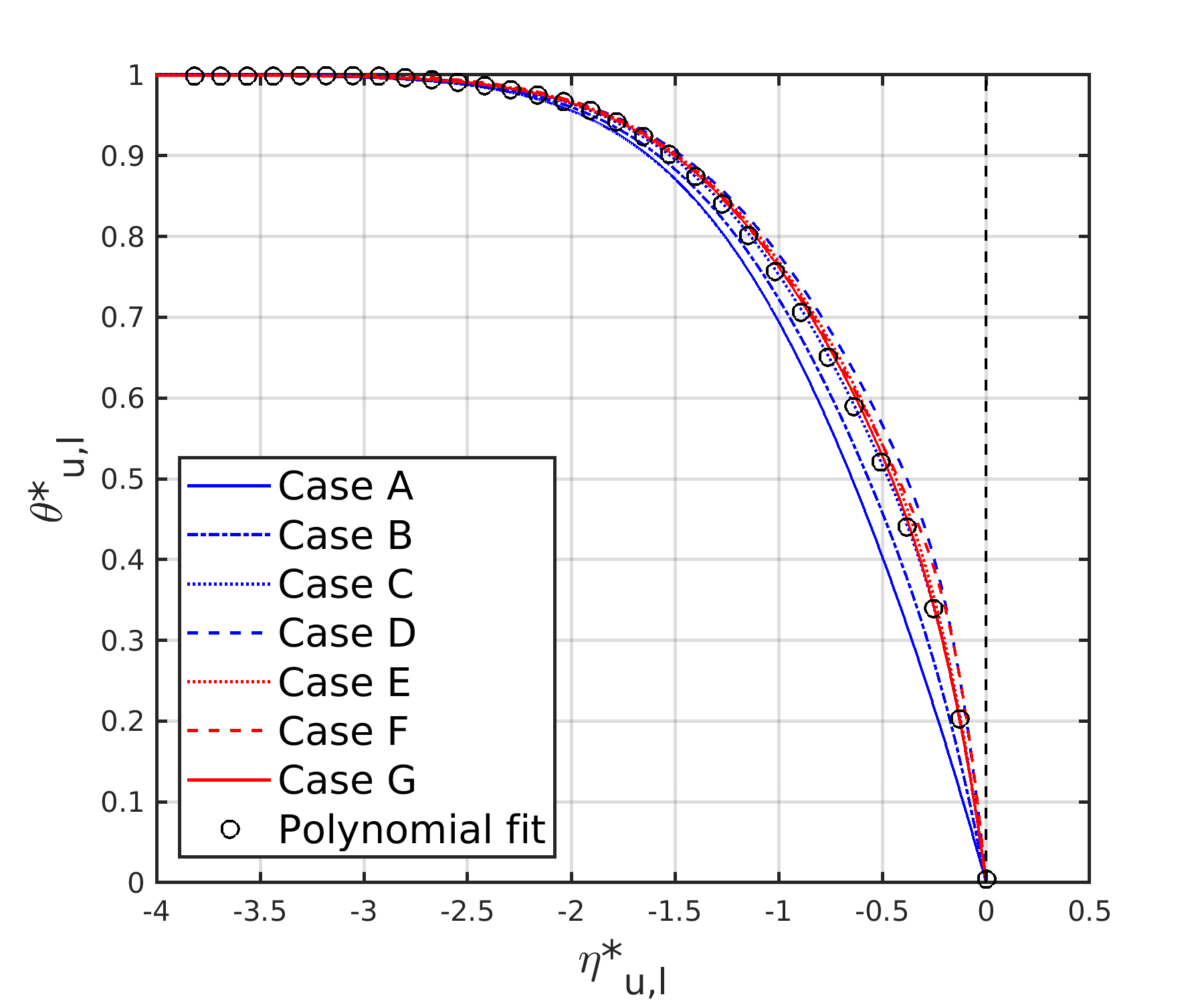}
  \caption{}
\end{subfigure}%
\begin{subfigure}{.5\textwidth}
  \centering
  \includegraphics[width=1.0\linewidth]{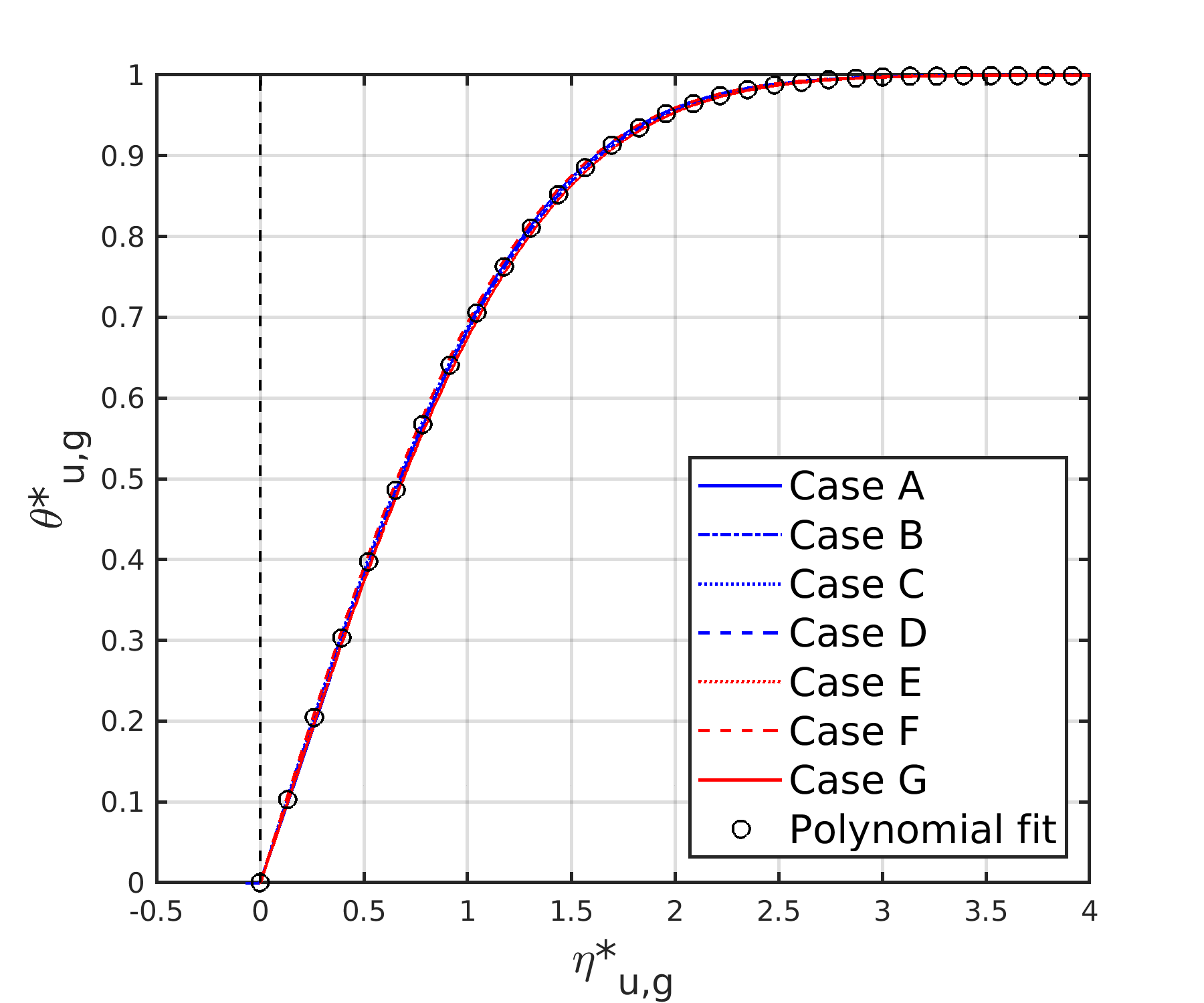}
  \caption{}
\end{subfigure}%
\caption{Momentum mixing layer thickness for different pressures, freestream velocities, freestream temperatures and mixture components. See Table~\ref{tab:thick_case} for reference. (a) liquid phase; (b) gas phase.}
\label{fig:thick_u}
\end{figure}

\begin{figure}[h!]
\centering
\begin{subfigure}{.5\textwidth}
  \centering
  \includegraphics[width=1.0\linewidth]{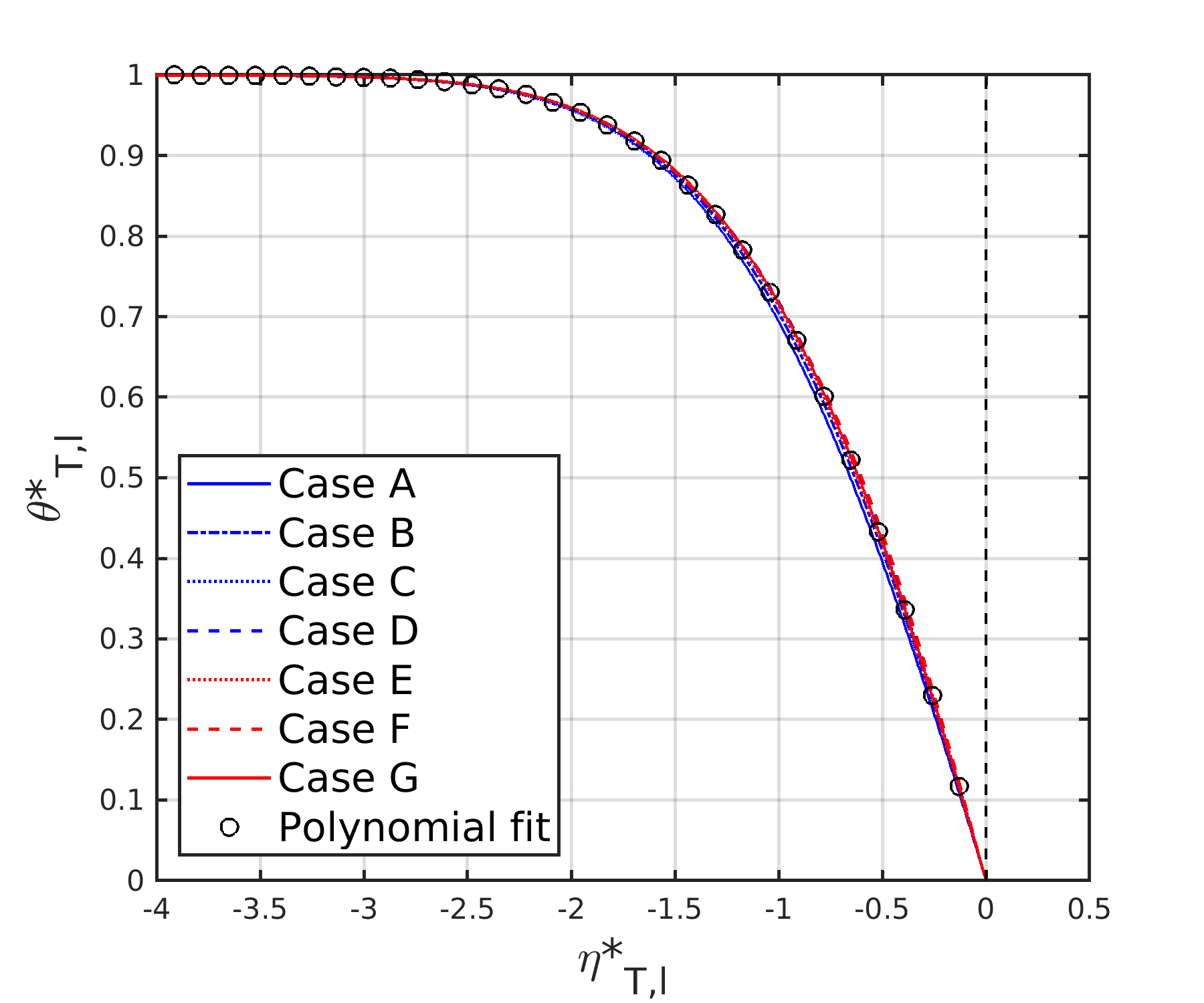}
  \caption{}
\end{subfigure}%
\begin{subfigure}{.5\textwidth}
  \centering
  \includegraphics[width=1.0\linewidth]{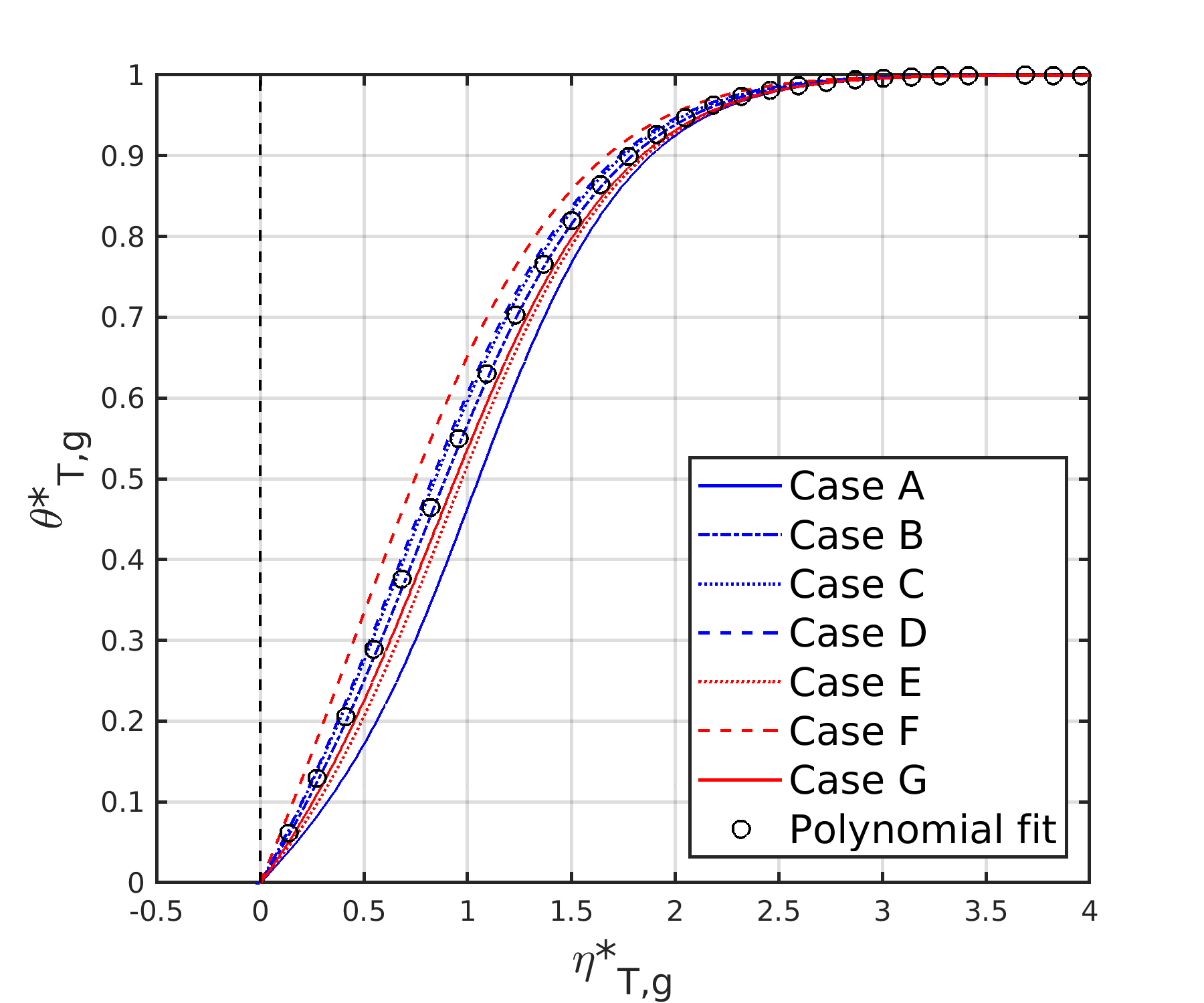}
  \caption{}
\end{subfigure}%
\caption{Thermal mixing layer thickness for different pressures, freestream velocities, freestream temperatures and mixture components. See Table~\ref{tab:thick_case} for reference. (a) liquid phase; (b) gas phase.}
\label{fig:thick_T}
\end{figure}

Davis et al.~\cite{davis2019development} report the non-similarity across different problem constraints. However, when the proper fluid properties are used to non-dimensionalize the similarity variable and the mixing profiles, the thickness of the mixing layers almost collapse onto each other for different configurations. The analyzed cases are defined in Table~\ref{tab:thick_case}. This is shown in Figures~\ref{fig:thick_Y},~\ref{fig:thick_u} and~\ref{fig:thick_T}. The profiles may also be nearly self-similar, although, in general, the variations in the behavior of the selected thermodynamic model creates different profiles as the problem configuration changes. The non-dimensional similarity variables used to plot the mass mixing layer are defined from freestream conditions for each phase as

\begin{equation}
\label{eqn:thick_Y_sim}
\eta_{Y,l}^* = \frac{\eta}{\sqrt{\frac{\rho_L^2D_L}{u_L}}} \quad ; \quad \eta_{Y,g}^* = \frac{\eta}{\sqrt{\frac{\rho_G^2D_G}{u_G}}}
\end{equation}

\noindent
and the non-dimensional similarity variables for the momentum mixing layer and the thermal mixing layer are defined, respectively, as

\begin{equation}
\label{eqn:thick_u_sim}
\eta_{u,l}^* = \frac{\eta}{\sqrt{\frac{\rho_L\mu_L}{u_L}}} \quad ; \quad \eta_{u,g}^* = \frac{\eta}{\sqrt{\frac{\rho_G\mu_G}{u_G}}}
\end{equation}

\noindent
and

\begin{equation}
\label{eqn:thick_T_sim}
\eta_{T,l}^* = \frac{\eta}{\sqrt{\frac{\rho_L\lambda_L}{c_{p_L} u_L}}} \quad ; \quad \eta_{T,g}^* = \frac{\eta}{\sqrt{\frac{\rho_G\lambda_G}{c_{p_G} u_G}}}
\end{equation}

The mixing layer thickness, \(\delta\), is obtained by using Eq.~(\ref{eqn:yvseta}). \(\delta = \sqrt{2\bar{x}}\tilde{I}(\eta)\) just represents the transverse coordinate evaluated at the \(\eta\) value defining the limit of the mixing layer. This limit is obtained from Eqs.~(\ref{eqn:thick_Y_sim}),~(\ref{eqn:thick_u_sim}) or~(\ref{eqn:thick_T_sim}), depending on the mixing layer being evaluated (i.e., mass, momentum or thermal). Note that the mixing layer thickness still evolves as a function of \(\sqrt{x}\) whether the fluid experiences ideal or real conditions. \par 

The diffusion layer thickness is defined at the location where \(\theta^*\) is 99\% of the difference between the freestream condition and the interface value. Using this definition, an exact value for the non-dimensional similarity variable, \(\eta^*\), can be obtained for each case. Together with the \(\rho(\eta)\) distribution obtained from the solution of the self-similar system of equations, the evaluation of \(\delta\) is straightforward and it is an exact representation of the flow evolution under the self-similar analysis. However, this approach still depends on the solution of Eqs.~(\ref{eqn:xmom_trans2})-(\ref{eqn:ene_trans2}) coupled with a thermodynamic model and does not yield an immediate estimate of the mixing layer thickness. \par 

In view of the quasi-collapse of the location where the mixing layer ends and the reasonably similar diffusion profiles, the following approach is suggested. The main goal is to reduce the complexity of the exact evaluation of the mixing layer thickness to a simpler model relying on freestream conditions and interface properties, coupled only with an equation of state. Therefore, it is no longer needed to solve the system of ODEs and the diffusion process within the mixing layer, as long as the interface solution is already known (e.g., it could be possible to tabulate interface solutions based on the thermodynamic model and problem configuration). \par 

A polynomial of degree 9 is fitted using a least-squares method among the data defining the distributions of mass fraction, velocity and temperature within the mixing layer for each case. That is, the following function is used to represent the diffusion profiles \par 

\begin{equation}
\label{eqn:polyfit}
p(\eta^*) = \sum_{i=1}^{n=10} p_i \eta^{*^{n-i}}
\end{equation}

\noindent
where \(p(\eta^*)\) represents \(\theta_{Y,l}^*(\eta_{Y,l}^*)\), \(\theta_{Y,g}^*(\eta_{Y,g}^*)\), \(\theta_{u,l}^*(\eta_{u,l}^*)\), \(\theta_{u,g}^*(\eta_{u,g}^*)\), \(\theta_{T,l}^*(\eta_{T,l}^*)\) or \(\theta_{T,g}^*(\eta_{T,g}^*)\). The polynomial reasonably predicts the diffusion profiles with the exception of the gas temperature profile and the liquid streamwise velocity profile, where larger variations are observed (see Figures~\ref{fig:thick_Y}-\ref{fig:thick_T}). The coefficients of Eq.~(\ref{eqn:polyfit}) to predict each distribution are given in Table~\ref{tab:thick_poly}.

\begin{table}[!h]
\centering
\begin{tabular}{ c c c c c c c }
\hline
Polynomial & \(\theta_{Y,l}^*(\eta_{Y,l}^*)\) & \(\theta_{Y,g}^*(\eta_{Y,g}^*)\) & \(\theta_{u,l}^*(\eta_{u,l}^*)\) & \(\theta_{u,g}^*(\eta_{u,g}^*)\) & \(\theta_{T,l}^*(\eta_{T,l}^*)\) & \(\theta_{T,g}^*(\eta_{T,g}^*)\) \\
\hline
\(p_1 \times 10^4\) & 0.15167  & 0.19023  & -3.29382 & 0.21758  & 0.10990  & 1.02187  \\
\(p_2 \times 10^4\) & -4.23064 & -4.66887 & -78.0620 & -5.33430 & 2.69978  & -26.3374 \\
\(p_3 \times 10^3\) & -4.90381 & 4.69201  & -78.5302 & 5.33306  & 2.92476  & 28.3791  \\
\(p_4 \times 10^2\) & -3.01070 & -2.41875 & -43.6676 & -2.70735 & 1.87074  & -16.4236 \\
\(p_5 \times 10^2\) & -10.1995 & 6.30927  & -146.378 & 6.69482  & 7.86786  & 54.0947  \\
\(p_6 \times 10^1\) & -1.74246 & -0.56551 & -30.2891 & -0.37532 & 2.17118  & -9.66246 \\
\(p_7 \times 10^1\) & -1.10601 & -0.41711 & -38.4520 & -1.37345 & 3.23844  & 7.32169  \\
\(p_8 \times 10^1\) & -1.44243 & -1.40570 & -30.5828 & 0.29545  & -0.27706 & -0.41950 \\
\(p_9 \times 10^1\) & -7.79459 & 8.99477  & -18.8875 & 7.85479  & -9.08996 & 4.50979  \\
\(p_{10} \times 10^4\) & -5.29066 & -0.24932 & 49.3394 & 0.27385 & -3.26408 & -4.58044 \\
\hline
\end{tabular}
\caption{Coefficients of the polynomial fit of \(\theta_{Y,l}^*(\eta_{Y,l}^*)\), \(\theta_{Y,g}^*(\eta_{Y,g}^*)\), \(\theta_{u,l}^*(\eta_{u,l}^*)\), \(\theta_{u,g}^*(\eta_{u,g}^*)\), \(\theta_{T,l}^*(\eta_{T,l}^*)\) and \(\theta_{T,g}^*(\eta_{T,g}^*)\).}
\label{tab:thick_poly}
\end{table}

A representative value of \(\eta^*\) defining the mixing layer thickness for all scenarios is obtained from the polynomial fitting of each profile (i.e., where each polynomial profile reaches 99\% of the freestream value). Table~\ref{tab:thick_res} provides information about these values for each non-dimensional similarity variable, as well as a range of observed values in all cases. Varying pressure, temperature freestream values, freestream velocities and mixture components resulted in very similar \(\eta^*\) values, with maximum relative deviations, \(\Delta\), from the polynomial solution below 5\%.

\begin{table}[!h]
\centering
\begin{tabular}{ c c c c c }
\hline
Mixing layer & \(\eta^{*}_{\text{poly}}\) & \(\eta^{*}_{\text{min}}\) & \(\eta^{*}_{\text{max}}\) & \(\Delta_{\eta^*,\text{max}}\) (\%) \\
\hline
\(\eta_{Y,l}^*\) & -2.672 & -2.785 & -2.589 & 4.229 \\
\(\eta_{Y,g}^*\) & 2.515  & 2.459  & 2.576  & 2.425 \\
\(\eta_{u,l}^*\) & -2.498 & -2.561 & -2.410 & 3.523 \\
\(\eta_{u,g}^*\) & 2.548  & 2.510  & 2.579  & 1.491 \\
\(\eta_{T,l}^*\) & -2.541 & -2.559 & -2.533 & 0.708 \\
\(\eta_{T,g}^*\) & 2.683  & 2.564  & 2.724  & 4.435 \\
\hline
\end{tabular}
\caption{Values of the non-dimensional similarity variables for mass, momentum and thermal mixing layer thicknesses obtained from the fitted polynomial. A range of values is also provided for the analyzed cases, together with a maximum relative deviation error from the polynomial value.}
\label{tab:thick_res}
\end{table}


Based solely on the freestream conditions and the interface properties, a distribution in terms of the similarity variable, \(\eta\), can be obtained for temperature and mass fraction. Using an equation of state, such as the volume-corrected SRK equation of state, a distribution for density or \(\rho(\eta)\) is found. The values for \(\eta^*\) obtained with the fitted polynomials define the integration limit of Eq.~(\ref{eqn:yvseta}); therefore, the equation can be solved and an estimate for the mixing layer thickness is obtained. \par 

Tables~\ref{tab:thick_poly_errors1} and~\ref{tab:thick_poly_errors2} show the maximum deviation errors between the exact solution obtained from the system of ODEs and the approximate diffusion profiles obtained from the polynomial fit. Relative errors are provided for streamwise velocity, temperature and density solutions, while absolute errors are provided for the mass fraction profiles. Errors are seen to be limited well below 5\% for the analyzed cases, being negligible in some scenarios. Note the small error in estimating the velocity profile. The profiles in the gas phase collapse almost perfectly, thus generating small deviations between exact solutions and estimated values. This situation also happens for the liquid temperature profile. However, the velocity profiles in the liquid phase showed greater variations between each other. Nevertheless, the interface streamwise velocity tends to be very close to the liquid freestream velocity for all analyzed cases. Therefore, the magnitude errors between the exact solution and the polynomial estimate are very small even though \(\theta_{u,l}^{*}\) profiles do not collapse onto each other. \par 

\begin{table}[!h]
\centering
\begin{tabular}{ c c c c c }
\hline
Case & \(\Delta_{Y_l,\text{max}}\) (\%) & \(\Delta_{Y_g,\text{max}}\) (\%) & \(\Delta_{u_l,\text{max}}\) (\%) & \(\Delta_{u_g,\text{max}}\) (\%) \\
\hline
A & 0.0153 & 3.1645 & 0.1563 & 0.3438 \\
B & 0.0345 & 0.4553 & 0.0529 & 0.0228 \\
C & 0.1082 & 0.0438 & 0.0031 & 0.0157 \\
D & 0.1652 & 0.3640 & 0.0362 & 0.0190 \\
E & 0.1879 & 0.3263 & 0.0198 & 0.0397 \\
F & 0.5118 & 0.1326 & 0.0267 & 0.0250 \\
G & 0.2575 & 0.1795 & 0.0107 & 0.0430 \\
\hline
\end{tabular}
\caption{Maximum relative deviations from the exact ODE solution within the diffusion layer using the fitted polynomial together with freestream conditions and interface properties to solve for \(Y(\eta)\) and \(u(\eta)\). For the mass fraction, the maximum absolute deviation is considered.}
\label{tab:thick_poly_errors1}
\end{table}

\begin{table}[!h]
\centering
\begin{tabular}{ c c c c c }
\hline
Case & \(\Delta_{T_l,\text{max}}\) (\%) & \(\Delta_{T_g,\text{max}}\) (\%) & \(\Delta_{\rho_l,\text{max}}\) (\%) & \(\Delta_{\rho_g,\text{max}}\) (\%) \\
\hline
A & 0.0229 & 2.3583 & 0.0044 & 4.3184 \\
B & 0.0145 & 0.2429 & 0.0546 & 0.5676 \\
C & 0.0089 & 0.3398 & 0.1200 & 0.3725 \\
D & 0.0292 & 0.4943 & 0.1355 & 0.6826 \\
E & 0.0083 & 1.1313 & 0.2992 & 1.3712 \\
F & 0.0400 & 1.5627 & 0.3429 & 1.7818 \\
G & 0.0090 & 0.7723 & 0.3439 & 0.9377 \\
\hline
\end{tabular}
\caption{Maximum relative deviations from the exact ODE solution within the diffusion layer using the fitted polynomial together with freestream conditions and interface properties to solve for \(T(\eta)\) and, coupled with the volume-corrected SRK equation of state, \(\rho(\eta)\).}
\label{tab:thick_poly_errors2}
\end{table}

\begin{figure}[h!]
\centering
\begin{subfigure}{.5\textwidth}
  \centering
  \includegraphics[width=1.0\linewidth]{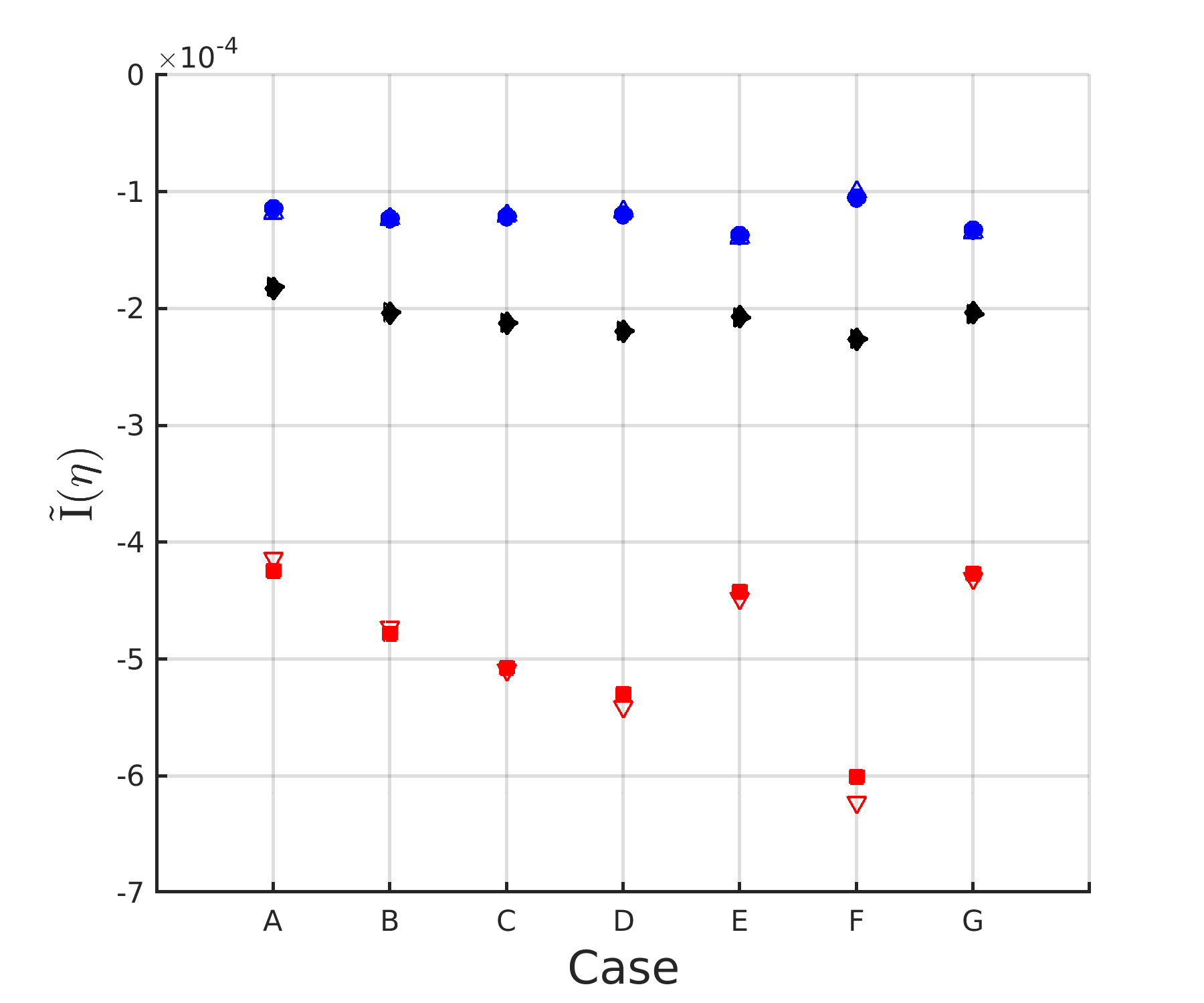}
  \caption{}
\end{subfigure}%
\begin{subfigure}{.5\textwidth}
  \centering
  \includegraphics[width=1.0\linewidth]{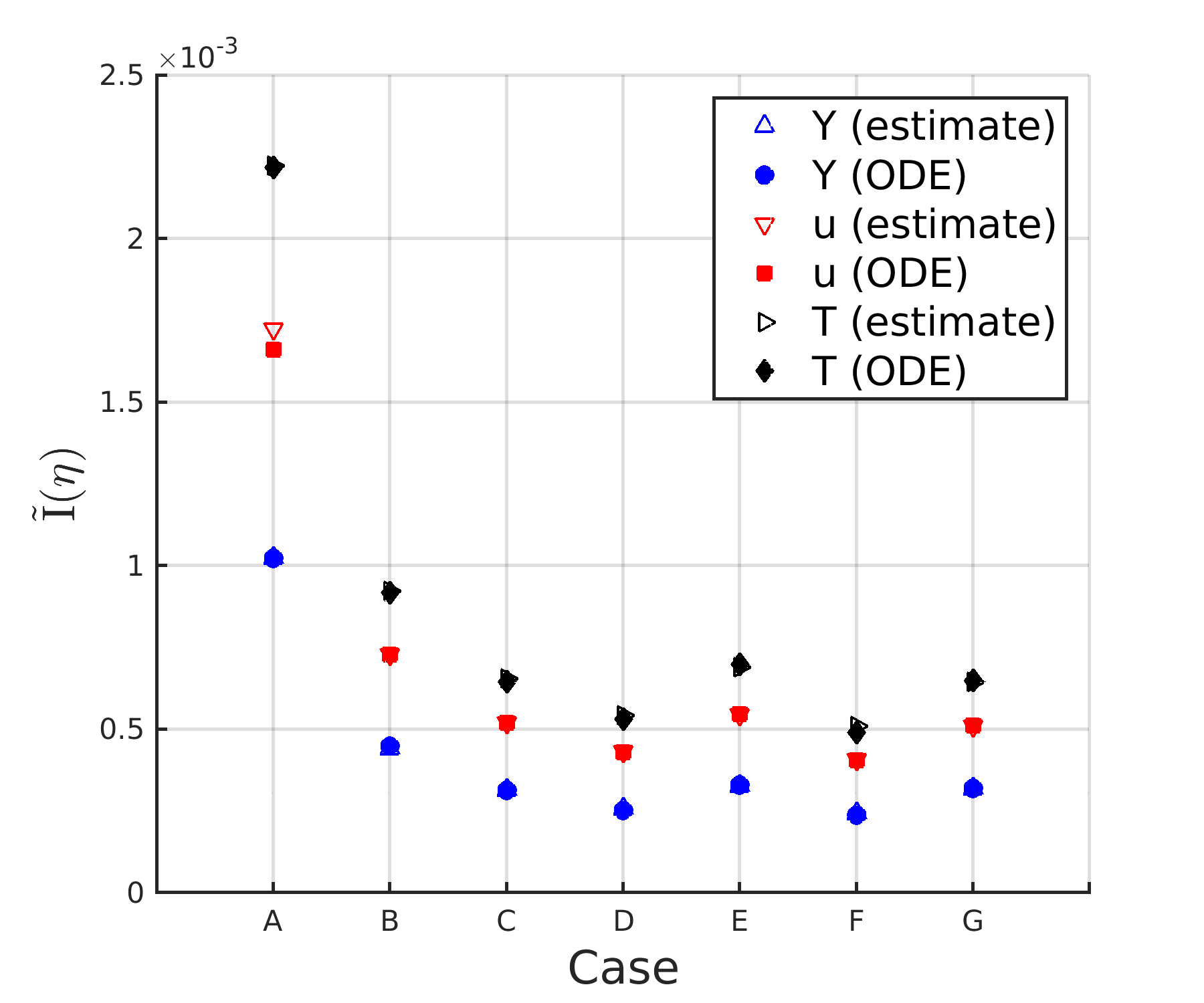}
  \caption{}
\end{subfigure}%
\caption{Comparison of the evaluation of mixing layer thickness between the exact solution of the ODE system and the approximate method based on the fitted polynomial. (a) liquid phase; (b) gas phase.}
\label{fig:thick_integral}
\end{figure}

\begin{figure}[h!]
\centering
\includegraphics[width=0.5\linewidth]{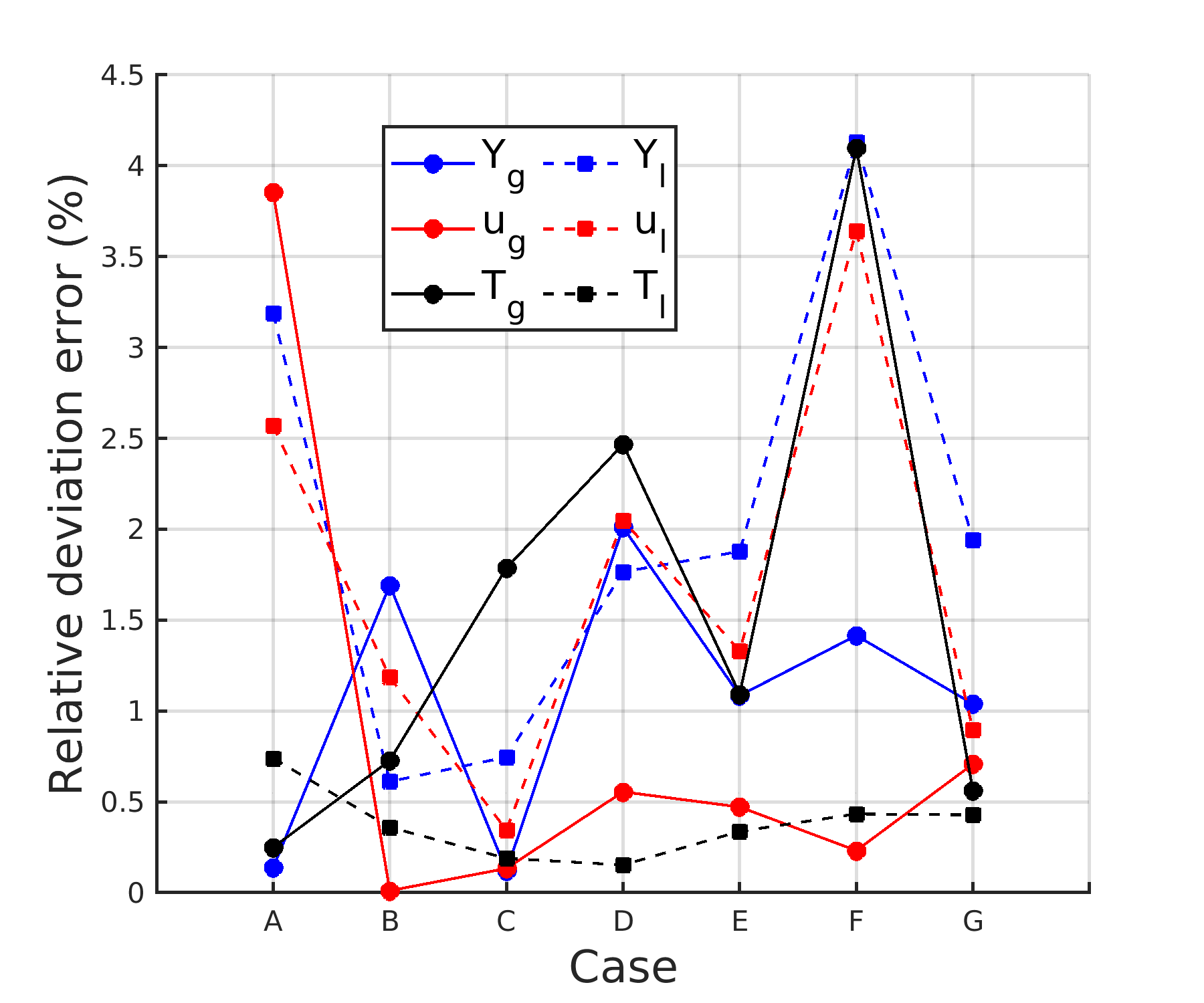}
\caption{Relative deviation error from the exact ODE solution of mixing layer thickness for the different mixing layers on each phase.}
\label{fig:thick_errors}
\end{figure}

The difference in the evaluation of the integral from Eq.~(\ref{eqn:yvseta}) or \(\tilde{I}(\eta)\), i.e., whether the exact solution of the ODE system or the approximate method is used, causes different mixing layer thicknesses. Figure~\ref{fig:thick_integral} shows the value of this integral where 99\% of the freestream value is obtained using both approaches. Overall, estimated values are in good agreement with the ODE solution. The relative deviation error of the mixing layer thickness value from the exact ODE solution is shown in Figure~\ref{fig:thick_errors}. In general, errors are small and below 2\%. However, some estimates present higher errors, but still below 5\%. \par 

Information on the mixing layer thickness can be directly obtained from Figure~\ref{fig:thick_integral}. As previously mentioned, the magnitude of \(\tilde{I}(\eta)\) is used to estimate the mixing layer thickness at any downstream location as \(\delta = \sqrt{2\bar{x}}\tilde{I}(\eta)\). It is interesting to note that while the thermal mixing layer is always larger in the gas phase, the momentum mixing layer dominates by a larger factor in the liquid phase. In both phases, the mass mixing layer is the smallest. \par 

The effects of increasing pressure on the mixing layer thickness are shown through cases A-D. The gas becomes more dense as pressure increases, reducing diffusivity. Thus, the gas mixing layer becomes thinner at larger pressures. On the other hand, the increased solubility of the gas into the liquid when pressure increases widens the momentum and thermal mixing layers in the liquid phase.

\begin{figure}[h!]
\centering
\includegraphics[width=0.5\linewidth]{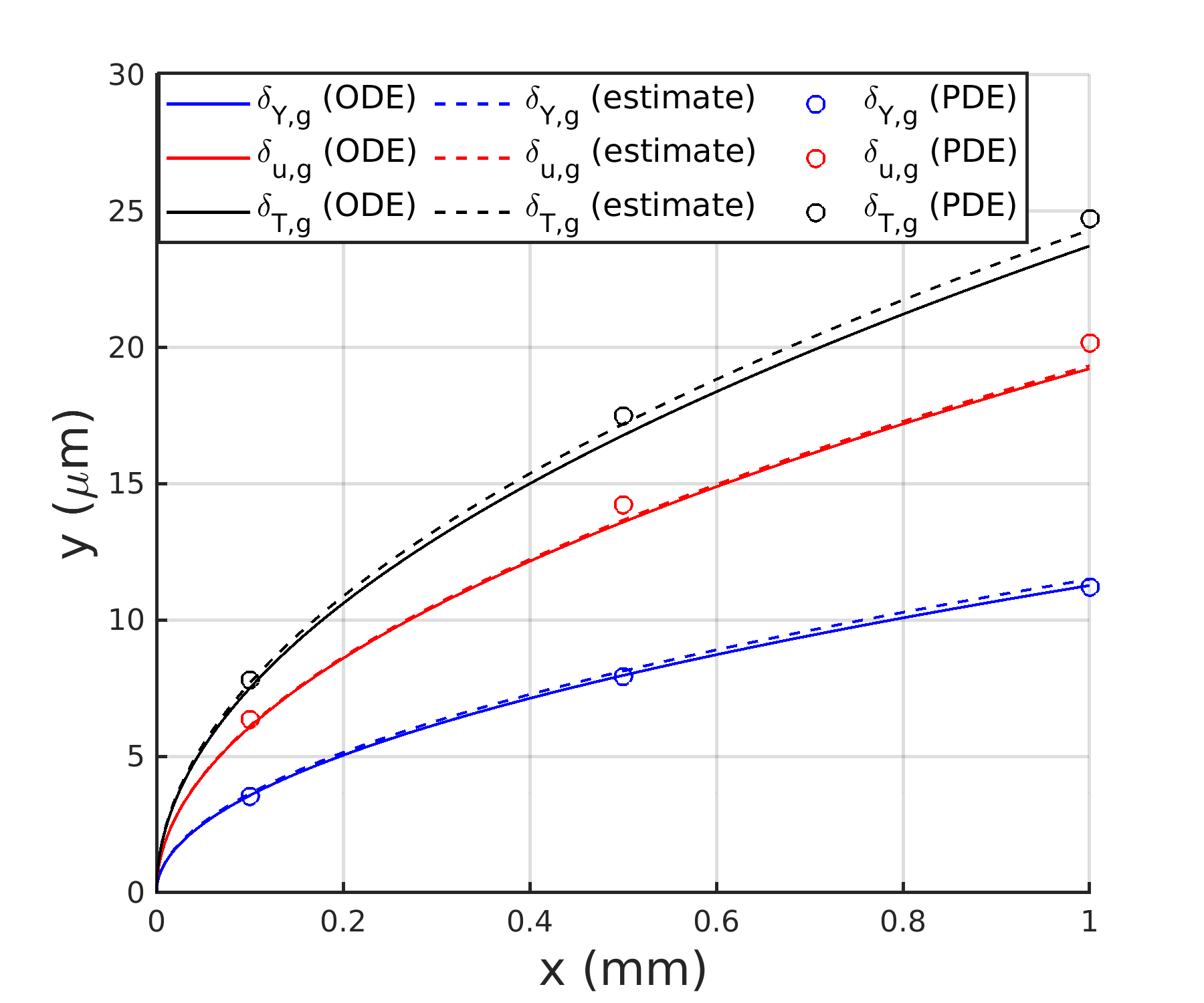}
\caption{Mass, momentum and thermal gas mixing layer thickness evolution for case D (see Table~\ref{tab:thick_case}). A comparison between the ODE solution, the polynomial approximation and the PDE solution is presented.}
\label{fig:thick_delta}
\end{figure}

Figure~\ref{fig:thick_delta} presents the evolution of the mass, momentum and thermal gas mixing layers for case D (i.e., a very high pressure case). The mixing layer thickness is computed using Eq.~(\ref{eqn:yvseta}) and a comparison between the solution obtained with the exact ODE solution and the polynomial estimation is shown. Furthermore, data points from the PDE solution obtained in~\cite{davis2019development} are overlapped. The mixing layer thicknesses based on the PDE solution are obtained as the transverse location, \(y\), where \(\theta_{Y,g}\), \(\theta_{u,g}\) and \(\theta_{T,g}\) become 0.99.

The evolution of the different mixing layer thicknesses is in accordance for the three methods used in this comparison. Errors between the ODE solution and the estimated solution based on the polynomial fit have already been discussed (see Figure~\ref{fig:thick_errors}). For both of them, the mixing layer must grow with downstream distance as \(\sqrt{x}\), which is corroborated with the PDE results from~\cite{davis2019development}. However, deviations between the self-similar model and the mixing layer equations are seen, especially for the evolution of the streamwise velocity and the temperature mixing layers, with errors of the same order as the relative deviations between the ODE solution and the approximate solution (\(< 5\%\)). \par 

Overall, the polynomial fitting approach is valid for the type of mixture and range of thermodynamic states evaluated in this work. It simplifies the evaluation of the mixing layer thickness with reasonable accuracy and only relies on the freestream conditions and the interface solution, which could be tabulated from previous simulations. Thus, it is not necessary to solve the system of ODEs or the system of PDEs for all cases and an estimate of the mixing layer thicknesses can be readily obtained. It is expected that other problem configurations not used in the development of the correlations could show larger errors than the ones reported for cases A-G.

\section{Summary and conclusions}
\label{summary}

The non-ideal, two-phase, laminar mixing-layer equations have been reduced to a system of ordinary differential equations in terms of a similarity variable, \(\eta\). The similarity transformation follows classical techniques used in compressible flows~\cite{williams2018combustion} and has been generalized to be implemented with any non-ideal thermodynamic model for the equation of state and phase equilibrium. In this work, the high-pressure thermodynamic model is based on a volume-corrected Soave-Redlich-Kwong cubic equation of state and other fundamental thermodynamic principles, coupled with high-pressure correlations to evaluate transport coefficients and phase equilibrium~\cite{davis2019development}. 

Good agreement between the self-similar solution and the solution of the system of partial differential equations~\cite{davis2019development} is obtained, proving the validity of the self-similar approach. This is an important step towards reducing the complexity of the analysis of supercritical two-phase flows, while still capturing the main physics involved (e.g., enhanced diffusion in the liquid phase, phase change reversal at the interface). This type of approach can be helpful when used to implement realistic initial conditions to more complex flow simulations (i.e., high-pressure atomization). The results show the existence of two phases at pressures above the critical pressure of any chemical component. It is seen that condensation can occur even though the gas is hotter and heat is conducting into the liquid, as reported in previous works~\cite{poblador2018transient,poblador2019axisymmetric}.

A correlation using a ninth order polynomial is used to fit data from various problem setups into a generalized approach to represent the mixing layer evolution. This reduces the set of ordinary differential equations to a set of functions which depend only on the freestream conditions, the interface equilibrium solution and a non-ideal equation of state. Good estimates of the evolution of the mixing layer thickness on both sides of the interface are obtained from this simplified model, with errors well below 5\% in most cases.

\section*{Conflict of interest}

The authors declared that there is no conflict of interest.

\section*{Acknowledgments}

The authors are grateful for the support of the NSF grant with Award Number 1803833 and Dr. Ron Joslin as Scientific Officer.


\newpage
\bibliography{journal_bib}

\end{document}